\documentclass[aps,prx,twocolumn,floatfix,nobibnotes]{revtex4-2}

\usepackage{hyperref}

\usepackage{amsmath}
\usepackage{braket}
\usepackage{graphicx}
\usepackage{graphics}
\usepackage{color}
\usepackage{stackengine}
\usepackage{array}
\usepackage{ulem}
\usepackage{verbatim}
\usepackage{soul}
\usepackage{float}

\begin{document}

\title{Interplay of disorder and interactions in a flat-band supporting diamond chain}

\author{Nilanjan Roy}
\affiliation{Department of Physics, Indian Institute of Science Education and Research, Bhopal, Madhya Pradesh 462066, India}
\author{Ajith Ramachandran}
\affiliation{Department of Physics, Indian Institute of Science Education and Research, Bhopal, Madhya Pradesh 462066, India}
\author{Auditya Sharma}
\affiliation{Department of Physics, Indian Institute of Science Education and Research, Bhopal, Madhya Pradesh 462066, India}
\date{\today}

\begin{abstract}
  We systematically study the effect of disorder and interactions on a
  quasi-one dimensional diamond chain possessing flat bands. Disorder
  localizes all the single particle eigenstates - while at low
  disorder strengths we obtain weak flat-band based localization
  (FBL), at high disorder strengths, we see conventional Anderson
  localization (AL). The compactly localized (CL) eigenstates of flat
  bands show a persisting oscillatory recurrence in the study of
  single-particle wavepacket dynamics. For low disorder a damped
  oscillatory recurrence behavior is observed which is absent for
  high disorder. Non-interacting many particle fermion states also
  follow the same trend except showing a delocalizing tendency at
  intermediate disorder due to the fermionic statistics in the
  system. As interactions are switched on, for the
    finite-sizes that we are able to study, a non-ergodic `mixed
    phase' is observed at low disorder which is separated from the MBL
    phase at high disorder by a thermal phase at
    intermediate-disorder. A study of many-body nonequilibrium
  dynamics reinforces these findings. \end{abstract}

\maketitle

\section{Introduction}
 Translationally invariant Hermitian systems possessing {\it flat
  bands} (FBs) \cite{flach2014detangling, toikka2018necessary,derzhko2015strongly} exhibit
remarkable properties, that emanate from the compactly localized
eigenstates (CLS) associated with them. These states are called
compact because they reside on a finite volume of the lattice and
strictly vanish
elsewhere~\cite{maimaiti2017compact,ramachandran2017chiral,mukherjee2015observation1,travkin2017compact,klembt2017polariton}.
The appearance of flat bands has been theoretically established for
lattice models in one~\cite{derzhko2010low},
two~\cite{takeda2004flat} and even three~\cite{goda2006inverse}
dimensions and realized experimentally with ultracold
atoms~\cite{jo2012ultracold,taie2015coherent}, in photonic crystal
waveguides~\cite{vicencio2015observation,mukherjee2015observation,mukherjee2018experimental},
and exciton-polariton condensates~\cite{masumoto2012exciton}. The
physics of such flat-band systems thus hinges on the twin properties
of large-scale degeneracy in the dispersion, coupled with extreme
localization of associated eigenstates.

Only a very delicate tuning allows for a system to admit exactly flat
bands, and the question of how the properties of such systems are
modified under different kinds of perturbations is of great
interest~\cite{pal2018nontrivial,zegadlo2017single}. The effect of
disorder, and interactions, is of particular importance because of the
pervasive appearance of each of these factors in real systems.  A set
of studies has been conducted on the effect of disorder in FB systems
and the localization-delocalization transitions therein have been
understood to some extent
\cite{shukla2018disorder1,bodyfelt2014flatbands,leykam2017localization,goda2006inverse,danieli2015flat,bodyfelt2014flatbands,leykam2017localization,radosavljevic2017light,gneiting2018lifetime}.
In particular, Ref.~\cite{shukla2018disorder1} points
  towards the possibility of a different origin for the different
  phases that emerge as disorder strength is changed. The fate of the
  FB disordered system when interactions are turned on is of interest
  too~\cite{sierra1999diagonal,jvidal2000prl,vidal2001disorder,gulacsi2007exact,mondaini2018pairing,kuno2019flat}. A
  complexity parametric formulation which could be applicable to a
  wide range of many-particle interacting flat band systems has been
  considered in Ref.~\cite{shukla2018disorder1}. It is also worth
  mentioning that localization properties in many-body flat
  band systems have been recently explored in the context of disorder-free
  systems
  ~\cite{kuno2019flat,danieli2020caging,danieli2020caging2,danieli2020many}.
In the present article, we carry out a systematic study of the effect
of disorder and interactions on the FB states in a particular system,
namely the diamond chain. The investigation of highly degenerate
systems under disorder and interactions is challenging, because
oftentimes, the degeneracy is only partially lifted, and the study of
a single observable is inadequate to explore the impact. Disorder and
interactions detune the energy levels as well as modify the
characteristics of eigenstates. We therefore conduct a thorough
investigation of both the energy spectrum and eigenstates.

We consider a quasi-one dimensional disordered interacting diamond
chain (Fig.~\ref{lattice}(a)). The band structure of the clean,
noninteracting diamond chain admits only three flat bands and the
system remains essentially in the insulating phase.  All the
eigenstates in the system are compactly localized on
two unit cells and any perturbation is expected to couple these states
and increase the volume spanned. We observe three distinct phases of
localization in the system: compact localization ($CL$) at zero
disorder, weak flat band-based localization ($FBL$) at low disorder,
and strong Anderson localization ($AL$) at high disorder with the
localization strength $CL>AL>FBL$. The non-interacting many-fermion
system also possesses the $FBL$ and $AL$ phases at low and high
disorder strengths respectively. In the simultaneous
  presence of disorder and interactions, we find evidence for three
  distinct phases: a many body localized (MBL) phase at high disorder,
  a thermal phase at intermediate disorder, and a nonergodic `mixed
  phase' at low disorder.

To explore the effect of disorder on single particle states, we
consider the level-spacing statistics which is obtained from the
inspection of eigenvalues and a variety of quantities: $IPR$,
fidelity, von Neumann entropy, and Shannon entropy which are obtained
from eigenstates. We also explore the dynamics of the system in terms
of revival probability and the imbalance parameter.  In the context of
noninteracting many fermion states, we study entanglement entropy in
the static case and growth dynamics of the entanglement entropy and
imbalance parameter. Finally, we study level statistics, many-body IPR
and nonequilibrium dynamics of the revival probability, entanglement
entropy and imbalance parameter to obtain insights into the properties
of many-particle states in the interacting system.

The article is arranged as follows. Section II describes the details
of the model. Section III is devoted to the study of disorder in
flat band systems which predicts various phases with different
localization properties. We then discuss many-particle states with
interactions switched off and explore the phases in the presence of
disorder in Section IV. In section V the effect of
disorder on the interacting many body system is discussed. The results
are summarized in section VI.

\section{The model}
\begin{figure}
 \includegraphics[width=0.9\columnwidth]{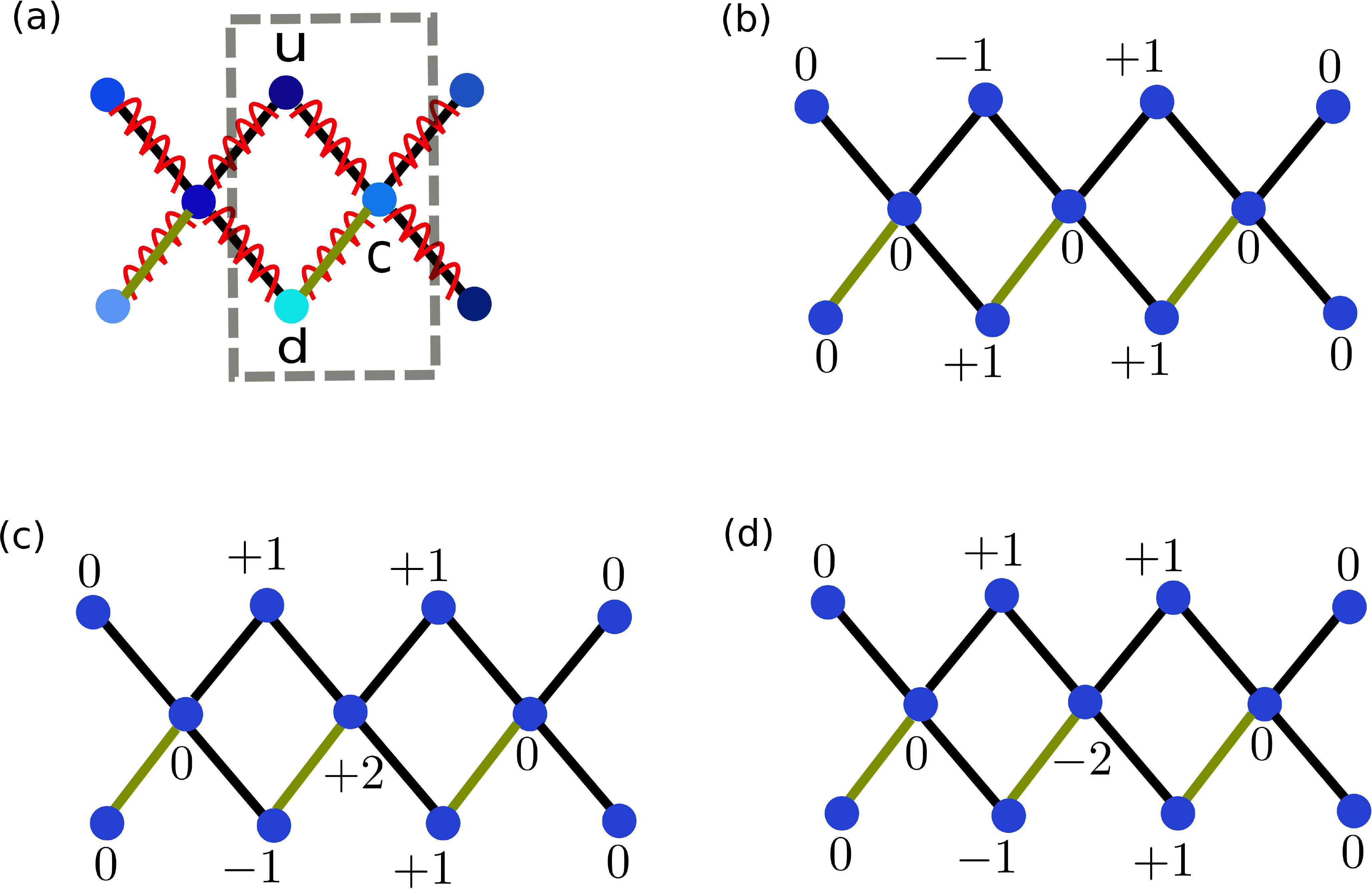}
 \caption{ (a) Schematic of the diamond chain: unit cell contains
   three sites: up (u), down (d) and central (c) as shown in a
   box. Nearest neighbor interaction $V$ is represented by red wiggly
   lines and on-site disorder by the variation in the shades of the
   color of the dots. (b-d)The CLS in the noninteracting,
   disorder-free limit corresponding to $E=0$, $E=-2J$ and $E=2J$
   respectively, where the numbers assigned to different sites are the
   probability amplitudes for a single particle.}
 \label{lattice}
 \end{figure}  

The Hamiltonian of the quasi-one dimensional diamond chain can be written as
\begin{eqnarray}
\label{ham}
\hat{H}=\hat{H}_{hop} + \hat{H}_{os} + \hat{H}_{int},
\end{eqnarray}
where
\begin{eqnarray}
\hat{H}_{hop}&&=-J\sum\limits_{i=1}^{N/3} (\hat{u}_i^\dagger\hat{c}_i - \hat{c}_i^\dagger\hat{d}_i + \hat{c}_i^\dagger\hat{u}_{i+1} + \hat{c}_i^\dagger\hat{d}_{i+1}) + h.c.,\nonumber\\
\hat{H}_{os}&&=\sum\limits_{i=1}^{N/3} (\zeta_i^u\hat{u}_i^\dagger\hat{u}_i + \zeta_i^c\hat{c}_i^\dagger\hat{c}_i + \zeta_i^d\hat{d}_i^\dagger\hat{d}_{i}),\nonumber\\  \hat{H}_{int}&&=V\sum\limits_{i=1}^{N/3} (\hat{u}_i^\dagger\hat{u}_i\hat{c}_i^\dagger\hat{c}_i +   \hat{c}_i^\dagger\hat{c}_i\hat{d}_i^\dagger\hat{d}_i + \hat{c}_i^\dagger\hat{c}_i\hat{u}_{i+1}^\dagger\hat{u}_{i+1} +\nonumber\\&& \hat{c}_i^\dagger\hat{c}_i\hat{d}_{i+1}^\dagger\hat{d}_{i+1}) + h.c.
\end{eqnarray}
Here $\hat{u}_i^\dagger$, $\hat{c}_i^\dagger$ and $\hat{d}_i^\dagger$
are the fermionic creation operators acting at the u (up), c (center)
and d (down) sites respectively in the $i\textsuperscript{th}$ unit
cell, as schematically shown in Fig.~\ref{lattice}(a). The total number of sites is denoted by $N$ and is taken as a multiple of $3$, because of the unit cell structure. The hopping
amplitude $J$ and interaction $V$ are nonzero for nearest neighbors
(represented as sites connected by lines in Fig. \ref{lattice}(a)) and
zero otherwise. $\zeta_i^\alpha$ $(\alpha=u,c,d)$ denotes the strength
of the on-site disorder chosen from a uniform random distribution
$[-\Delta,\Delta]$.

\subsection{Compact Localization}
In the absence of disorder and interaction
  ($\zeta_i^\alpha = 0, V = 0$), the band structure of the diamond
  chain contains three flat bands at $E = 0,\pm 2J$. Lattices with all
  bands flat are rare and the diamond chain is carefully tuned to
  admit this intriguing feature~\cite{jvidal2000prl}. Conventional
  wisdom implies that the eigenstates of a translationally invariant
  system are Bloch states which span the entire lattice, and which
  carry the delocalization characteristics associated with the
  dispersion of the energy bands. The dispersionless property of the
  flat bands on the one hand implies zero group velocity for the
  states and on the other hand provides a convenient representation
  where the eigenstates are compactly localized. Due to the large
  scale degeneracy, the basis can be chosen in multiple ways, however
  one can always look for a basis such that the eigenstates occupy the
  smallest volume in the lattice. Once one such state has been
  identified, the translational invariance of the Hamiltonian can be
  utilized to obtain others states belonging to the flat
  bands. Although suitable linear combinations of the compact
  localized eigenstates can of course recover the Bloch state
  representation, it is the properties of the CLS that make these
  systems most interesting. 

 The states belonging to all three flat bands in the
  diamond chain can be represented as different compact localized
  states. Fig. \ref{lattice}(b-d) shows one CLS belonging to each band
  where $\pm1$ corresponds to the amplitudes (not normalized) of the
  states on the sites. The origin of compactness of these states is
  the destructive interference at the neighboring sites which
  effectively confines the states within a finite volume. For the
  diamond chain Hamiltonian in Eq. \ref{ham} with no disorder and
  interaction, the smallest possible volume for a CLS is two unit
  cells as shown in Fig. \ref{lattice}(b-d). The other states
  corresponding to each flat band can be obtained by shifting these
  states along the lattice utilizing the translational invariance of
  the Hamiltonian. It can be shown that the states identified in this
  particular case lack orthogonality although they form a complete
  set. The localization properties of the CLS are special as they are
  observed in a translationally invariant system and span very small
  volume of the lattice with an abrupt boundary without any tail in
  spread.  This is to be contrasted with Anderson localization (AL)
  observed in disordered systems where the states exhibit an
  exponential tail. Hence the CLS possess stronger localization
  compared to that seen in AL.


\section{Effect of disorder on single-particle properties}
In this section, we analyze the eigenvalues and single-particle
eigenfunctions to study the effect of disorder. The
level-spacing ratio and level-spacing distribution are obtained from
the single-particle eigenvalues while the inverse participation ratio
($IPR$), fidelity, von Neumann entropy and Shannon entropy are
calculated from the eigenfunctions. Nonequilibrium dynamics throws
further light on some of these single-particle properties.
\subsection{Level-spacing statistics}
\begin{figure}
\centering
\includegraphics[width=0.494\columnwidth]{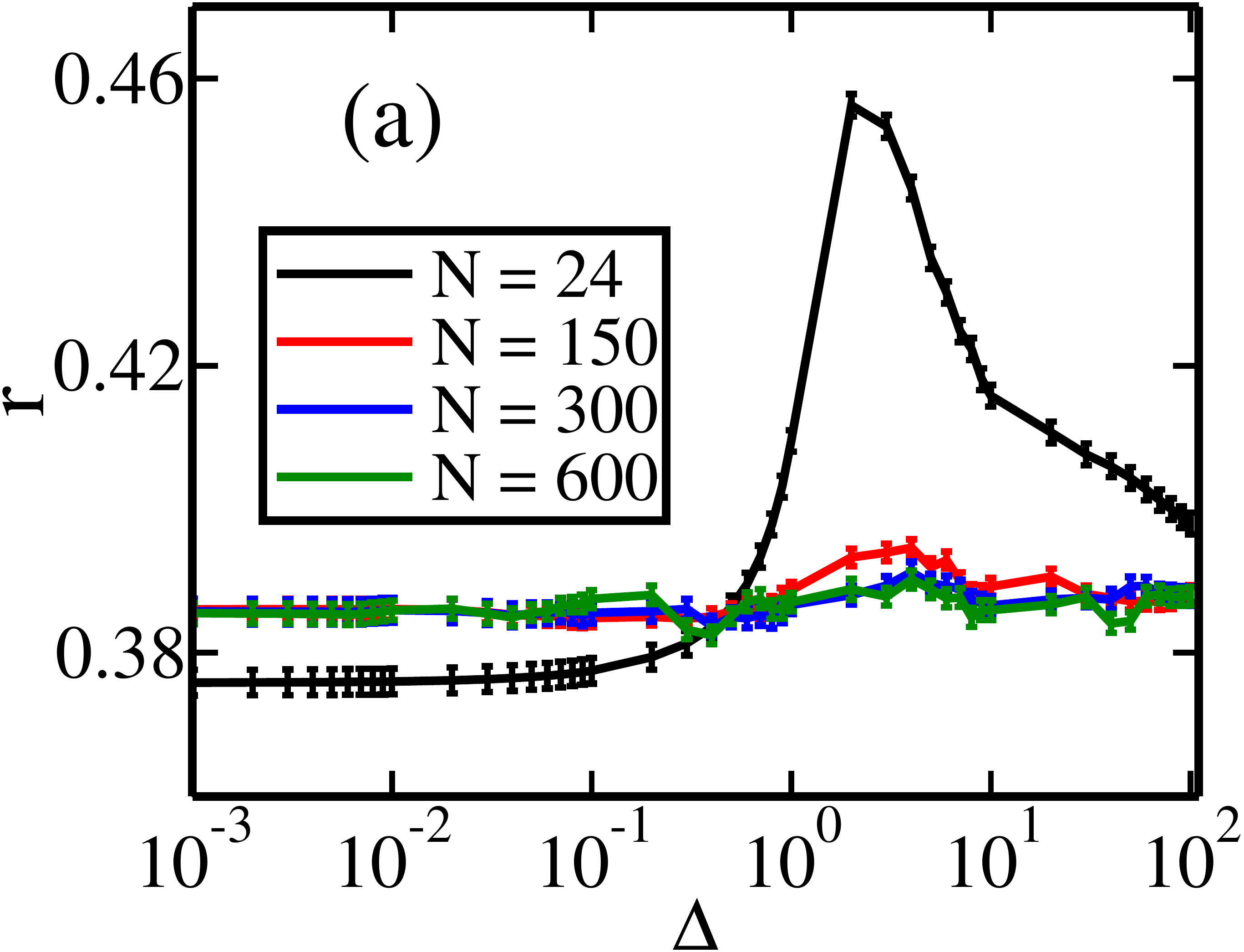}
\includegraphics[width=0.494\columnwidth]{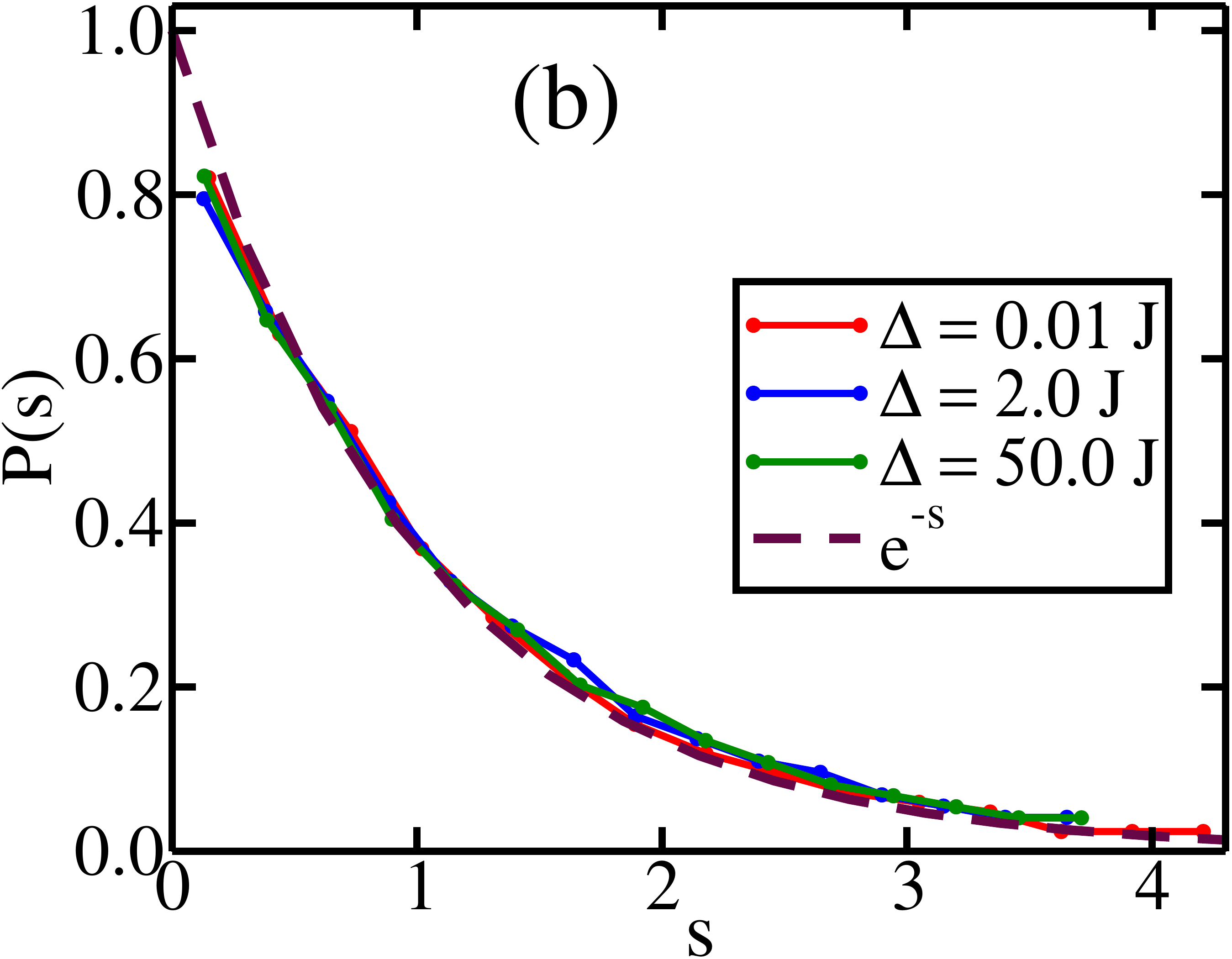}
\caption{(a) Level-spacing ratio $r$ vs disorder strength $\Delta$ (in units of $J$) for
  increasing values of system size $N$. (b) Level-spacing distribution
  for increasing disorder strength for $N=300$. The dimensionless
  level-spacings $s$ are obtained by dividing the original
  level-spacings by the mean level-spacing of the spectrum. The number of disorder realizations varies for different system sizes, but all of them have at least $100$ samples of disorder.}
\label{single_levelstat}
\end{figure}

The level-spacing ratio $r$ is calculated from the eigenvalues of the disordered Hamiltonian and is defined as
\begin{eqnarray}
r=\frac{1}{N-1}\sum\limits_{k=1}^{N-1}\frac{\min[s_{k},s_{k+1}]}{\max[s_{k},s_{k+1}]},
\label{eq_level}
\end{eqnarray}
where energy level-spacing $s_k=E_{k+1}-E_k$ with single particle
energies $E_k$'s organized in the ascending order for a system size of
$N$ sites and with an implicit average over realizations of
random disorder assumed. In the delocalized and localized phases $r$ is
expected to be approximately $0.528$ and $0.386$, respectively. 
Fig.~\ref{single_levelstat}(a) shows the dependence of $r$ on the strength of disorder, for various system sizes. 
For a small system size $N=24$, $r$ is close to $0.386$ at very small
disorders $\Delta$ ($\Delta<<J$) and seems to show a tendency to deviate from $0.386$ for disorder strength
comparable to hopping strength $J$. However, this is seen to be a finite-size effect and for larger system sizes, the
level-spacing ratio $r$ remains at $0.386$ in the full range of the disorder
strength $\Delta$. This indicates that the system is localized in
the presence of disorder. In the absence of disorder, the level-spacing ratio becomes unintelligible due to the
massive degeneracy.

We also study the energy level-spacing distribution function $P(s)$,
where ${s_i}$'s are made dimensionless by dividing the original level
spacings by the mean level-spacing of the spectrum such that $\int
P(s) ds=1$. For a real Hamiltonian, in the delocalized phase the
spectral statistics is $GOE$, i.e., $P(s)=\frac{\pi}{2}s
e^{-\frac{\pi}{4}s^2}$, which yields $ r\approx0.528$. In the
localized phase, the spectral statistics is Poissonian $P(s)=e^{-s}$
yielding $ r\approx0.386$. In Fig.~\ref{single_levelstat}(b), we
observe that the energy level-spacing distribution follows Poissonian
spectral statistics for all disorder strengths confirming that the
system is localized irrespective of the strength of disorder.
The introduction of disorder into the insulating
  all-band flat diamond chain keeps the system in the insulating phase
  itself. We further explore the localization properties of the system
  at different disorder strengths in detail through a study of the
  eigenstates.

\subsection{Inverse participation ratio}
The inverse participation ratio ($IPR$) is defined as
\begin{equation}
I_k = \sum_{i=1}^{\frac{N}{3}}\sum_{\alpha={u,c,d}} |\psi_k(i,\alpha)|^4
\end{equation}
where the $k\textsuperscript{th}$ normalized single-particle
eigenstate $\ket{\psi_k}=\sum_{i,\alpha}\psi_k(i,\alpha)
\ket{i,\alpha}$ is written in terms of the Wannier basis
$\ket{i,\alpha}$, representing the state of a single particle
localized at the site $\alpha$ $(\alpha=u,c,d)$ in the
$i\textsuperscript{th}$ unit cell of the lattice. For a perfectly
delocalized eigenstate, $I_k$ scales as $1/N$ while for a single-site
localized eigenstate $I_k=1$ and at criticality $I_k$ shows
intermediate behavior.
\begin{figure}
    \centering
    \includegraphics[width=0.7\columnwidth]{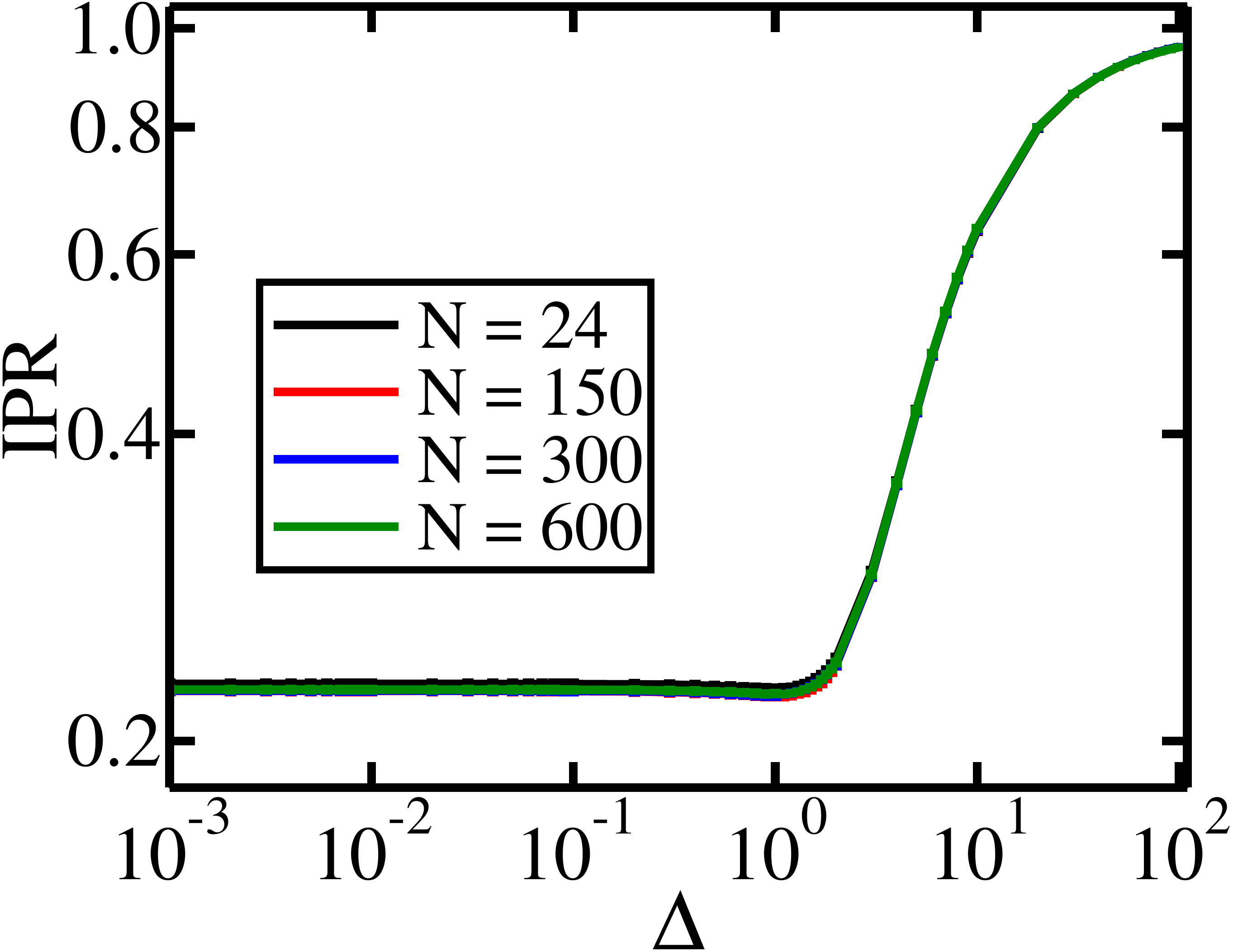}
    \caption{The average $IPR$ of all the single-particle eigenstates vs disorder for increasing values of system size $N$.  The number of disorder realizations varies for different system sizes, but all of them have at least $100$ samples of disorder.}
    \label{single_ipr}
\end{figure}

The $IPR$ averaged over all the eigenstates plotted in
Fig. \ref{single_ipr} reveals two distinct localization phases: a weak
flat band based localization ($FBL$) at low disorder with low $IPR$
and a strong Anderson localization ($AL$) at high disorder with an
$IPR$ close to one. We observe a cross-over from the $FBL$ to $AL$ at
$\Delta=2J$. In the low disorder region, the states within one band
hybridize leading to the $FBL$ and as disorder strength increases
states from all bands hybridize leading to the conventional
$AL$. Thus, even though the system is localized, there
  could be substantial difference in the localization properties at
  different disorder strengths. 
  How different are the states at low
  and high disorder regions is an interesting question and we explore
  this through the study of fidelity in the next section.

 \subsection{Fidelity}
\begin{figure}
  \centering
  \includegraphics[width=0.7\columnwidth]{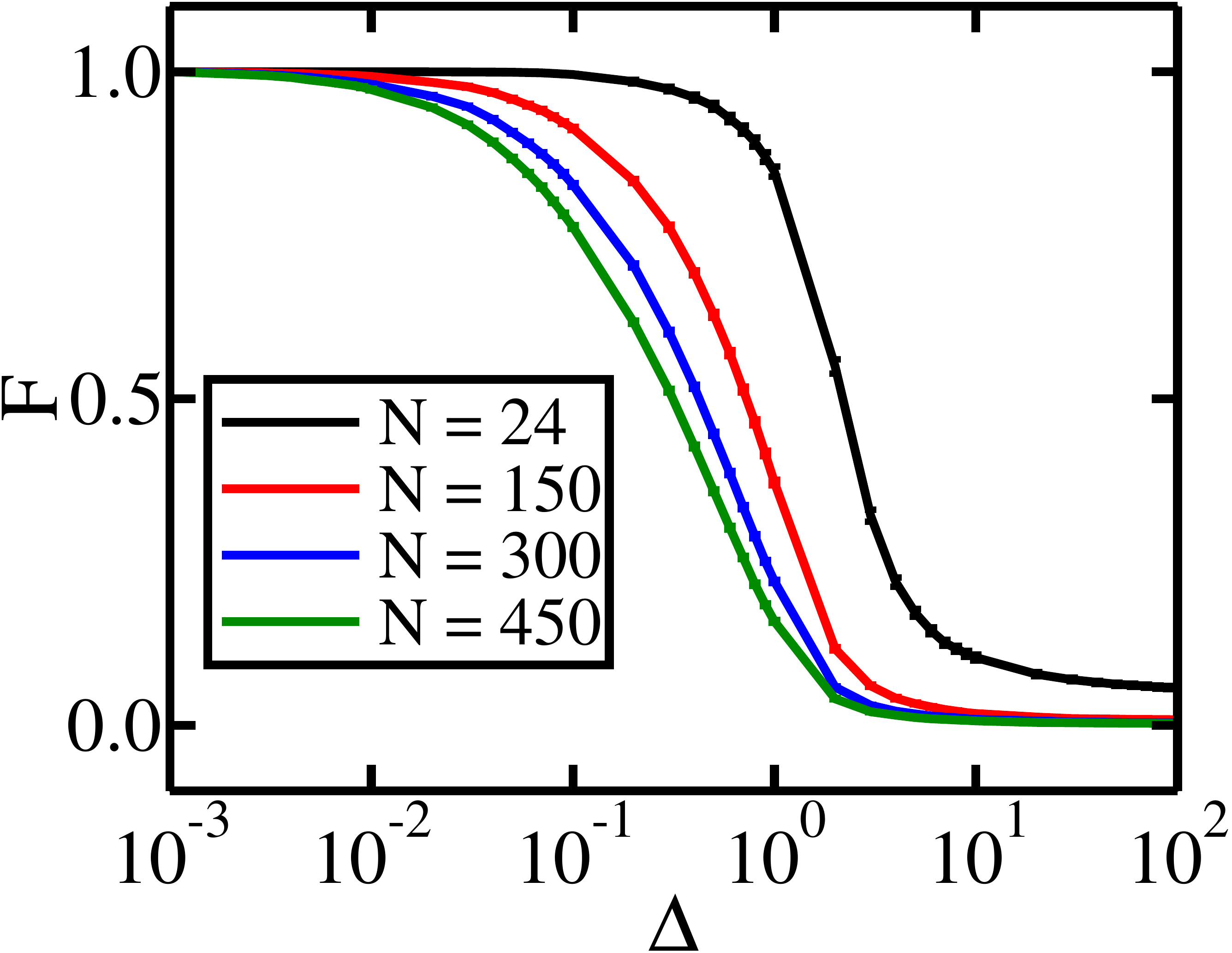}
  \caption{Spectrum averaged fidelity $F$ as a function of the disorder strength $\Delta$ (in units of $J$) for increasing system sizes $N$. The number of disorder realizations varies for different system sizes, but all of them have at least $100$ samples of disorder.}
  \label{single_fidel}
  \end{figure}

To understand the nature of the single-particle eigenstates in the
presence of disorder, the average fidelity or overlap between the
eigenstates corresponding to $\Delta=0.001J$ and those corresponding
to higher values of $\Delta>0.001J$ is calculated. Fidelity between
the $k\textsuperscript{th}$ eigenstates corresponding to two values of
$\Delta$ is defined as
 \begin{eqnarray}
 F_{12}^k=|\bra{\psi_k(\Delta_1)}\psi_k(\Delta_2)\rangle|^2.
 \end{eqnarray}
 The two different phases of localization $FBL$ and $AL$ are evident
 from the spectrum averaged fidelity $F=\langle F^k\rangle$ shown in
 Fig.~\ref{single_fidel}. The characteristics of eigenstates in the
 $FBL$ phase and the $AL$ phase are different from one
 another. Fidelity provides a comparison between each eigenstate at
 different disorder strengths. One main observation we make through
 this study is that, even though the $IPR$ in the weak $FBL$ states
 are obtained to be independent of disorder
 (Fig.~\ref{single_ipr}(a)), a closer examination of fidelity
 (Fig.~\ref{single_fidel}) reveals that the nature of eigenstates in
 fact changes substantially with disorder strength. This is in
 contrast to conventional $AL$ where the eigenstates are frozen such
 that the variation of $F$ with respect to disorder strength is
 minimal. Moreover, in the $FBL$ phase, the fidelity seems to have a
 strong dependence on system size $N$, in sharp contrast to the $AL$
 phase.
  
\subsection{von Neumann entropy and Shannon entropy}  
The von Neumann entropy is a good measure to explore localization phenomena in quantum systems~\cite{gong}. In this work we aim to calculate von Neumann entropy connected to a single-site. A single particle has two local states $\ket{0}_{i,\alpha}$ and $\ket{1}_{i,\alpha}$ at the site $\alpha$ of the $i$\textsuperscript{th} unit cell and hence the local density matrix  $\rho_{k}^{i\alpha}$ for the $k\textsuperscript{th}$ eigenstate can be written as~\cite{SChakravarty}
  \begin{eqnarray}
  \rho_{k}^{i\alpha}= |\psi_k(i,\alpha)|^2 \ket{1}_{i\alpha}\bra{1}_{i\alpha} + (1 - |\psi_k({i,\alpha})|^2) \ket{0}_{i\alpha}\bra{0}_{i\alpha}.\nonumber\\
  \end{eqnarray}
  The von Neumann entropy associated with site $i$ is given by
  \begin{eqnarray}
  S_{k}^{i\alpha} = &&-|\psi_k({i,\alpha})|^2 \log_2(|\psi_k({i,\alpha})|^2) \nonumber \\
  &&- (1 - |\psi_k({i,\alpha})|^2) \log_2 (1 - |\psi_k({i,\alpha})|^2).
  \end{eqnarray}
\begin{figure}
    \centering
    \includegraphics[width=0.49\columnwidth]{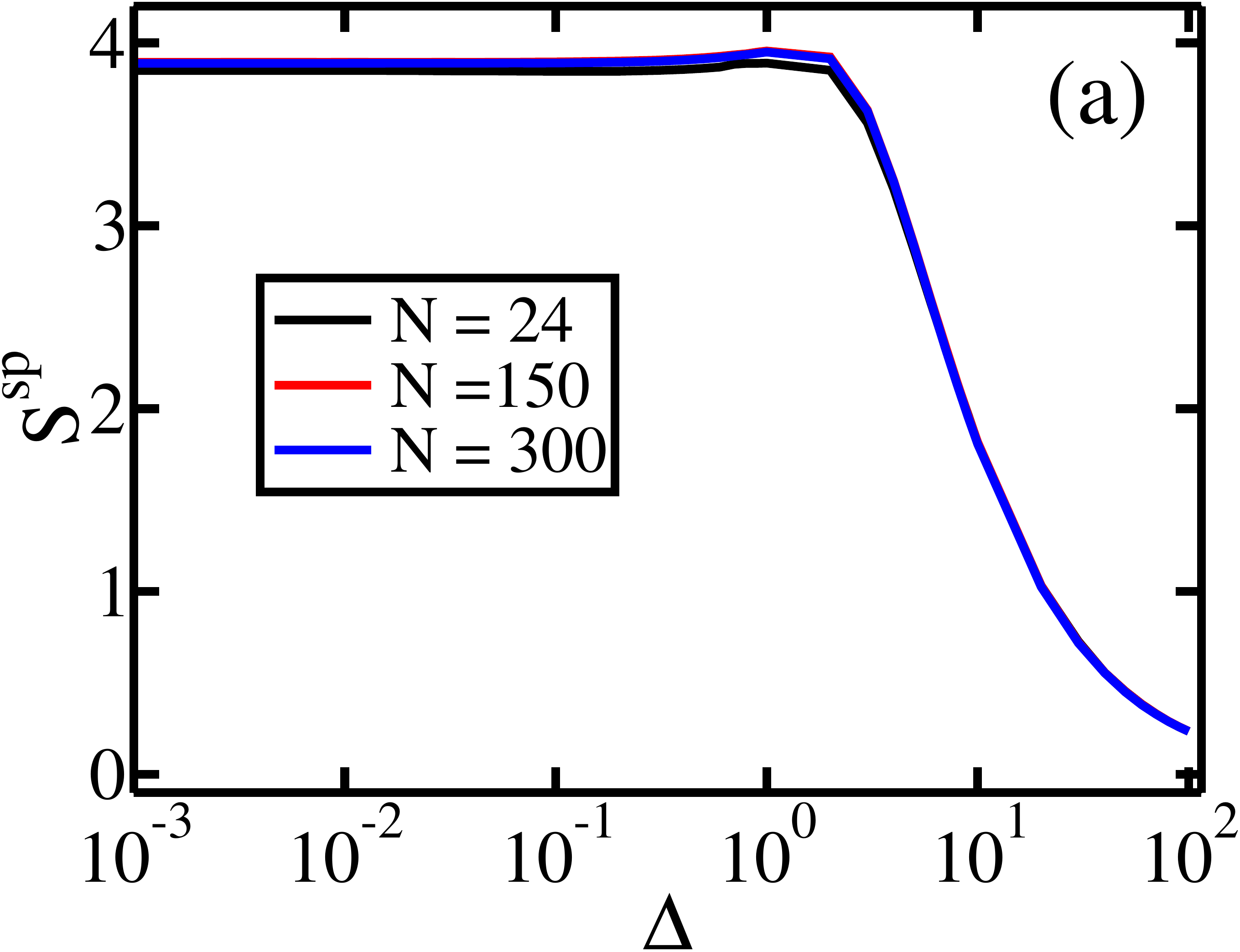}
     \includegraphics[width=0.49\columnwidth]{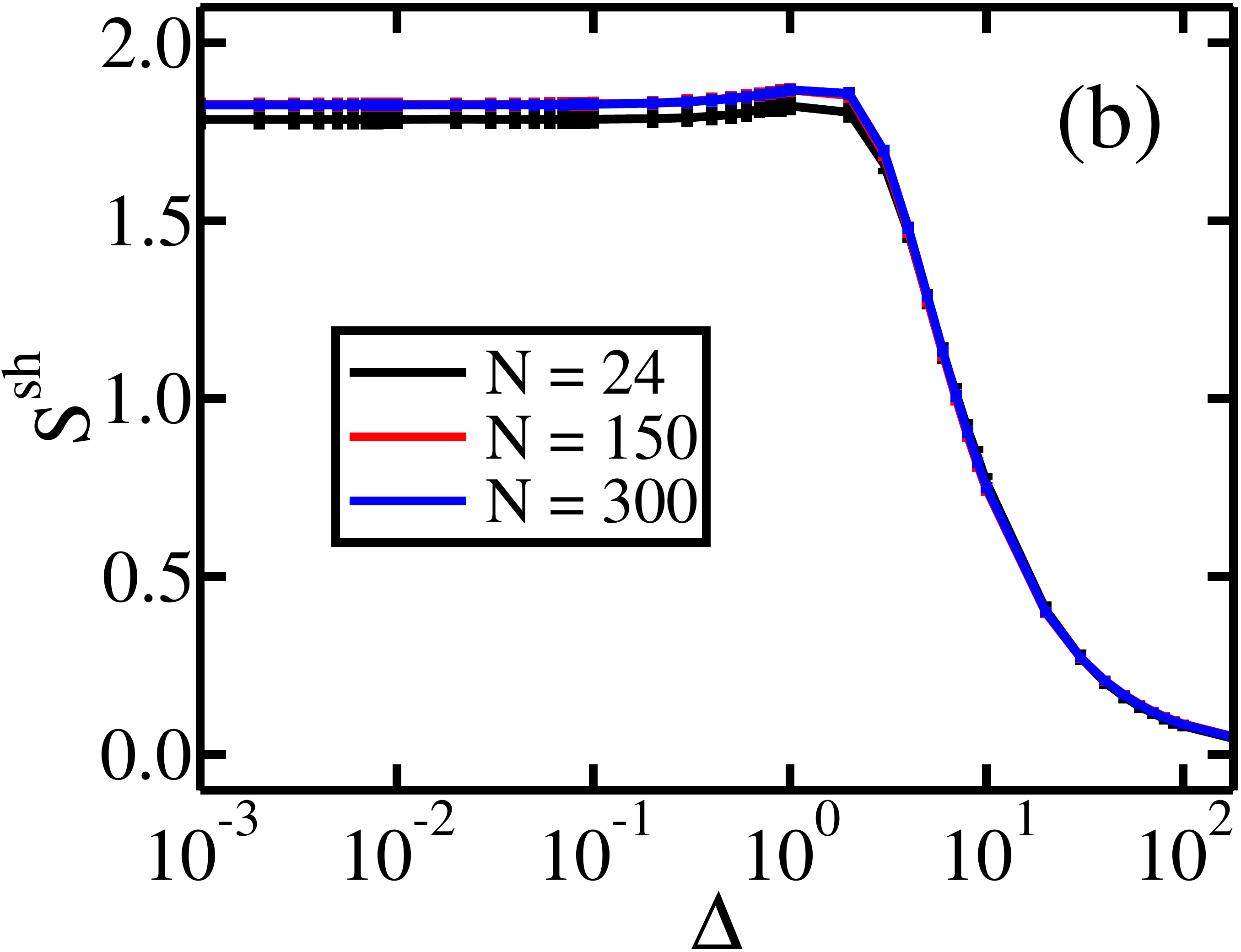}
    \caption{(a) The von Neumann entropy averaged over all the eigenstates vs disorder for increasing system sizes $N$. (b) Shannon entropy $S^{sh}$, averaged over all the
    eigenstates as a function of disorder strength $\Delta$ (in units of $J$) for
    increasing $N$. The number of disorder realizations varies for different system sizes, but all of them have at least $100$ samples of disorder.}
    \label{single_spee}
  \end{figure}    
In a delocalized eigenstate $|\psi_k({i,\alpha})|^2=1/N$ and
hence $S_{k}^{i\alpha}\approx \frac{1}{N}\log_2 N + \frac{1}{N}$
for large values of $N$ whereas for a single-site localization
$S_{k}^{i\alpha}=0$.  The contributions from all sites for a
particular eigenstate are given by $S_k=\sum_{i,\alpha}
S_{k}^{i\alpha}$.  Thus, the average von Neumann
entropy over all the eigenstates is defined as:
\begin{eqnarray}
S^{sp}=\frac{\sum_{k=1}^{N} S_k}{N}.
\end{eqnarray}
For large values of $N$, $S^{sp}\approx(\log_2 N + 1)$ in the
delocalized phase whereas $S^{sp}\approx0$ in an extremely
(single-site) localized phase. For the critical phases, $S^{sp}$ can
take any intermediate values. From Fig.~\ref{single_spee}, which shows
the von Neumann entropy $S^{sp}$ as a function of disorder $\Delta$,
it can be seen that $S^{sp}$ is less than $(\log_2 N + 1)$ for all
disorder strengths and system sizes, pointing towards
localization. However, the higher values of von Neumann entropy at low
disorder strengths compared to those at high disorder strengths
indicate the weakly localized $FBL$ phase for small
$\Delta$ as compared to the strong $AL$ phase for large $\Delta$. Also we
note that similar to $IPR$, the value of spectrum averaged $S^{sp}$
remains disorder strength independent in the $FBL$ phase.

For the $k\textsuperscript{th}$ eigenstate, the Shannon entropy is
given by
\begin{eqnarray}
  S^{sh}_k = -\sum\limits_{i,\alpha} |\psi_k({i,\alpha})|^2 \ln
  |\psi_k({i,\alpha})|^2.
\end{eqnarray}  
The Shannon entropy averaged over all the eigenstates $S^{sh}=\langle
S^{sh}_k\rangle$ is shown in Fig.~\ref{single_spee}(b). For a
perfectly delocalized state, $S^{sh}=\ln N$ whereas for a perfectly
localized state $S^{sh}$ is $0$, where the system size independence
implies localization. The figure shows the presence of two localized
phases: $FBL$ phase with higher $S^{sh}$ and $AL$ phase with lower
$S^{sh}$, consistent with the observation from $IPR$. 
%
        
\subsection{Non-equilibrium dynamics of a single particle}
Next we study non-equilibrium
properties of the system by keeping a single particle initially at some
lattice site $m_0$, i.e. $\ket{\psi_{in}}=\ket{m_0}$. Here we arbitrarily choose
$m_0$ to be a $d$ site, although the other choices should yield similar results. We calculate various dynamical quantities such as the entanglement entropy, revival probability and width of the single particle wavepacket
in the following.

\begin{figure}
           \includegraphics[width=0.494\columnwidth]{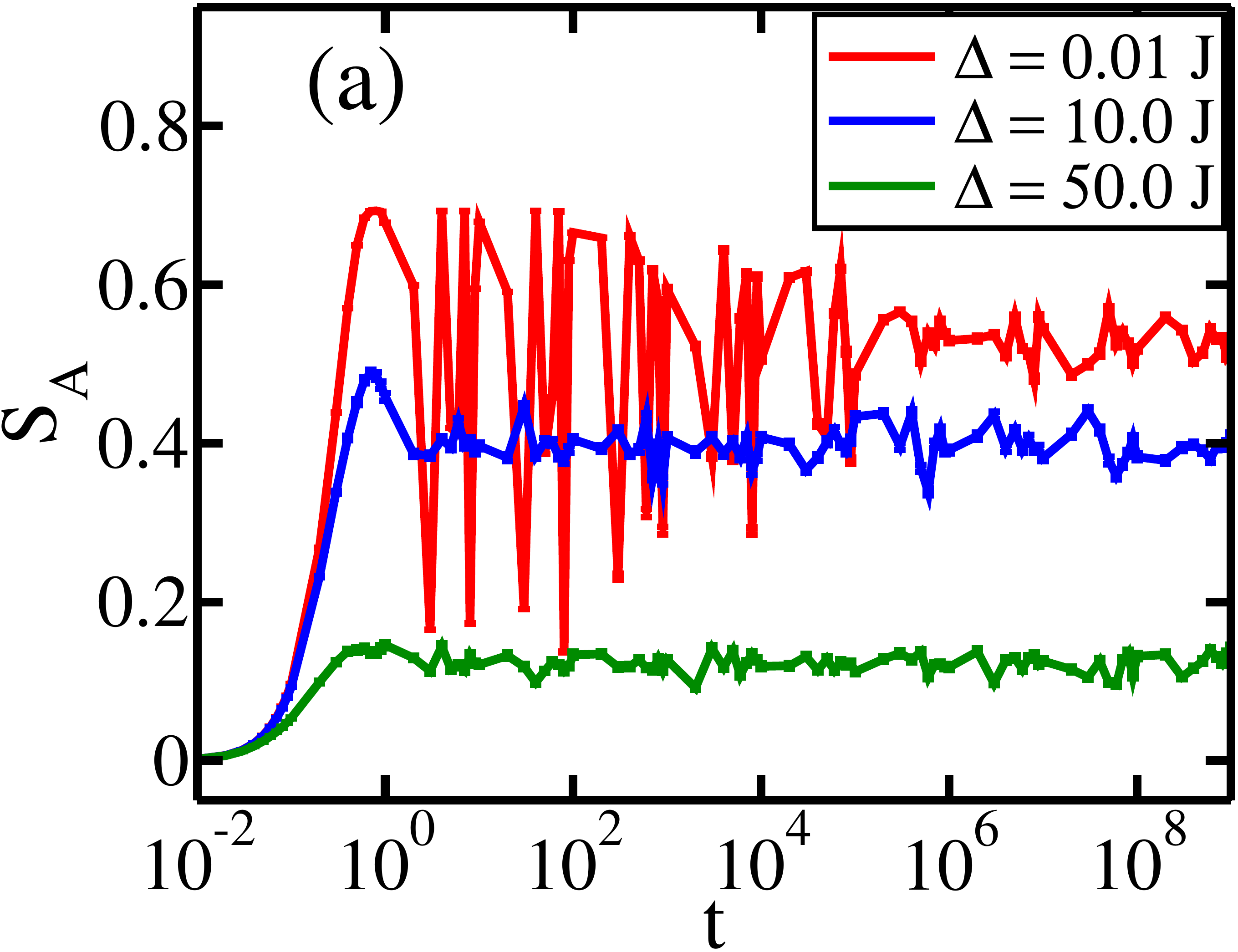}
           \includegraphics[width=0.494\columnwidth]{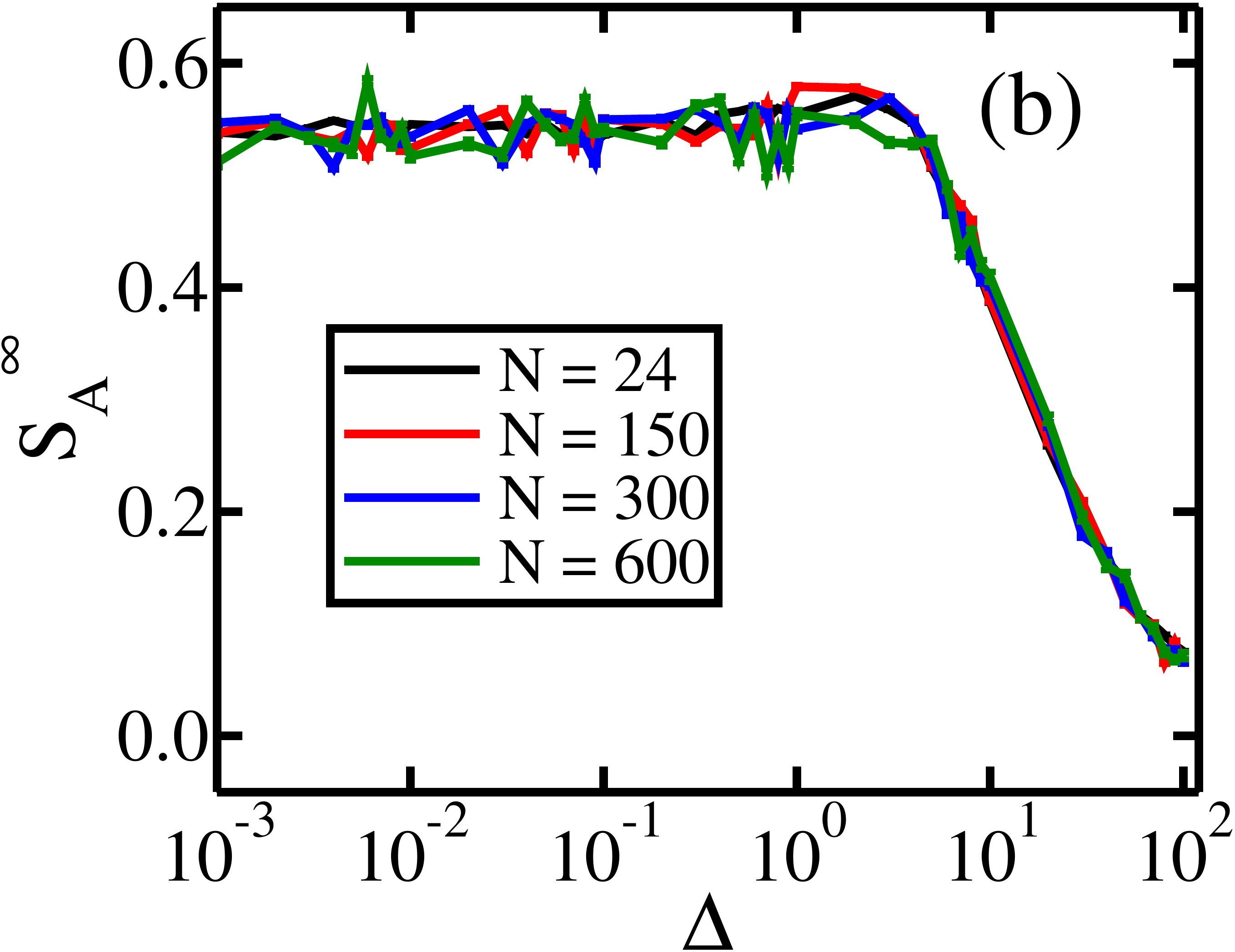}
           \caption{(a) The entanglement entropy $S_A$ as a function of time (in units of $J^{-1}$) for a single particle for system size $N=600$ and different disorder strengths.(b) The saturation value $S_A^{\infty}$ as a function of $\Delta$ (in units of $J$) for increasing values of $N$. The number of disorder realizations varies for different plots, but all of them have at least $100$ samples of disorder.}
           \label{single_ee}
   \end{figure}

The entanglement entropy of a single particle is given
by~\cite{SChakravarty,prb} $S_A(t) = -p_A(t) \ln p_A(t) + (1-p_A(t)) \ln (1-p_A(t))$,
where $p_A(t)=\sum\limits_{{i,\alpha}\in A} |\psi_({i,\alpha})|^2$, is the instantaneous
probability of finding the particle inside subsystem A which is taken to be half
of the full system. The dynamics of entanglement entropy is shown in
Fig.~\ref{single_ee}(a) for different disorder strengths
$\Delta$.  For very low
disorder $(\Delta=0.01 J)$, after a super-ballistic transient, a damped oscillatory behavior is observed for the entanglement entropy (see Appendix~\ref{appA} for the zero-disorder case) 
followed by saturation. This oscillatory behavior vanishes at high disorder strength (e.g. $\Delta\ge 10 J$). The
saturation value $S_A^{\infty}$ is plotted as a function of $\Delta$ for
increasing system sizes in Fig.~\ref{single_ee}(b), which clearly
reflects the presence of two localized regimes; weak flat-band based localization and strong Anderson localization.

\begin{figure}
              \includegraphics[width=0.494\columnwidth]{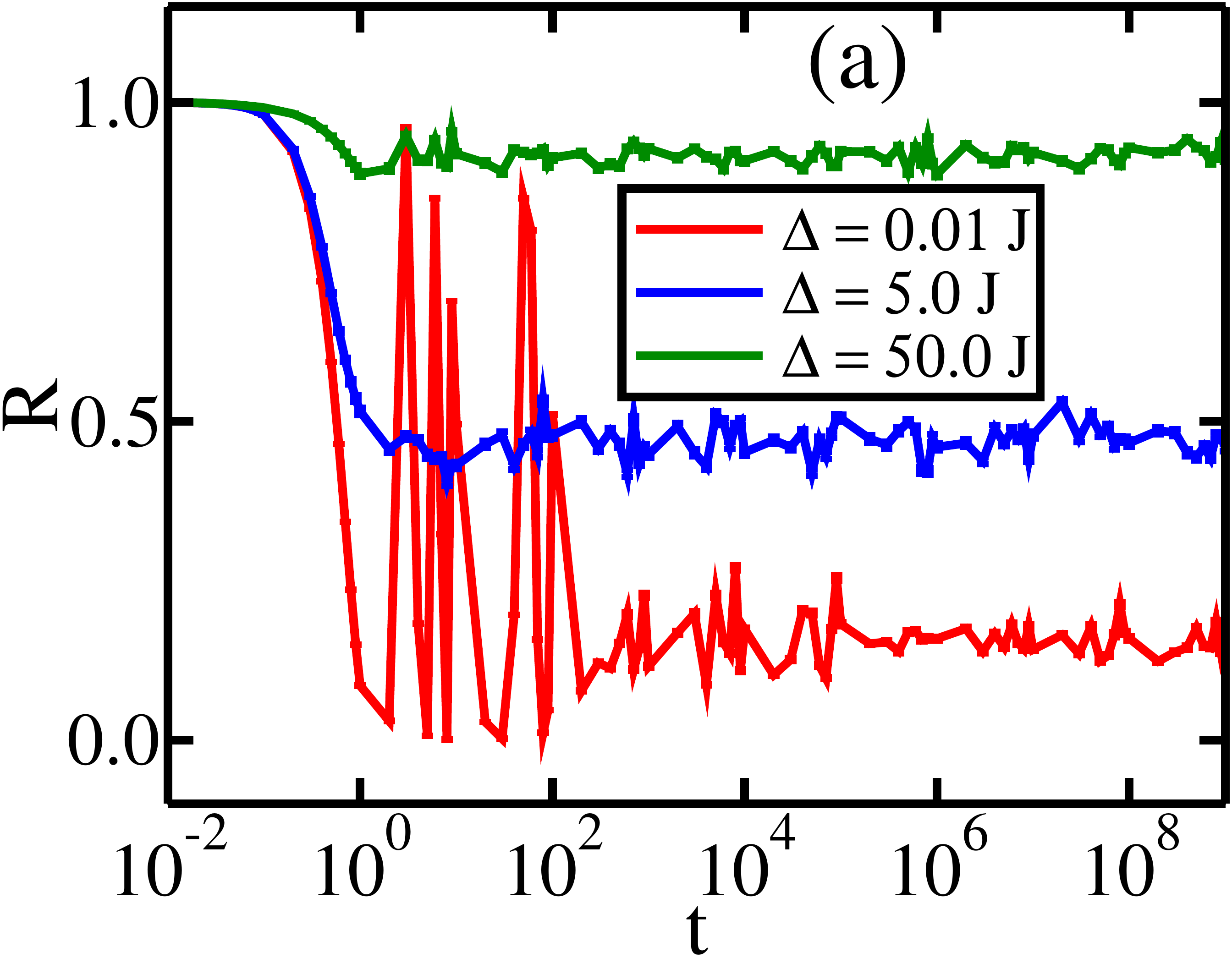}
              \includegraphics[width=0.494\columnwidth]{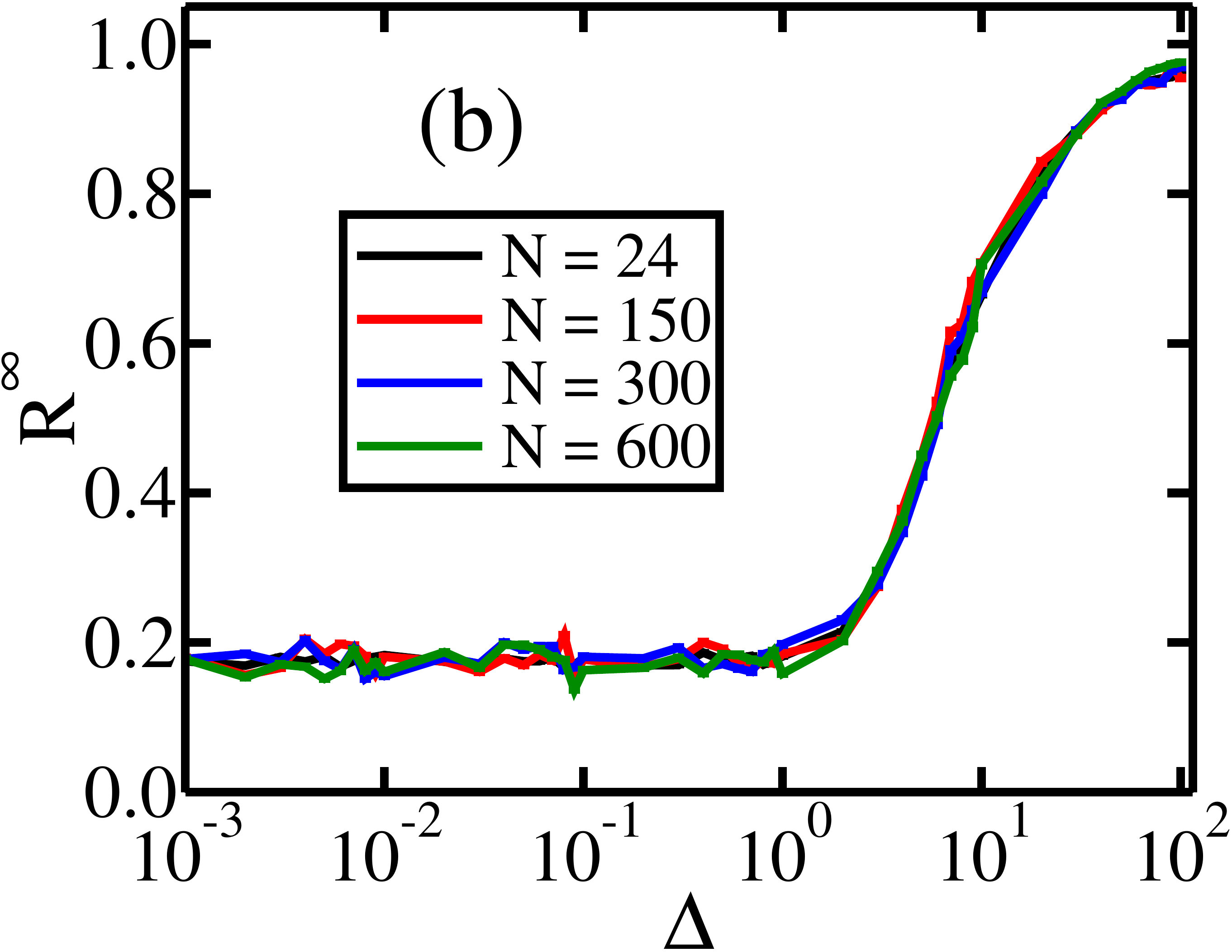}
              \caption{(a) The revival probability $R$ as a function of time (in units of $J^{-1}$) for a single particle for system size $N=600$ and increasing disorder strength $\Delta$. (b) The saturation value $R^{\infty}$ as a function of $\Delta$ (in units of $J$) for increasing values of $N$. The number of disorder realizations varies for different plots, but all of them have at least $100$ samples of disorder.}
              \label{single_revival}
    \end{figure}

We also study revival probability  $R(t) = |\braket{\psi_{in}|\psi_t}|^2$ of finding the particle at the initial site. The dynamics of the revival probability $R$  shown in Fig.~\ref{single_revival}(a) 
indicates damped oscillatory behavior in the tiny disorder regime $\Delta=0.01J$ whereas this characteristic vanishes in the high disorder regime $\Delta>5.0J$. The saturation value $R^{\infty}$ of revival probability (Fig.~\ref{single_revival}(b) )
turns out to be independent of system size $N$ for all disorder strengths, which is characteristic of a localized phase. The figure clearly shows the presence of two localized phases thereby supporting the results obtained in the static calculations.
       
    \begin{figure}
                  \includegraphics[width=0.494\columnwidth]{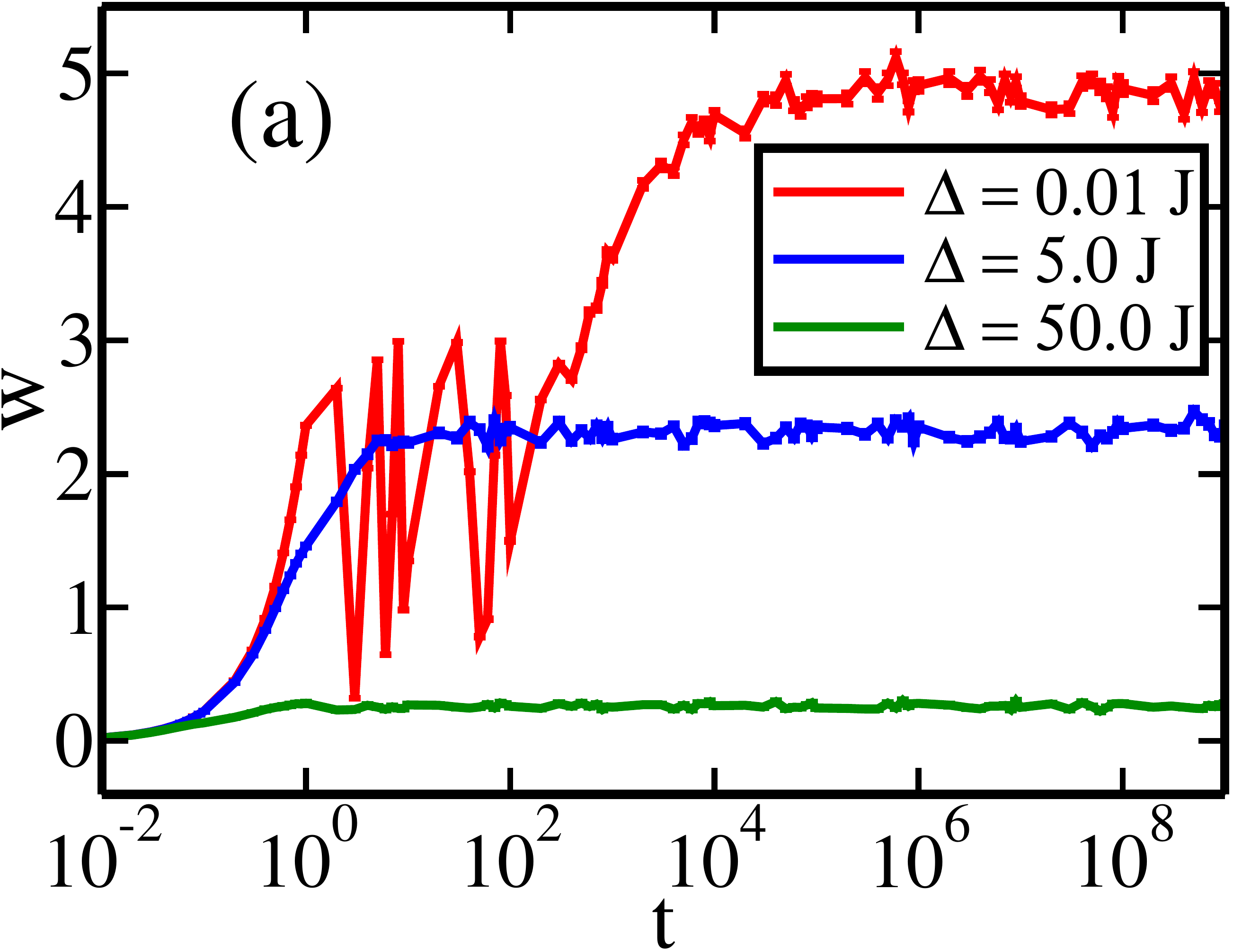}
                  \includegraphics[width=0.494\columnwidth]{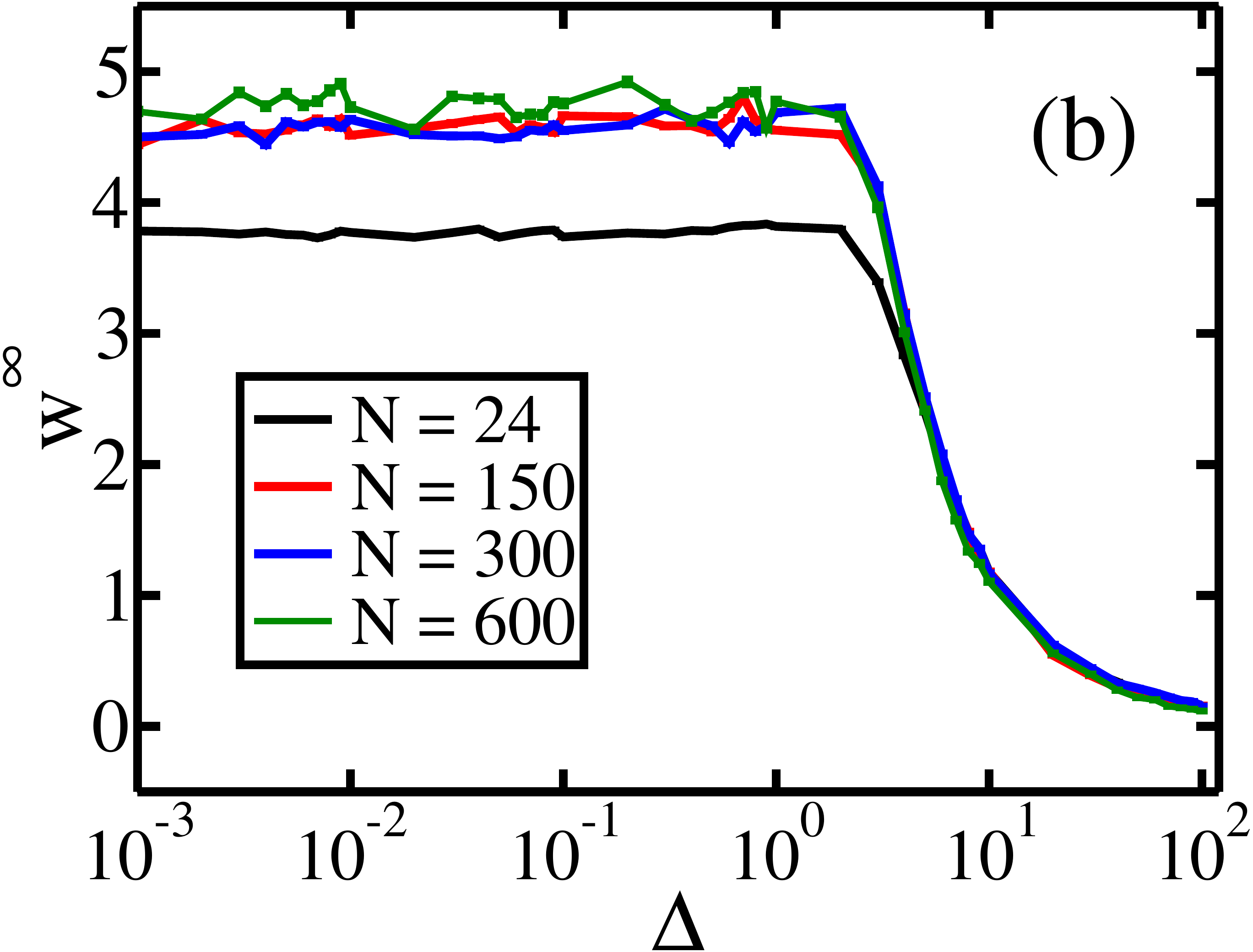}
                  \caption{(a) The dynamics of width of the wavepacket $w$ as a function of time (in units of $J^{-1}$) for a single particle for system size $N=600$ and different disorder strengths. (b) The saturation value $w^{\infty}$ as a function of $\Delta$ (in units of $J$) for increasing values of $N$. The number of disorder realizations varies for different plots, but all of them have at least $100$ samples of disorder.}
                  \label{single_wid}
        \end{figure}
The width of the wavepacket is defined as
$w=\sqrt{\langle\hat{x}^2\rangle - \langle\hat{x}\rangle^2}$, where
$\hat{x}$ is the position operator. The dynamics of $w$ is shown in
Fig.~\ref{single_wid}(a).  After a ballistic transient there is a
damped oscillation due to recurrence for $\Delta=0.01J$, followed by
an extended sub-diffusive regime before it reaches saturation. For
large values of disorder $\Delta>5.0J$, after the ballistic transient
there is a short sub-diffusive regime followed by saturation. The
oscillatory recurrence regime is absent in this case. The saturation
value $w^{\infty}$ as a function of $\Delta$ shown in
Fig.~\ref{single_wid}(b) clearly indicates weak flat band based
localization for small disorder and strong Anderson localization for
large disorder respectively. We note that $w^{\infty}$ becomes
independent of $N$ for large values of $N$, for all values of
$\Delta$, which is a signature of localization.

\begin{figure}
  \includegraphics[width=0.8\columnwidth]{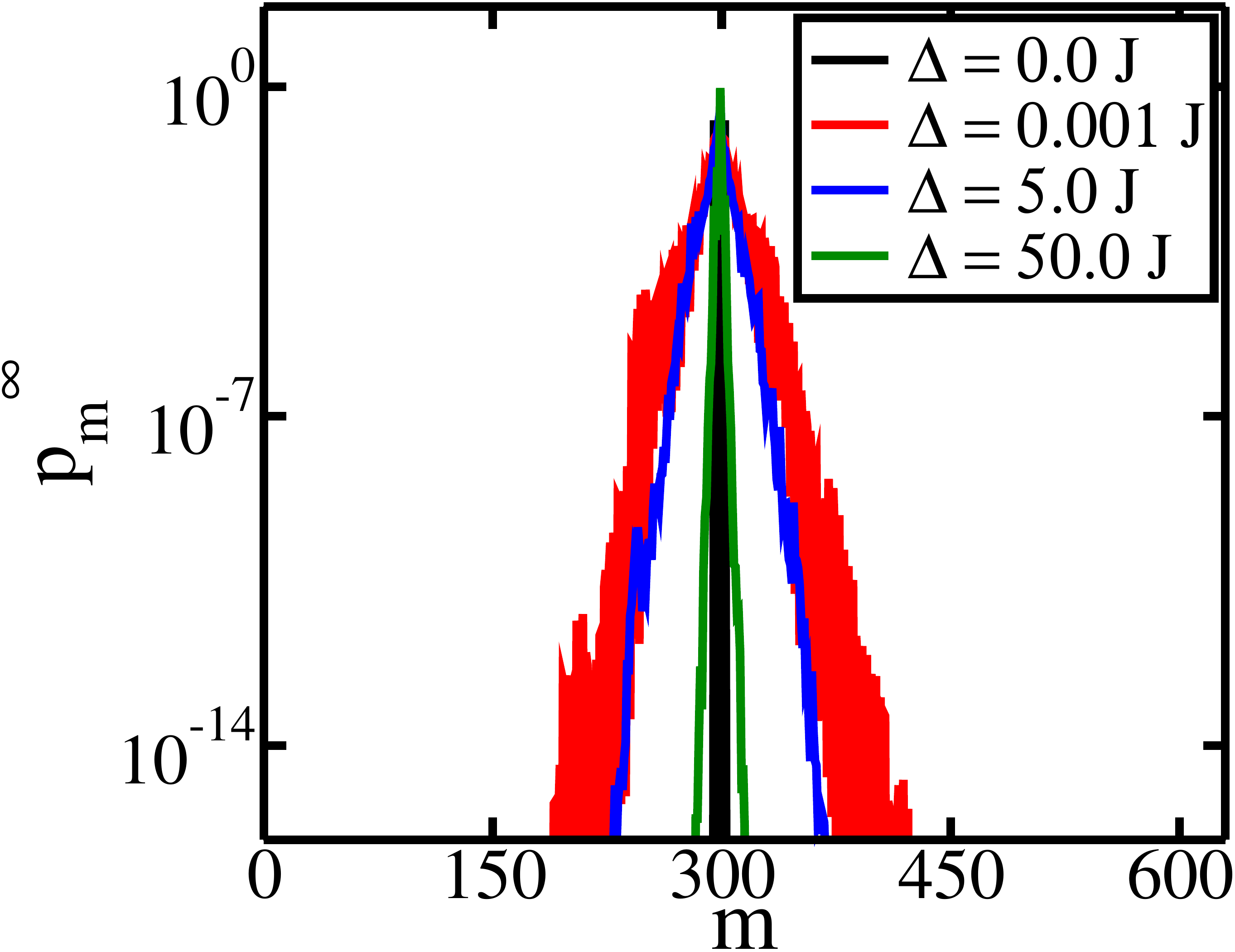}
  \caption{The on-site occupation probability of a single particle in the long time limit, denoted as $p_m^{\infty}$ (in log-scale) at site $m$ for $N=600$ and disorder strength $\Delta=0,0.001J,5.0J$ and $50.0J$ respectively. In this plot, number of disorder realizations is $200$.}
  \label{single_wfn}
\end{figure}

In Fig.~\ref{single_wfn}, we show the on-site occupancy of a
single particle in the long time limit. For
$\Delta=0$, the wavefunction is extremely localized within very few
sites reflecting the localization properties of CLS. Turning on a tiny
disorder $\Delta=0.001J$ makes the occupation probability function $p_m^{\infty}$ extended in space with
exponential tails. Increasing the disorder further to
$\Delta>5.0J$ shrinks the function with sharper exponential
tails. This establishes the hierarchy for localized phases: $CL > AL > FBL$ in
terms of the strength of localization.

\section{Non-interacting spinless fermions}
In this section, we report the effect of disorder on
noninteracting spinless fermions in the diamond chain. This should be
viewed as a bridge between the single-particle physics of the previous
section, and interacting many particle physics of the following section. In this entire section, we set
the interaction strength to zero ($V = 0$), and the filling fraction to $\frac{1}{6}$. Using correlation matrix techniques, we compute
entanglement entropy and imbalance parameter, which help in revealing the
characteristics of the system under disorder.

\subsection{Fermionic entanglement entropy of the ground state}
Entanglement entropy is a useful quantity to distinguish between
different many-body quantum phases in the context of localization~\cite{eisert,laflorencie}. The entanglement entropy between
two subsystems is given by $S_A=-Tr(\rho_A \log \rho_A)$, where the
reduced density matrix $\rho_{A}=Tr_{B}(\rho)$ is obtained from the
many body ground state $\Ket{\Psi_0}$ through the density matrix
$\rho=\Ket{\Psi_0} \Bra{\Psi_0}$.  The computation of entanglement
entropy for many-body ground states of noninteracting spinless
fermions is greatly simplified with the help of free fermionic
correlation matrix techniques~\cite{peschel2003calculation,peschel2009}.
For a single Slater determinant ground state, using Wick's theorem,
the reduced density matrix can be represented as
$\rho_{A}=\frac{e^{-H_{A}}}{Z}$, where $H_{A}$ is the entanglement
Hamiltonian (that is guaranteed to have a quadratic form), and $Z$ is
obtained from the condition $Tr (\rho_{A}) = 1$. The information
contained in the reduced density matrix of size $2^L\times 2^{L}$ can
be captured in terms of the correlation matrix $C$ of size $L\times
L$~\cite{peschel2003calculation} within the subsystem A, where
$C_{ij}^{\alpha\beta}=\left\langle \alpha_{i}^{\dagger}\beta_{j} \right\rangle$ with $\alpha,\beta \in \{u,c,d\}$, and $i,j$ denoting the
unit cell index. The
correlation matrix and the entanglement Hamiltonian are related
by~\cite{peschel2003calculation,peschel2009,peschel2012special}:
$C=\frac{1}{e^{H_A}+1}$. Using this relation, the entanglement entropy
for free fermions is given by~\cite{peschel2009,peschel2012special}:
\begin{equation}
S_A=-\sum\limits_{m=1}^{L} [\zeta_m \log \zeta_m + (1-\zeta_m) \log (1-\zeta_m)],
\end{equation}
where $\zeta_m$'s are the eigenvalues of the correlation matrix
$C$. 

\begin{figure}
	 \includegraphics[width=0.494\columnwidth]{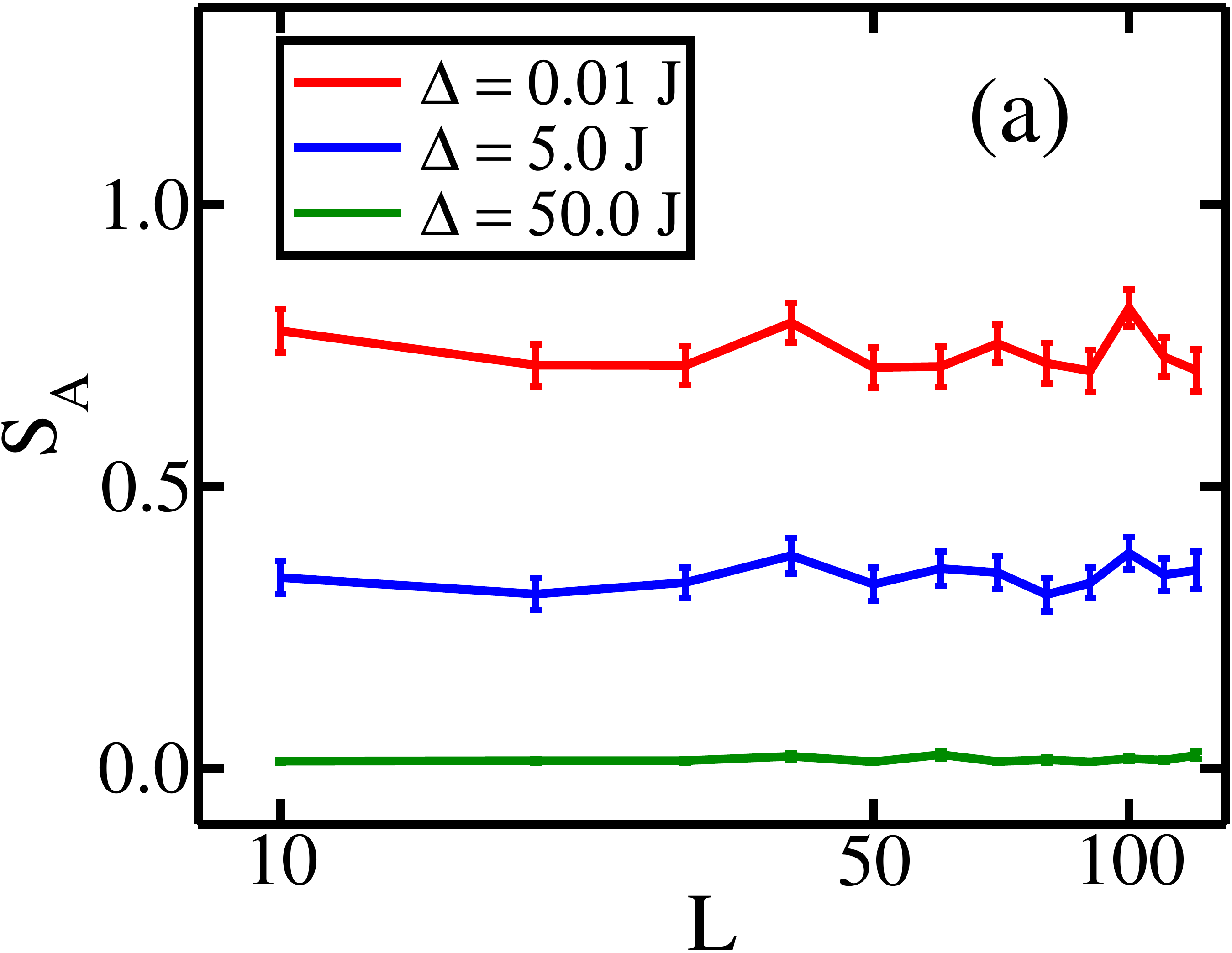}
	 \includegraphics[width=0.494\columnwidth]{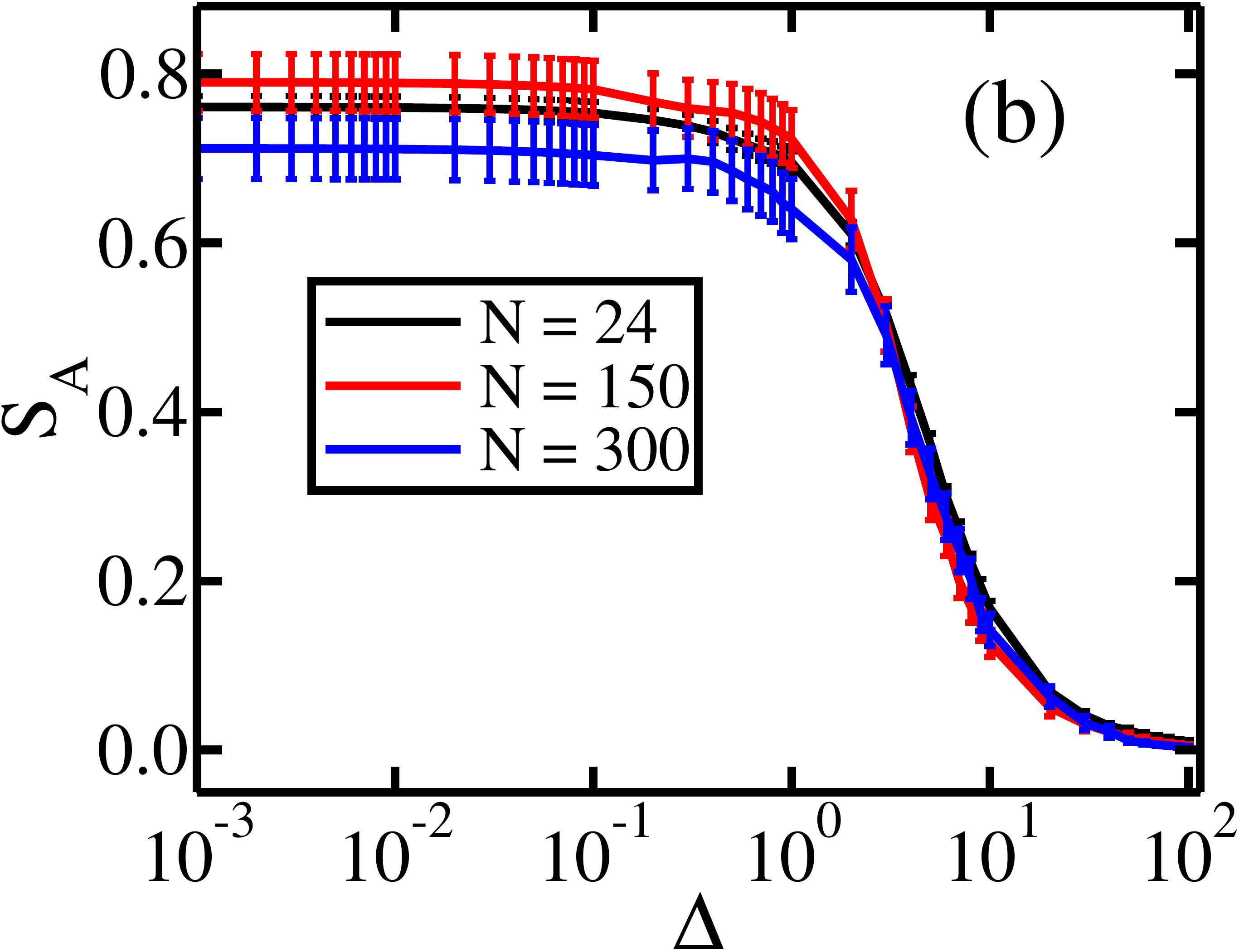}
	\caption{(a) The subsystem size $L$ dependence of the entanglement entropy $S_A$ of the fermionic ground state for increasing values of disorder strength $\Delta$ for system size $N=300$. (b) The $S_A$ as a function of $\Delta$ (in units of $J$) for increasing values of $N$ with $L=N/2$. For all the plots, the number of disorder realizations is $100$ and filling fraction of the noninteracting fermions is $1/6$. }
	 \label{noint_eent}
\end{figure}

Subsystem scaling of entanglement entropy has been used to distinguish
quantum
phases~\cite{eisert,laflorencie,roy2019quantum,roy2019study}. For free
fermions in $d$ dimensions, typically $S_A\propto L^{d-1}\log
L$~\cite{swingle} in metallic phases whereas $S_A\propto L^{d-1}$
(also known as area-law) in localized phases in the presence of
disorder. Since localized systems do not care about the boundary of
two subsystems, generally the `area-law' can be used as the defining
signature to detect such phases. For the present system, we show the
subsystem size dependence of $S_A$ of noninteracting fermions at $1/6$
filling in Fig.~\ref{noint_eent}(a).
Here subsystem size $L$ is the
number of sites belonging to subsystem A. All the plots abide by the
area-law for all values of $\Delta$, indicating localization.  As
$\Delta$ is increased $S_A$ shows (Fig.~\ref{noint_eent}(b)) a
crossover from a high value to a low value independent of system size
pointing towards the presence of two localized phases in the
many-particle ground state. We observe that the low-$\Delta$ region
corresponding to $FBL$ has a small (compared to thermal value) finite
value of entanglement, which is independent of system size, and hence
characteristic of a localized phase. However at large
$\Delta$, the entanglement entropy is almost zero, indicating the $AL$ phase.

\subsection{Nonequilibrium dynamics of noninteracting fermions}
Next, we study the non-equilibrium transport properties of
noninteracting fermions in the system. We consider an initial
many-body state of the following form:
$\ket{\Psi_{in}}=\prod\limits_{i=1}^{N/3}\hat{c}_{2i}^\dagger\ket{0}$,
which is a product state of fermions in which $c$ site of alternating unit cells is occupied by a single particle leading to a filling fraction of $1/6$. For this type of a choice of the initial state, Wick's theorem
works and hence the free fermionic
techniques~\cite{peschel2003calculation,peschel2009,peschel2012special} can be applied
to study the growth dynamics of the entanglement entropy.
After a super-ballistic transient (Fig.~\ref{noint_eentdyn}(a)) an
oscillatory regime (see Appendix~\ref{appA}) is observed for $\Delta=0.01J$ which
is absent in case of $\Delta=3.0J, 50.0J$. It is notable that the
long-time saturation value of $S_A$ is higher for $\Delta\approx3.0J$ as
compared to those for $\Delta=0.01J$ and $50.0J$.  The hump in the
saturation value $S_A^{\infty}$ (Fig.~\ref{noint_eentdyn}(b)) in
the intermediate range of values of $\Delta$ is due to
fermionic statistics, and is absent in the single-particle case.

We have also studied the imbalance parameter, a more experimentally relevant quantity, which is given by
\begin{equation}
I_b(t)=\frac{\sum\limits_{i=1}^{N/3} (-1)^i n_i^c - n_i^u - n_i^d}{\sum\limits_{i=1}^{N/3 }n_i^c + n_i^u + n_i^d},
\label{imb}
\end{equation}
where $n_i^\alpha$ is the occupancy of fermions at the site $\alpha$
of the $i\textsuperscript{th}$ unit cell where $\alpha\in{u,c,d}$.
Essentially the numerator is the difference in occupancies between the
initially occupied sites and the initially unoccupied sites whereas
the denominator simply sums up to the total number of particles $N_p$. For
a perfectly delocalized many-body state $I_b=-2/3$ whereas for a
perfectly localized many-body state $I_b=1$.

\begin{figure}
  \includegraphics[width=0.494\columnwidth]{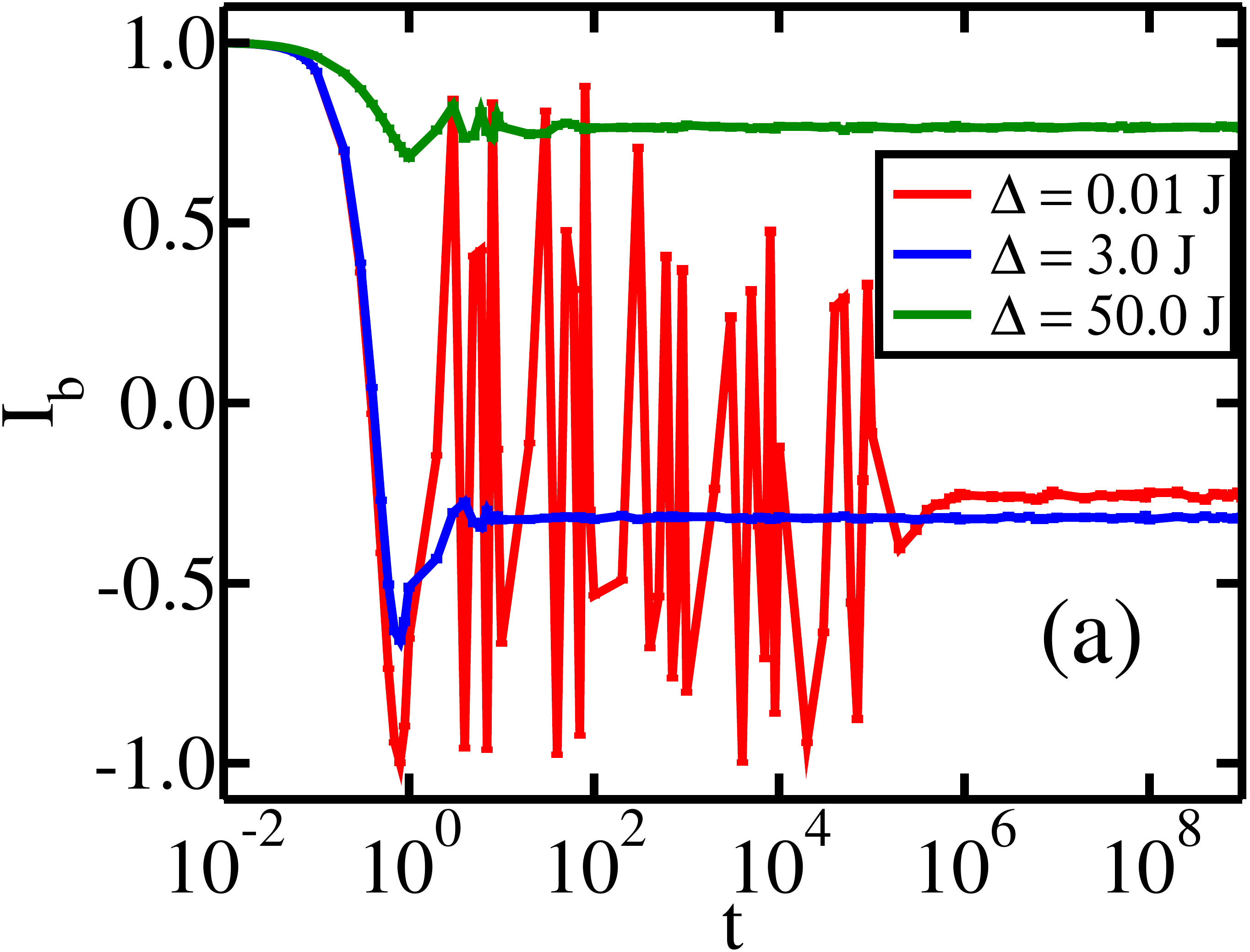}
  \includegraphics[width=0.494\columnwidth]{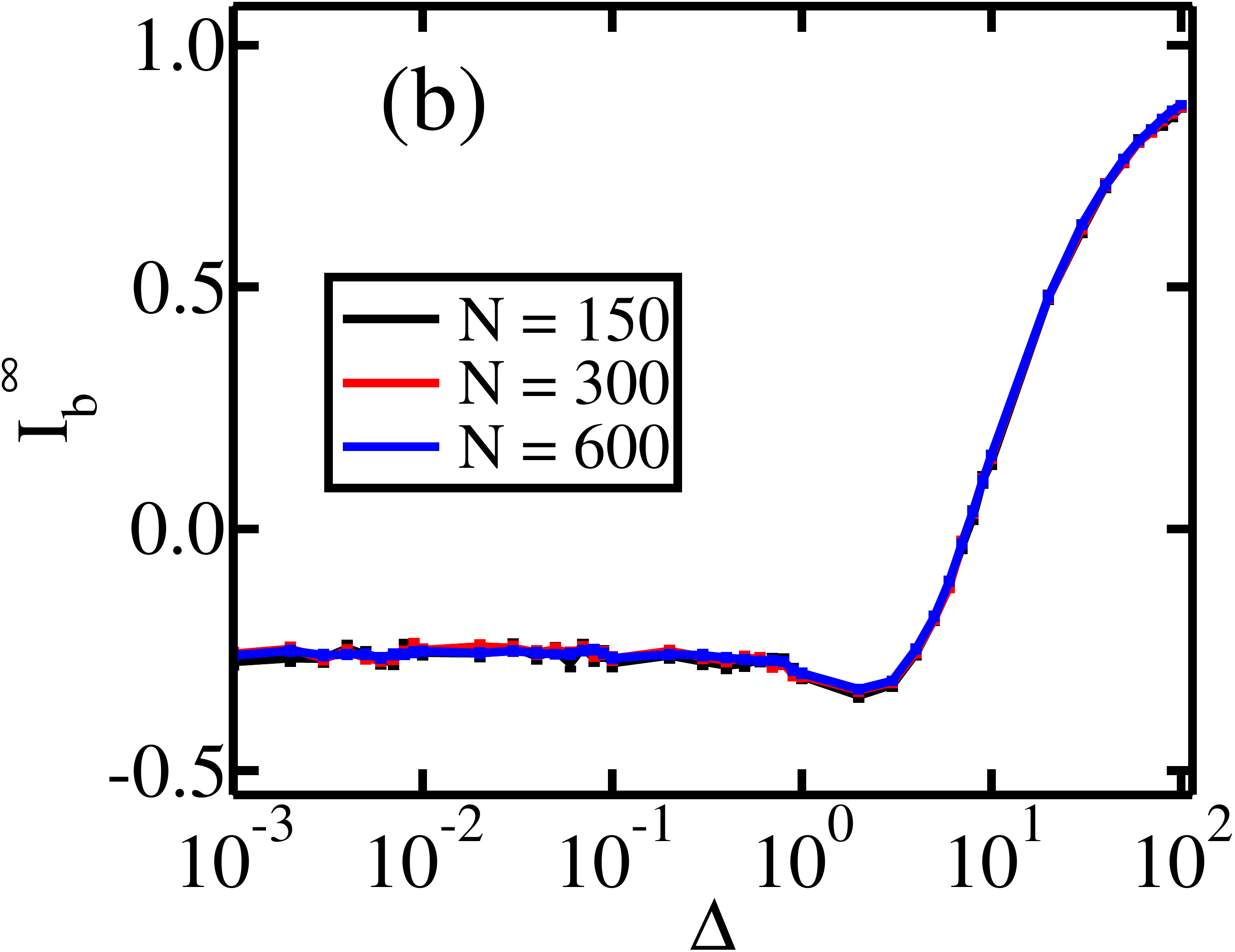}
  \caption{(a) The imbalance parameter $I_b$ as a funtion of time (in units of $J^{-1}$) for system size $N=600$ and disorder strength $\Delta=0.01J,3.0J$ and $50.0J$ respectively. (b) The saturation value $I_b^{\infty}$ as a function of $\Delta$ (in units of $J$) for increasing values of $N$. The number of realizations of disorder is $200$.}
  \label{noint_imbdyn}
\end{figure}

The dynamics of the imbalance parameter $I_b(t)$
(Fig.~\ref{noint_imbdyn}(a)) shows a damped oscillatory regime for
$\Delta=0.01J$ and this behavior is absent for $ \Delta = 3.0J,50.0J$
pointing out the characteristic difference between the small and large
$\Delta$ regimes. The saturation value of $I_b$ for $\Delta=3.0J$ is
low compared to those for $\Delta=0.01J$ and $50.0J$ indicating the
possibility of an intermediate phase. To explore this further, the
saturation value $I_b^{\infty}$ as a function of disorder strength
$\Delta$ for different system sizes $N$ is shown in
Fig.~\ref{noint_imbdyn}(b). A dip is observed in the intermediate
range of disorder strength $\Delta$, where a hump is observed for the
entanglement entropy. 
Such dips may be signatures of delocalization
but on the other hand, system size $N$-independence of $I_b^{\infty}$
at dips is a signature of localization. In order to make a conclusive
comment on delocalization/localization one needs to look at the energy
level-statistics, which is rather difficult to perform for large
system sizes $N$ using the exact diagonalization method. Such dips are
absent in single-particle properties, but seem to show up in the
noninteracting many-particle properties, and will also be encountered
in the interacting many-particle properties in the next
section. Therefore, as we might expect, the non-interacting
many-particle properties show features in between those of
single-particle and interacting many-particle states.
\begin{figure}
  \includegraphics[width=0.494\columnwidth]{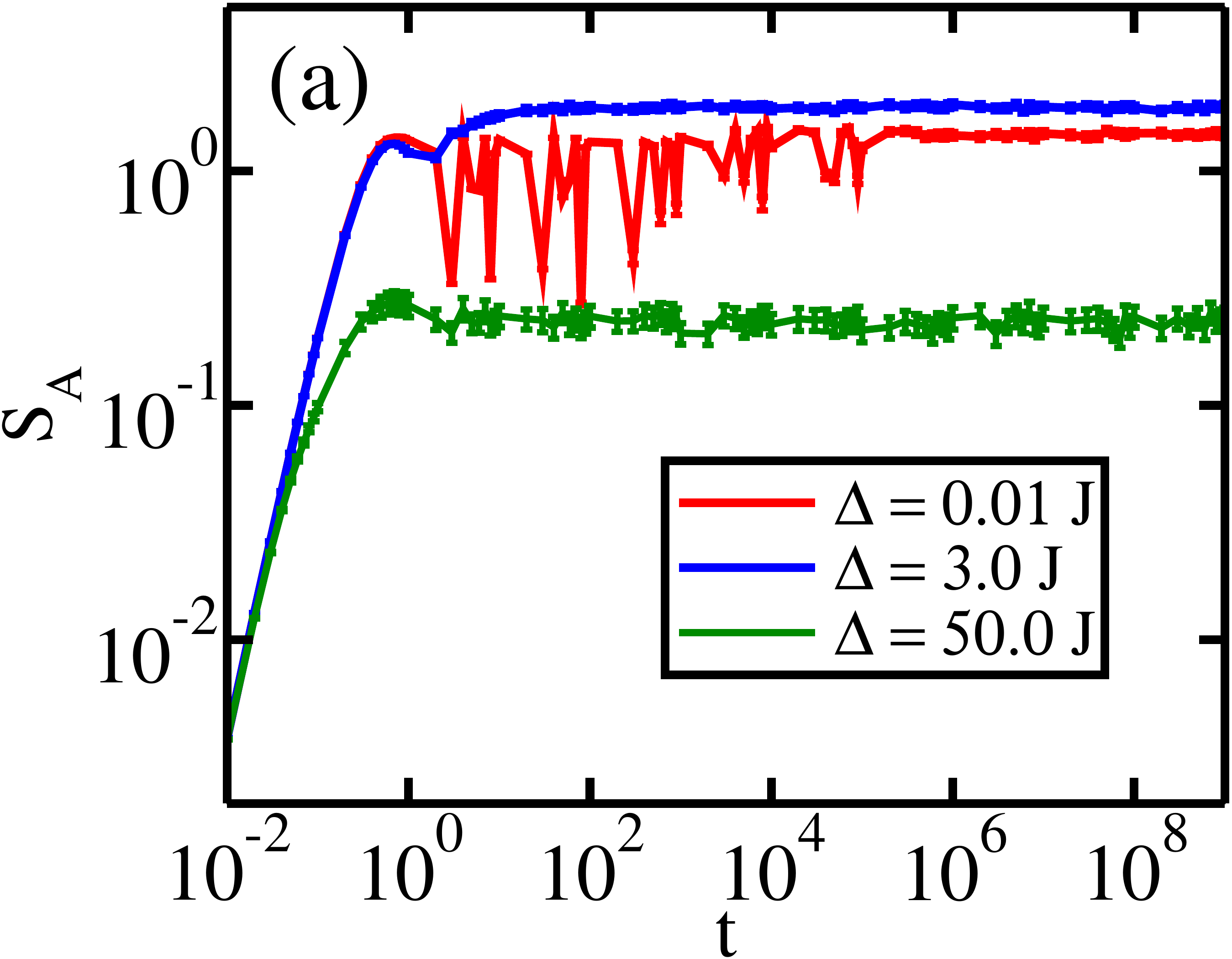}
  \includegraphics[width=0.494\columnwidth]{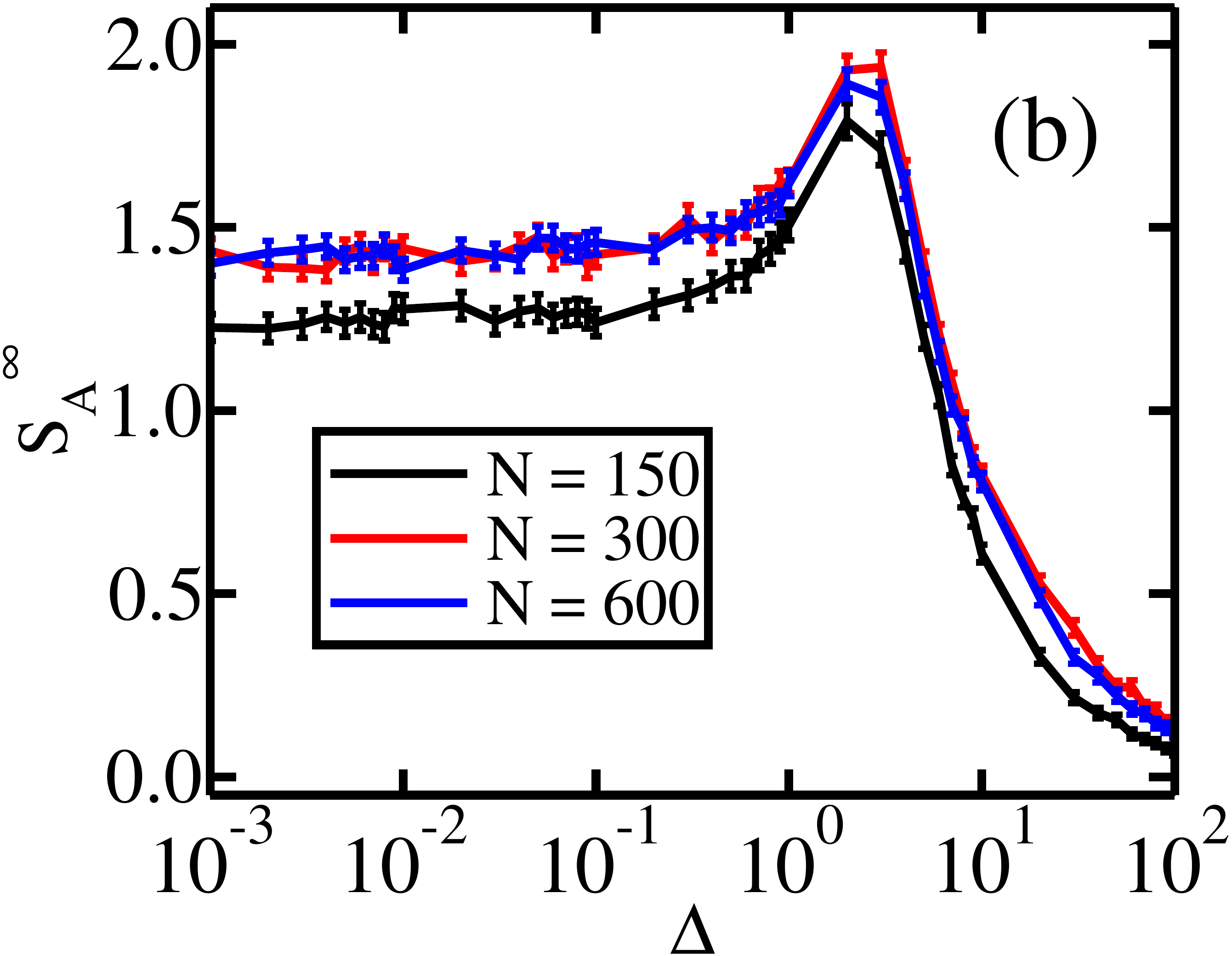}
  \caption{(a) The growth of the entanglement entropy $S_A$ with time (in units of $J^{-1}$) for system size $N=600$ and disorder strength $\Delta=0.01J,3.0J$ and $50.0J$ respectively. (b) The saturation value $S_A^{\infty}$ as a function of $\Delta$ (in units of $J$) for increasing values of $N$. The number of realizations of disorder is $200$.}
  \label{noint_eentdyn}
\end{figure}

\section{Interacting spinless fermions}
\begin{figure*}
  \includegraphics[width=0.6\columnwidth]{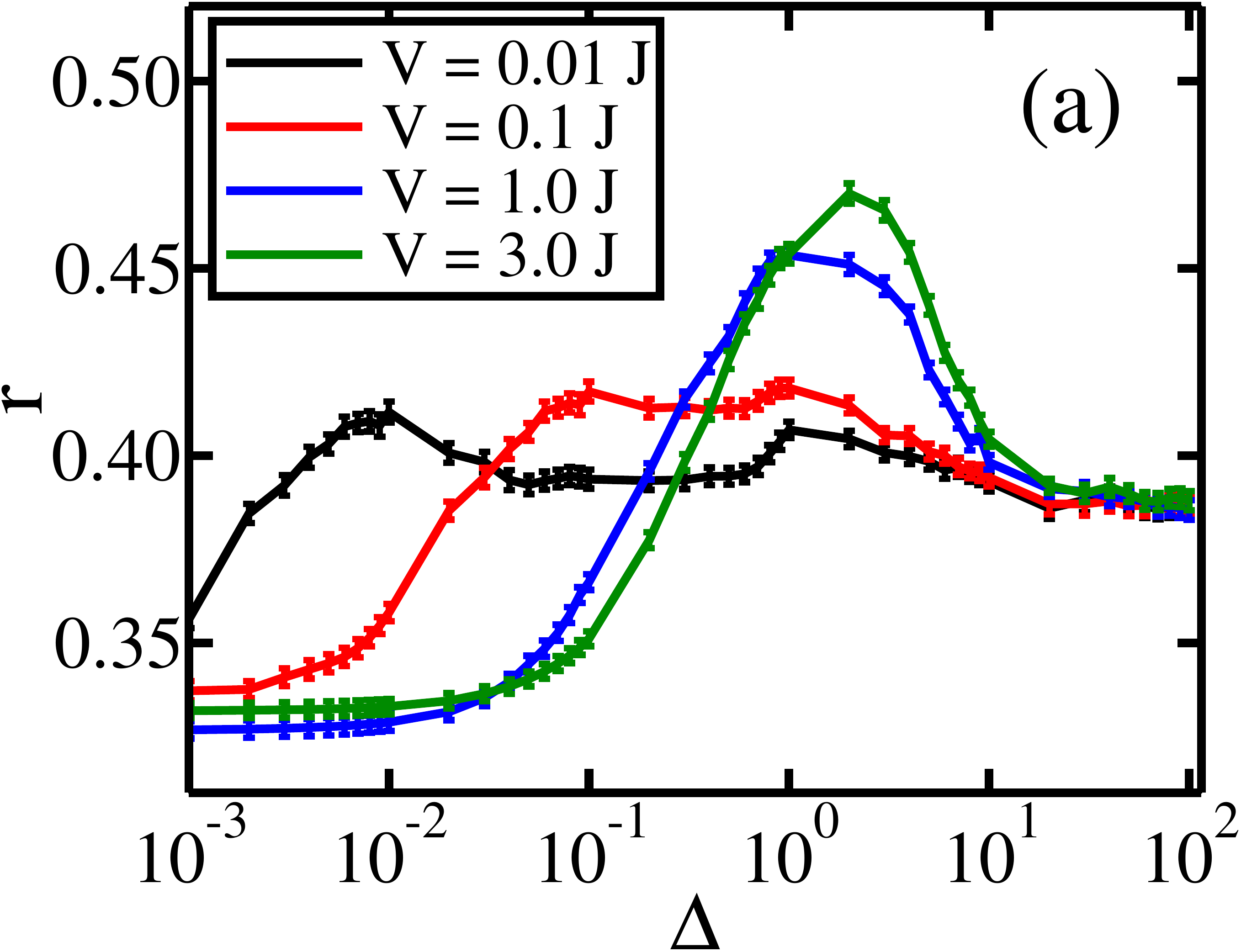}
  \includegraphics[width=0.6\columnwidth]{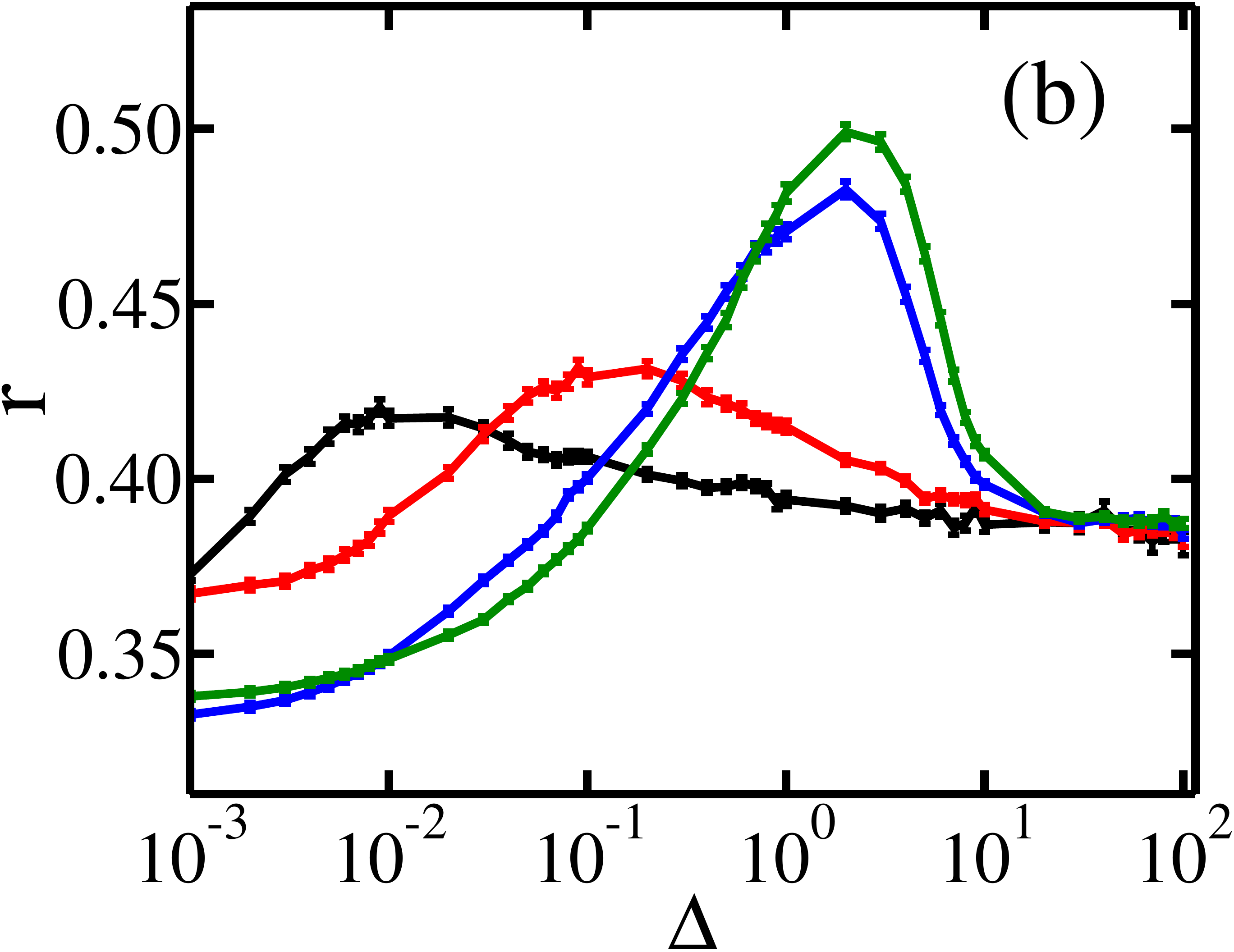}
  \includegraphics[width=0.6\columnwidth]{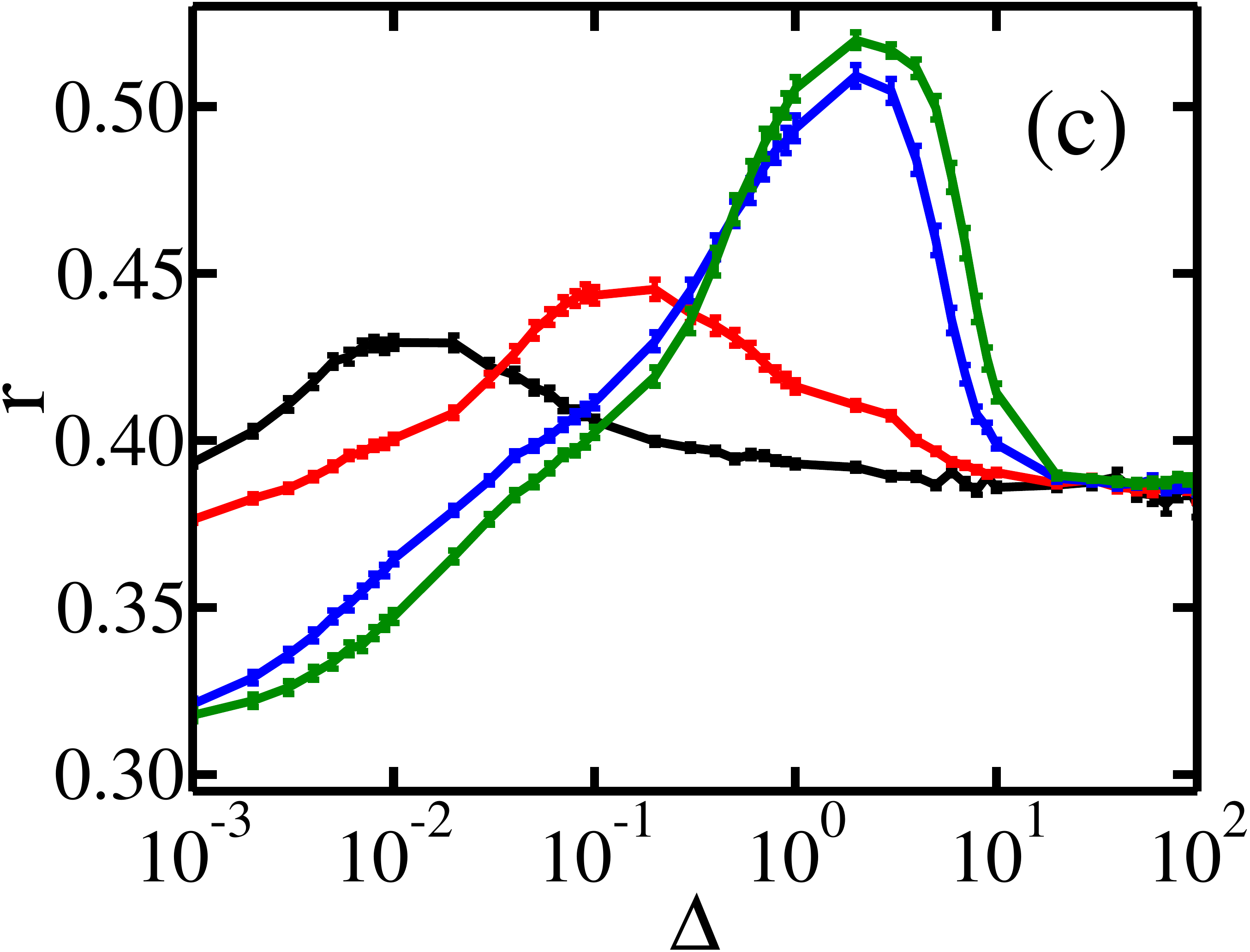}
  \caption{(a-c) The level-spacing ratio $r$ as a function of disorder strength $\Delta$ (in units of $J$) for increasing interaction strength $V$ for fermionic filling fraction (a) $1/9, (b)1/6$, and (c) $2/9$ respectively. For all the plots system size $N=18$. Number of disorder realizations are $500$, $200$ and $100$ for $1/9,1/6$ and $2/9$ fillings respectively.}
  \label{int_ravg}
\end{figure*} 
In this section we develop an understanding of the transport
properties of interacting spinless fermions in the disordered diamond
chain. We employ the level-spacing ratio and
spectrum averaged many-particle inverse participation ratio to capture
the interaction induced $(V\neq0)$ crossovers in this
disordered model.

\subsection{Level-spacing statistics}
\begin{figure}
			  	\includegraphics[width=0.7\columnwidth]{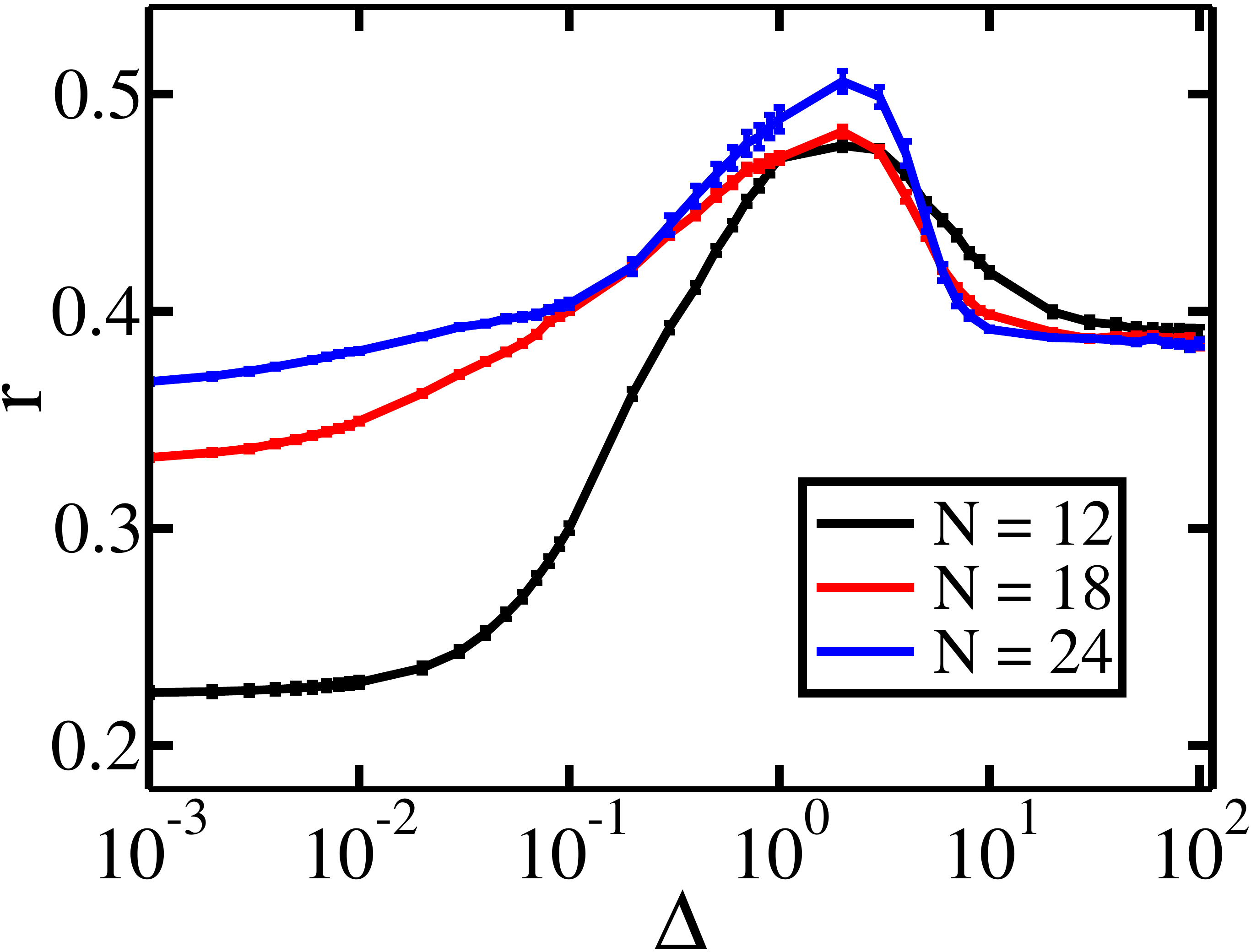}
			  	\caption{The level-spacing ratio $r$ as a function of disorder strength $\Delta$ (in units of $J$) for fixed $V=1.0 J$ and system sizes $N=12,18$ and $24$ for filling fraction of fermions $\nu=1/6$. The quantities are averaged over $500,200$ and $30$ disorder realizations for $N=12,18$ and $24$ respectively for fermion filling $\nu=1/6$.}
			  	\label{rint_size}
\end{figure}  
Level-spacing ratio defined in Eq.~\ref{eq_level} is studied using the
energy spectra
$(\mathcal{E}_1,\mathcal{E}_2,\mathcal{E}_3,\dots,\mathcal{E}_D)$ of
the interacting Hamiltonian via exact diagonalization, where $D={N
  \choose N_p}$ is the dimension of the particle-number constrained
Hilbert space and $\nu=N_p/N$ with $N_p$ and $N$ being the number of
fermions and the system size respectively.  The level-spacing ratio
$r$ as a function of disorder strength $\Delta$ for different $V$ is shown in Fig.~\ref{int_ravg}(a),(b) and (c) for $N=18$ and filling fraction of fermions $\nu=1/9,1/6$ and $2/9$
respectively. We observe that the three phases become most distinct for the highest filling as shown in
Fig.~\ref{int_ravg}(c): an $MBL$ phase at high disorder strengths
$\Delta>10.0J$ with $r = 0.386$, a thermal phase for intermediate
disorder strengths $\Delta=\mathcal{O}(J)$ and for $V=\mathcal{O}(J)$ with $r$ approaching $0.528$, and a
`mixed phase' at low disorder strengths $\Delta<<J$ with $r$ neither
$0.528$ nor $0.386$. In the low disorder region, when the interaction
$V\approx \Delta$, the amount of delocalization increases in the
system in the mixed phase, as $r$ shows a peak (see subsection~\ref{resolved}). 
For the intermediate disorder strengths, the system delocalizes with increasing
filling fraction and the peak of $r$ (Fig.~\ref{int_ravg}(b) and Fig.~\ref{int_ravg}(c) ) almost approaches $0.528$ at $V\approx \Delta$ as the number of fermions is increased. We also observe
that the peak corresponding to delocalization moves towards a lower
value of disorder strength $\Delta$ as the number of fermions increases.
Thus, the filling fraction appears to aid the repulsive effects of
Pauli exclusion. What happens in the low $\Delta$ regime as a function of
filling fraction will be addressed in detail later.
\begin{figure*}
			  	\includegraphics[width=0.6\columnwidth]{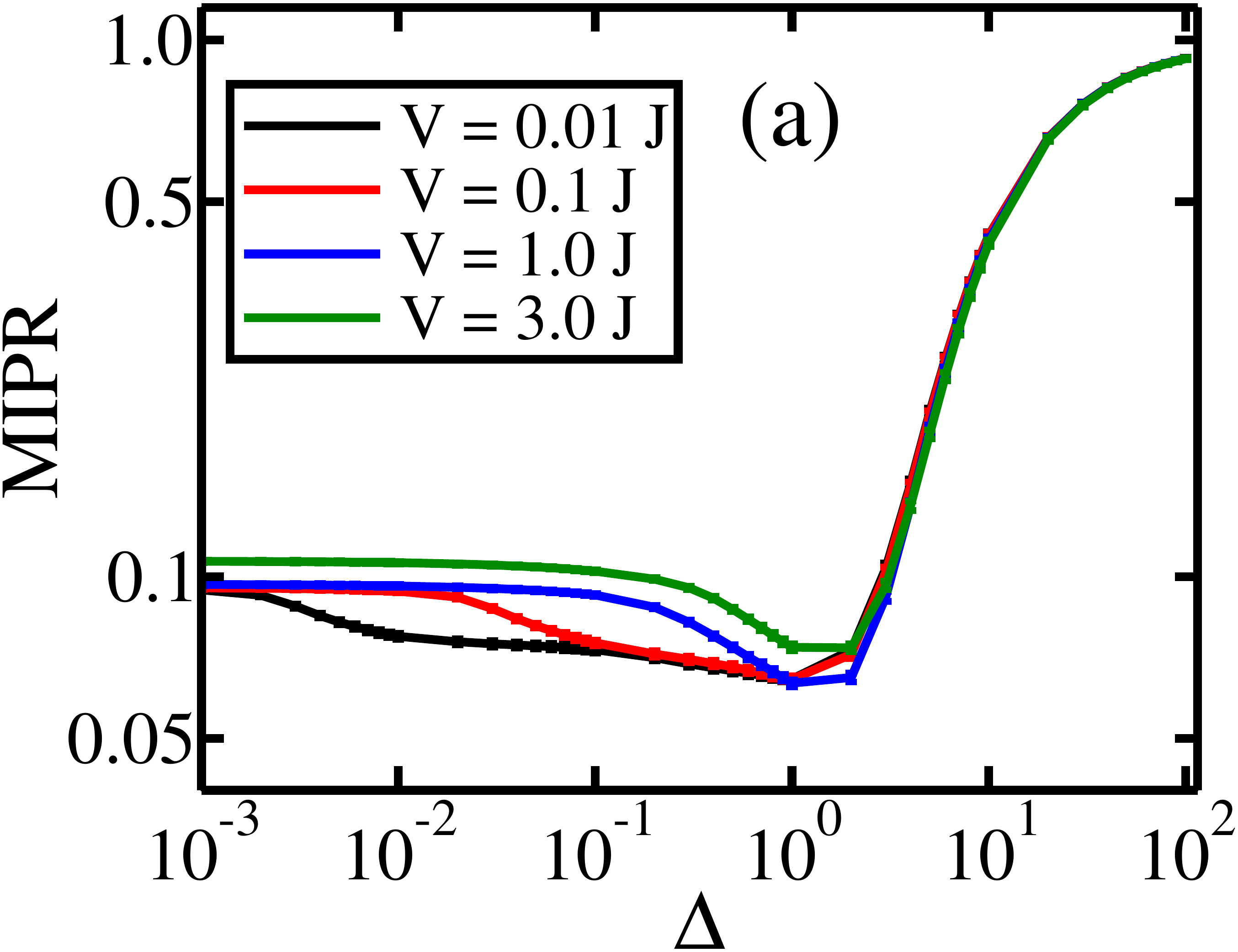}
			  	\includegraphics[width=0.6\columnwidth]{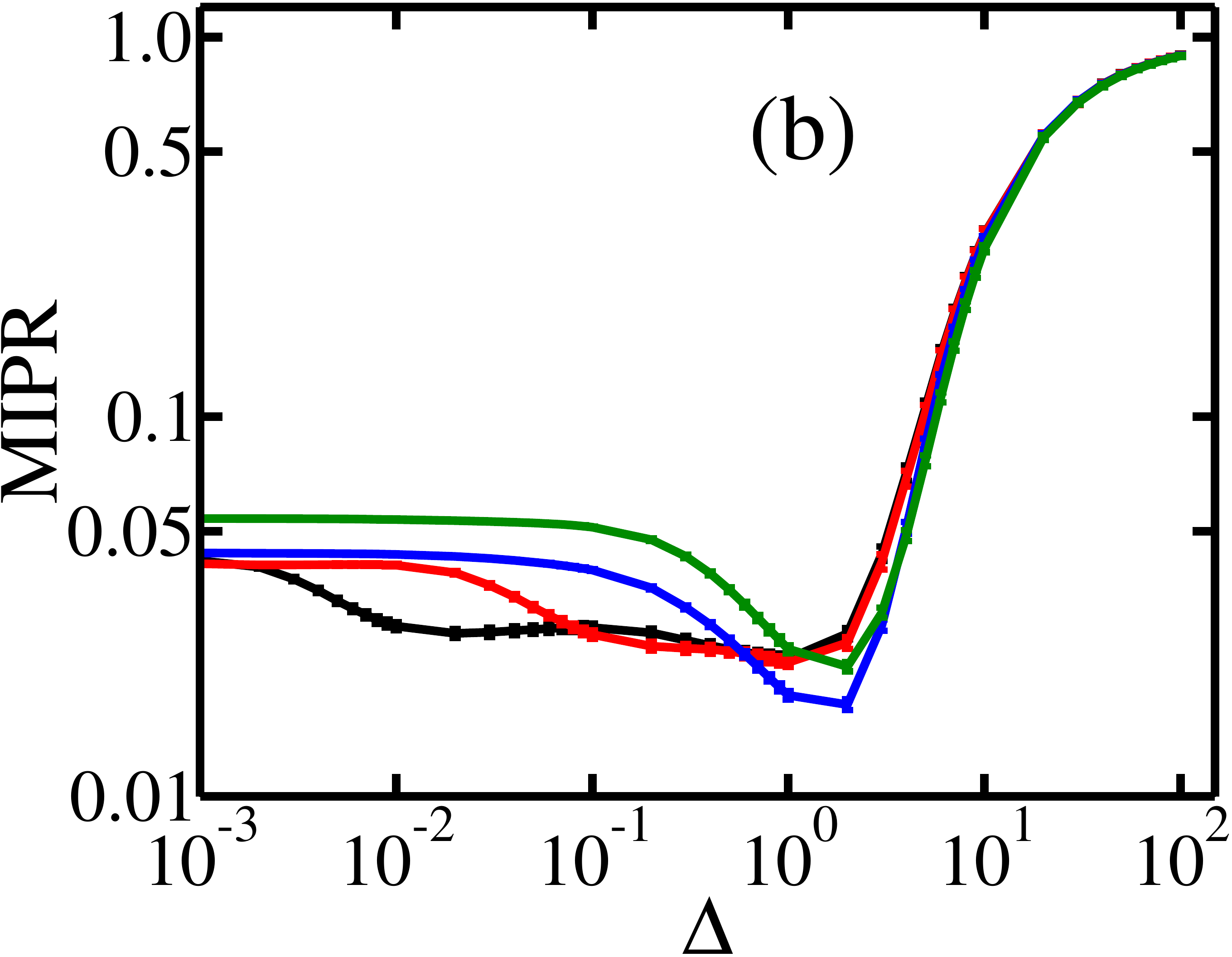}
			  	\includegraphics[width=0.6\columnwidth]{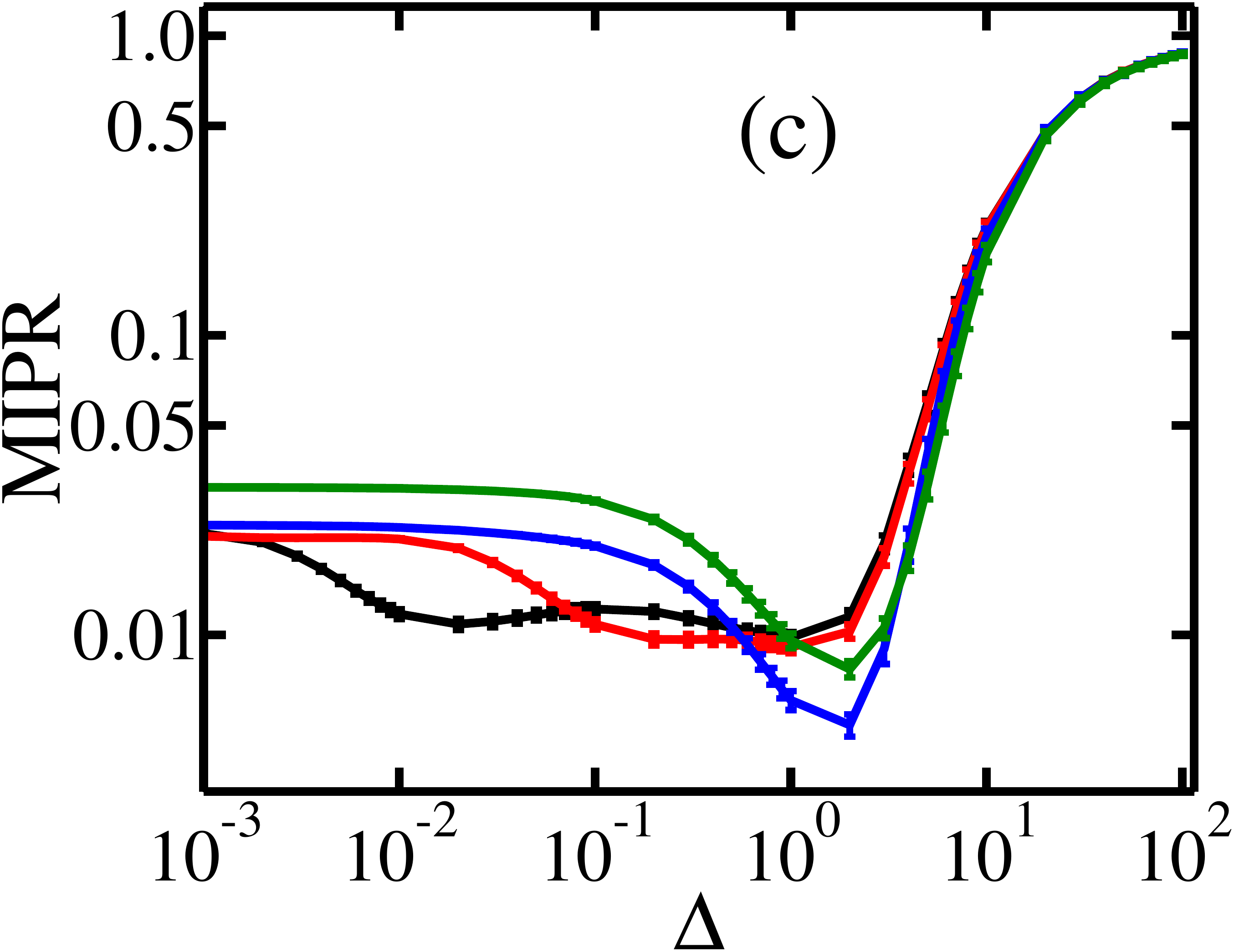}
			  	\caption{The log-log plots of the
                                  spectrum-averaged $MIPR$ as a
                                  function of $\Delta$ (in units of $J$) for different
                                  interaction strengths $V$ for
                                  fermionic filling fraction (a)$1/9,
                                  (b)1/6$, and (c) $2/9$
                                  respectively. For all the plots
                                  system size $N=18$. Number of
                                  disorder realizations are $500$,
                                  $200$ and $100$ for $1/9,1/6$ and
                                  $2/9$ fillings respectively.}
			  	\label{int_mipr}
\end{figure*} 

We also show the system size dependence of level-spacing ratio $r$ vs
$\Delta$ for $V=1.0 J$ and $\nu=1/6$ in Fig.~\ref{rint_size}. Here, as
the system size $N$ increases, the particle number $N_p$ also
increases to maintain $\nu=1/6$. As $N$ increases, we observe that the phase at 
  high disorder strengths gets more sharply defined. A substantial delocalization is observed in the intermediate disorder region and $r$ seems to approach $0.528$ with increase in system size. In the low $\Delta$
regime, values of $r$ are much lower than $0.386$ for $N=12$. This
happens typically due to restoration of the crystal momentum
conservation~\cite{oganesyan2007localization}, which has been recently
addressed in the literature for two interacting
particles~\cite{flach2019taming}. As system size increases ($N_p$
increases accordingly to fix $\nu$), $r$ increases and becomes $0.386$
for $N=24$, although it is not very clear whether $r$ will increase
even further with $N$. We infer that it is a mixed phase with the
localized eigenstates being dominant (see Fig.~\ref{iprscaling_size}
and the discussion in the subsection~\ref{resolved}).
 The three phases become more well established with
  increasing $\nu$ as can be seen from Fig.~\ref{int_ravg}.
  
\subsection{Many-particle inverse participation ratio}
Expanding the normalized eigenstate $\ket{\Psi}$ in the
particle-number constrained space as
$\ket{\Psi}=\sum\limits_{n=1}^{D}C_n \ket{n}$, the many-particle
inverse participation ratio $(MIPR)$ is defined as:
\begin{eqnarray}
MIPR = \sum\limits_{n=1}^{D} |C_n|^4.
\end{eqnarray}
For an extremely localized eigenstate $MIPR=1$ whereas for a perfect
delocalized eigenstate $MIPR=1/D$. The spectrum-averaged $MIPR$ as a
function of $\Delta$ for different interaction strengths $V$ is shown
in Fig.~\ref{int_mipr}(a),(b) and (c) for system size $N=18$ and
filling $\nu=1/9,1/6$ and $2/9$ respectively. The plots also confirm
the three distinct phases: $MBL$, thermal and mixed observed during
the examination of level-spacing ratio $r$. The $MBL$ phase possesses
high $MIPR\approx1$ while the thermal phase has a low $MIPR$
representing a dip at $\Delta\approx2J$. The mixed phase is observed
at the low disorder strengths $\Delta<<J$. In the mixed phase, another
small dip is observed at $V\approx\Delta$  corresponding to an increment
in the amount of delocalization.
\begin{figure}
			  	\includegraphics[width=0.494\columnwidth]{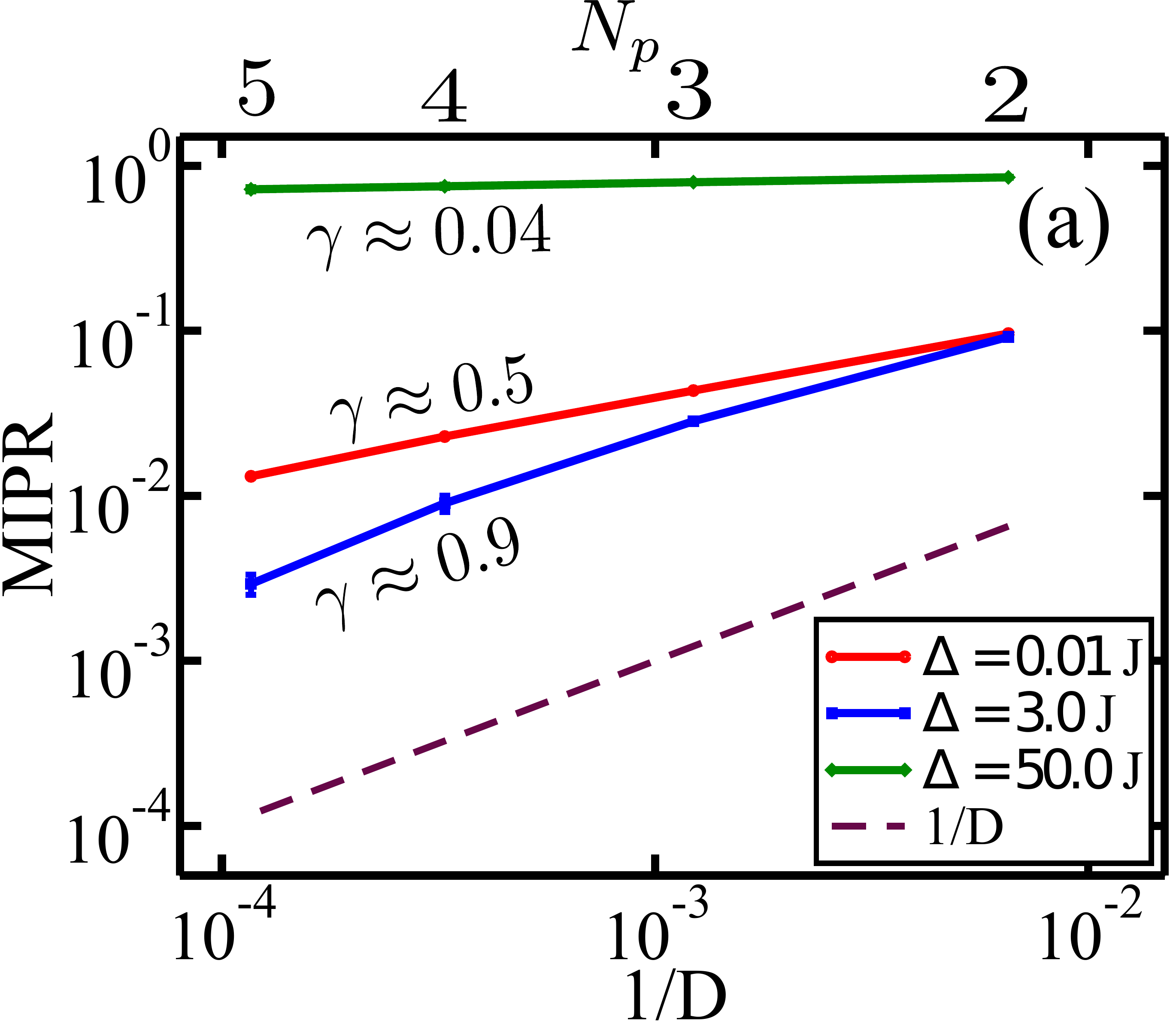}
			  	\includegraphics[width=0.494\columnwidth]{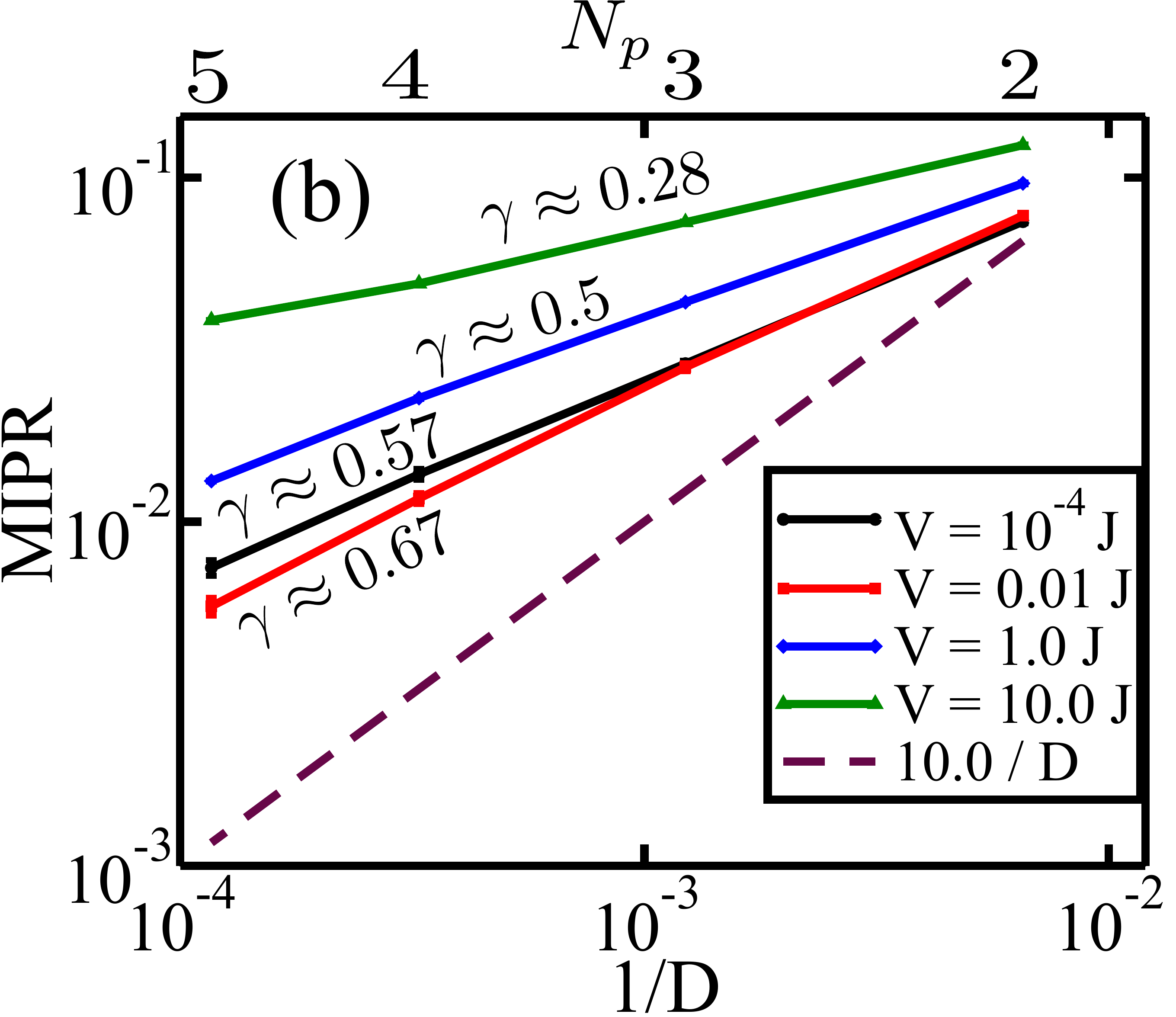}
			  	\caption{The log-log plot of the spectrum-averaged $MIPR$ with $1/D$ (where $D={N
			  	  \choose N_p}$), keeping the system size $N=18$ and increasing the number of fermions from $N_p=2,3,4$ and $5$ for (a) fixed interaction strength $V=1.0J$ and different disorder strength $\Delta$ and (b) for fixed $\Delta=0.01J$ and different interaction strength $V$. The dashed line shows a linear relationship between $MIPR$ and $1/D$.}
			  	\label{iprscaling_filling}
\end{figure} 

Apparently the dips of $MIPR$ at the intermediate disorder strength become deeper with increasing filling
fraction indicating more delocalization. To confirm this we carry out
a careful scaling analysis with $D$. 

{\it Effect of filling fraction: }First we explore the effect of
filling fraction $\nu$ by looking at the scaling relation $MIPR\propto
\frac{1}{D^\gamma}$. For a perfectly delocalized many-body phase
$\gamma=1$ whereas for a non-
ergodic many-body phase $0<\gamma<1$. Deep in the non-ergodic MBL phase $\gamma$ can be as small as close to $0$~\cite{mace2019multifractal}. Fixing $V= 1.0J$, we show the $MIPR$ as a function of
$1/D$ for different disorder strengths in
Fig.~\ref{iprscaling_filling}(a).  The scaling analysis shown in the
figure indicates that, as filling fraction increases, the many-body
system is in the $MBL$ phase for large $\Delta$ and is delocalized in
the intermediate range of $\Delta$. In the low $\Delta$ regime, the
system is a mixed one with $\gamma=0.5$.  To explore
  the behavior of the mixed phase, in
  Fig.~\ref{iprscaling_filling}(b), we plot $MIPR$ vs $1/D$ for
  different $V$ by fixing $\Delta=0.01J$. The increasing values of
  $\gamma$ w.r.t interaction indicates that, in a finite system, this
  phase is a mixed one and the amount of delocalization is high (red
  curve) when $V\approx\Delta$. For $V=10.0J$ in
  Fig.~\ref{iprscaling_filling}(b), the value of $\gamma$ is much
  smaller due to the strong repulsive interaction aiding
  localization.
\begin{figure}
			  	\includegraphics[width=0.7\columnwidth]{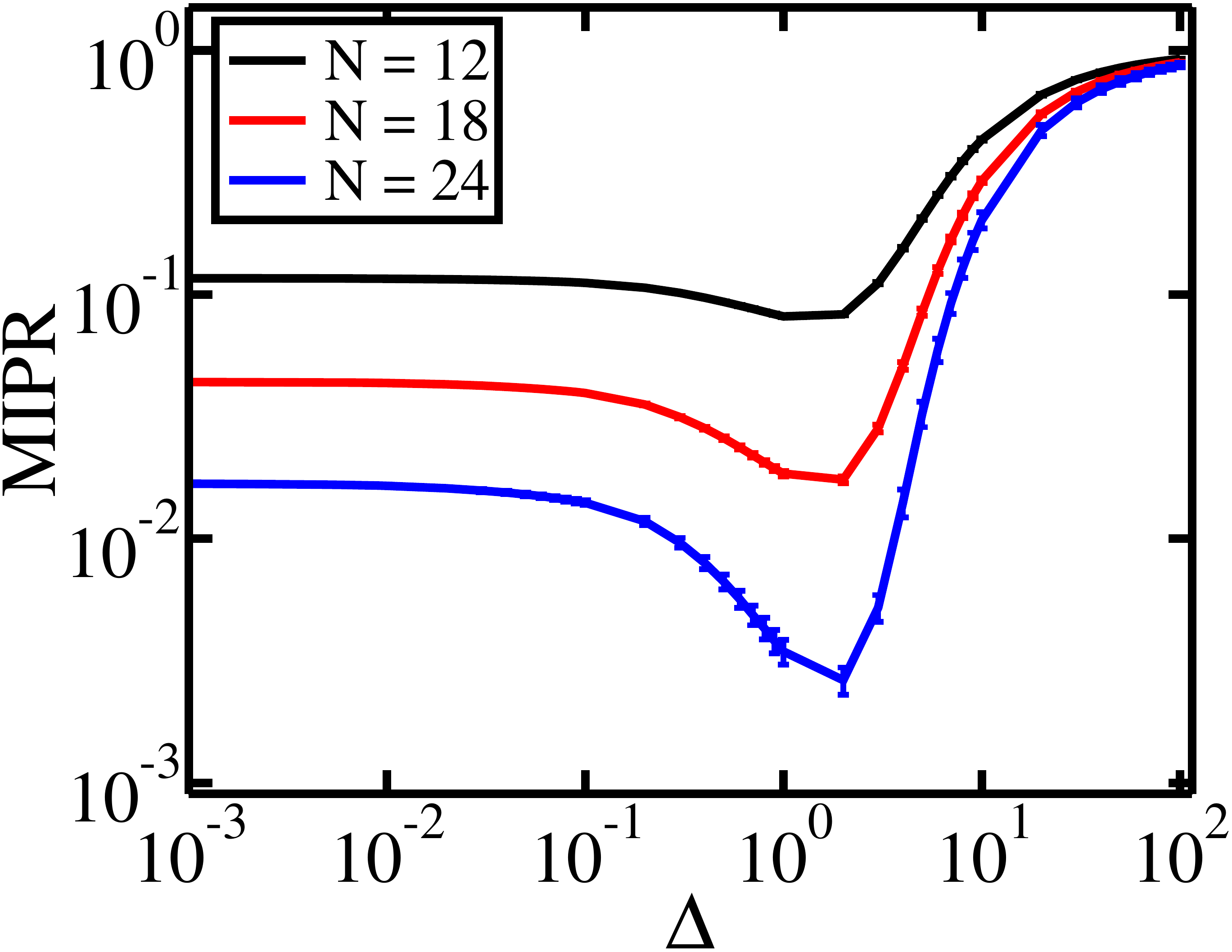}
			  	\caption{The spectrum averaged $MIPR$ as a function of disorder strength $\Delta$ (in units of $J$) for fixed $V=1.0 J$ and system sizes $N=12,18$ and $24$ for filling fraction of fermions $\nu=1/6$. The quantities are averaged over $500,200$ and $30$ disorder realizations for $N=12,18$ and $24$ respectively for fermion filling $\nu=1/6$.}
			  	\label{mipr_size}
\end{figure}  
\begin{figure}
			  	\includegraphics[width=0.6\columnwidth]{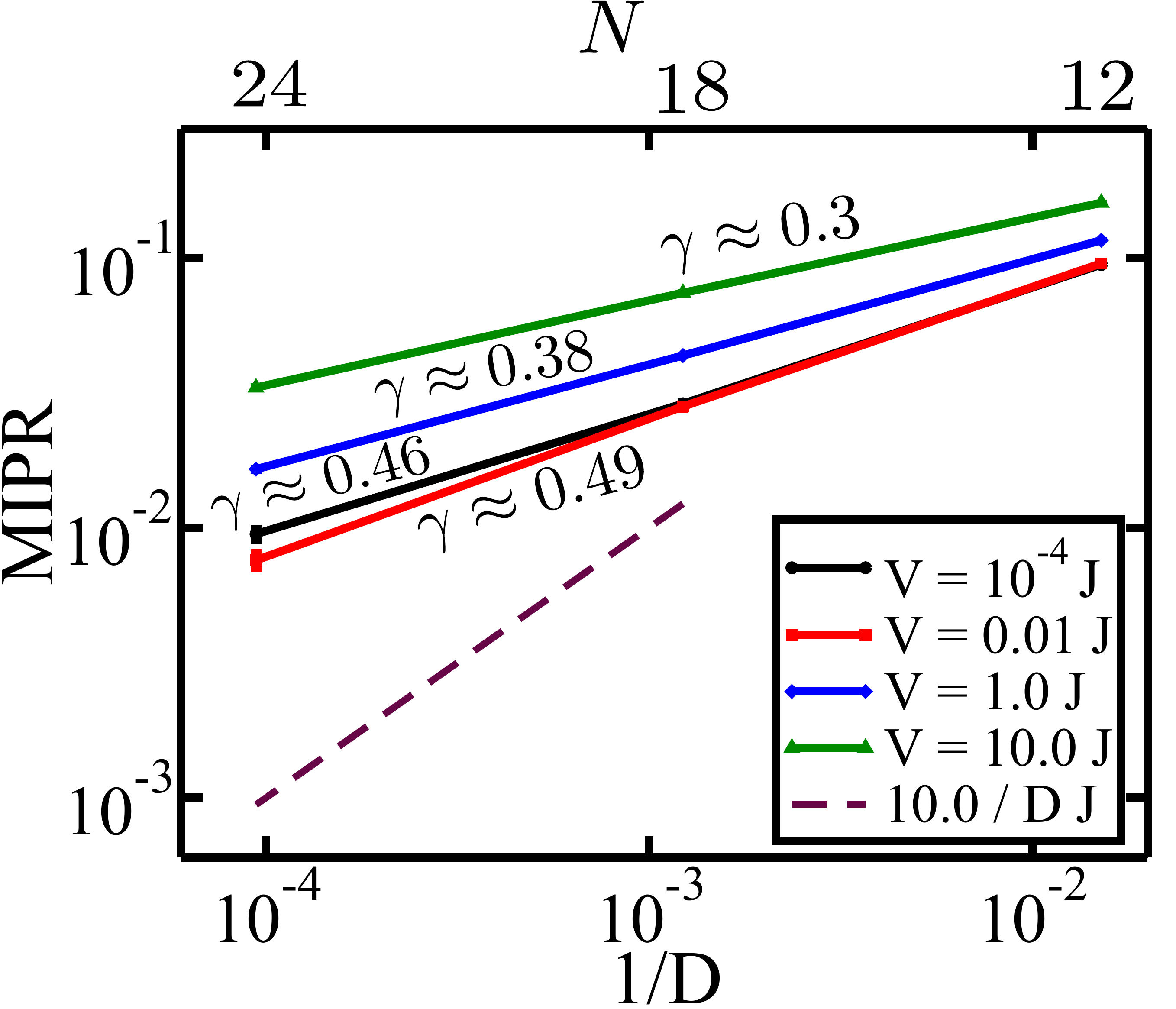}
			  	\caption{The log-log plot of the spectrum-averaged $MIPR$ with $1/D$ (where $D={N
			  	\choose N_p}$), keeping the system filling fraction $\nu=1/6$ and increasing the system size from $N=12,18$ and $24$ for fixed $\Delta=0.01J$ and increasing interaction strength $V$. The dashed line shows a linear relationship between $MIPR$ and $1/D$.}
			  	\label{iprscaling_size}
\end{figure}

{\it Effect of system size: } In Fig.~\ref{mipr_size}
  we show the spectrum averaged $MIPR$ as a function of disorder
  strength $\Delta$ for system sizes $N=12,~18,~24$ and for filling
  fraction $\nu=1/6$, and interaction strength $V=1.0 J$. $MIPR$
  decreases rapidly with $N$ for the intermediate values of $\Delta$
  indicating a delocalization in the many-body system. The trend of
  the plots suggests that the many-body system may thermalize for
  intermediate values of $\Delta$ as $N$ increases. However, for large
  $\Delta$, $MIPR$ barely changes implying a possible many-body
  localization in the system. In the small $\Delta$ regime $MIPR$
  decreases with increase in system size, but not as fast as compared
  to the intermediate $\Delta$ regime. This implies that for the small
  disorder strengths $\Delta$, the system does not attain
  thermalizaion and appears to be in a distinct nonergodic phase as the system size increases. Next we analyze the effect of system size $N$ in the
  low-disorder phase. To explore the characteristics of the phase at
low disorder strengths, we fix $\Delta=0.01J$, and plot $MIPR$ vs
$1/D$ for different $V$ in Fig.~\ref{iprscaling_size}.  The plots
affirm the conclusions made from Fig.~\ref{iprscaling_filling}(b); the
values of $\gamma$ indicate that the phase is neither thermal nor MBL
and hence it is dubbed as the `mixed phase'. The amount of delocalization in the mixed phase increases when
$V=\Delta$. For $V=1.0J$, $\gamma=0.38$ implying localization-dominated mixed phase. What happens to the fate of the `mixed phase' in the
thermodynamic limit may need a further study which goes
beyond exact diagonalization.

\subsection{Energy-resolved level statistics, inverse participation ratio and entanglement entropy}\label{resolved}

Here we carry out an energy-resolved study of the level-spacing ratio,
many-body inverse participation ratio and entanglement entropy to get
some further insights into the system, especially when it is
in the mixed phase. We divide the many-body energy spectrum into a
number of equal segments and calculate the local average of the
quantities for each segment of the spectrum. We denote the energy
resolved quantities as $r(\varepsilon)$, $MIPR(\varepsilon)$ and
$S_A(\varepsilon)$ respectively, where $\varepsilon$'s are the
fractional eigenstate index at the middle of each segment.
 
\begin{figure}
  \includegraphics[width=0.494\columnwidth]{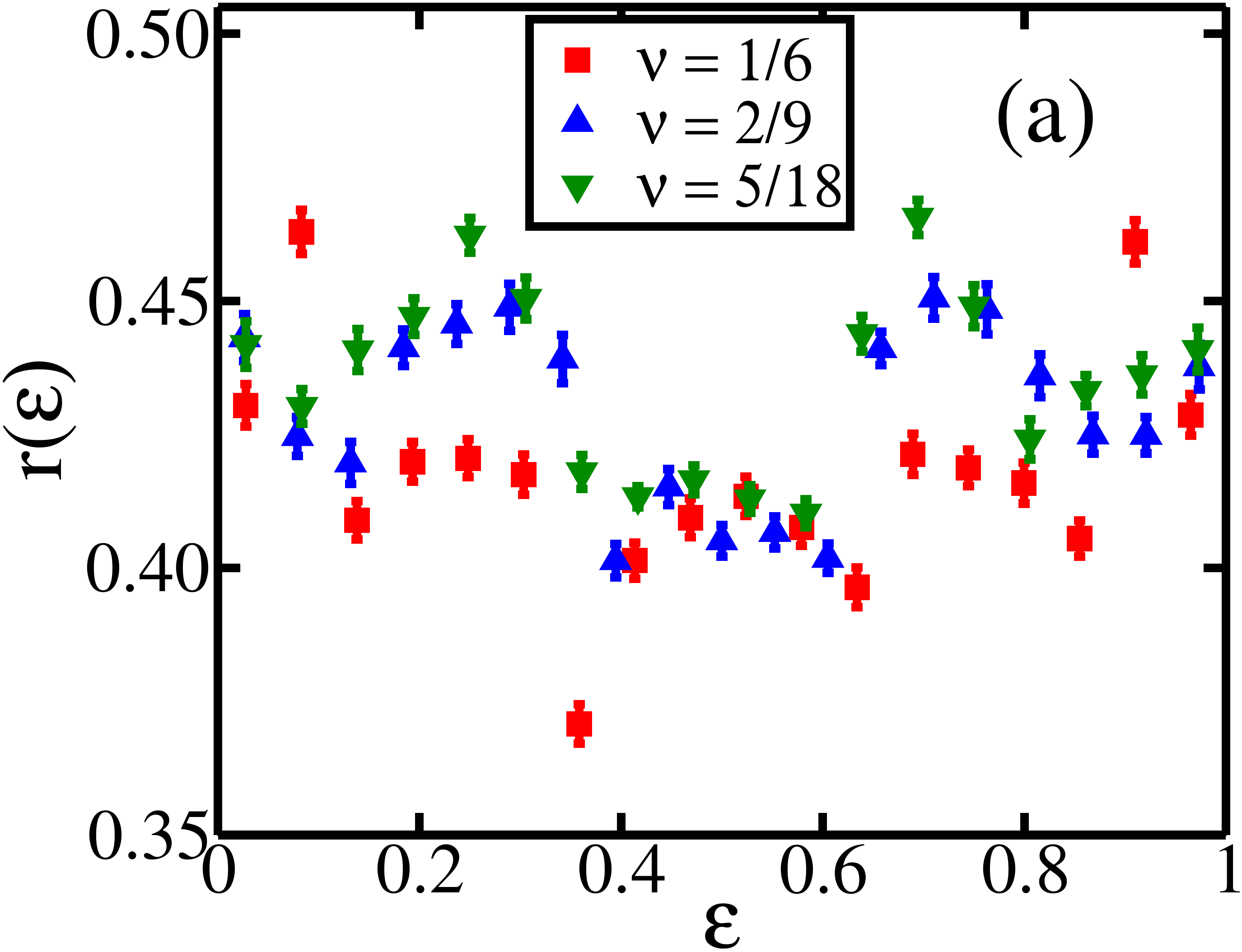}
  \includegraphics[width=0.494\columnwidth]{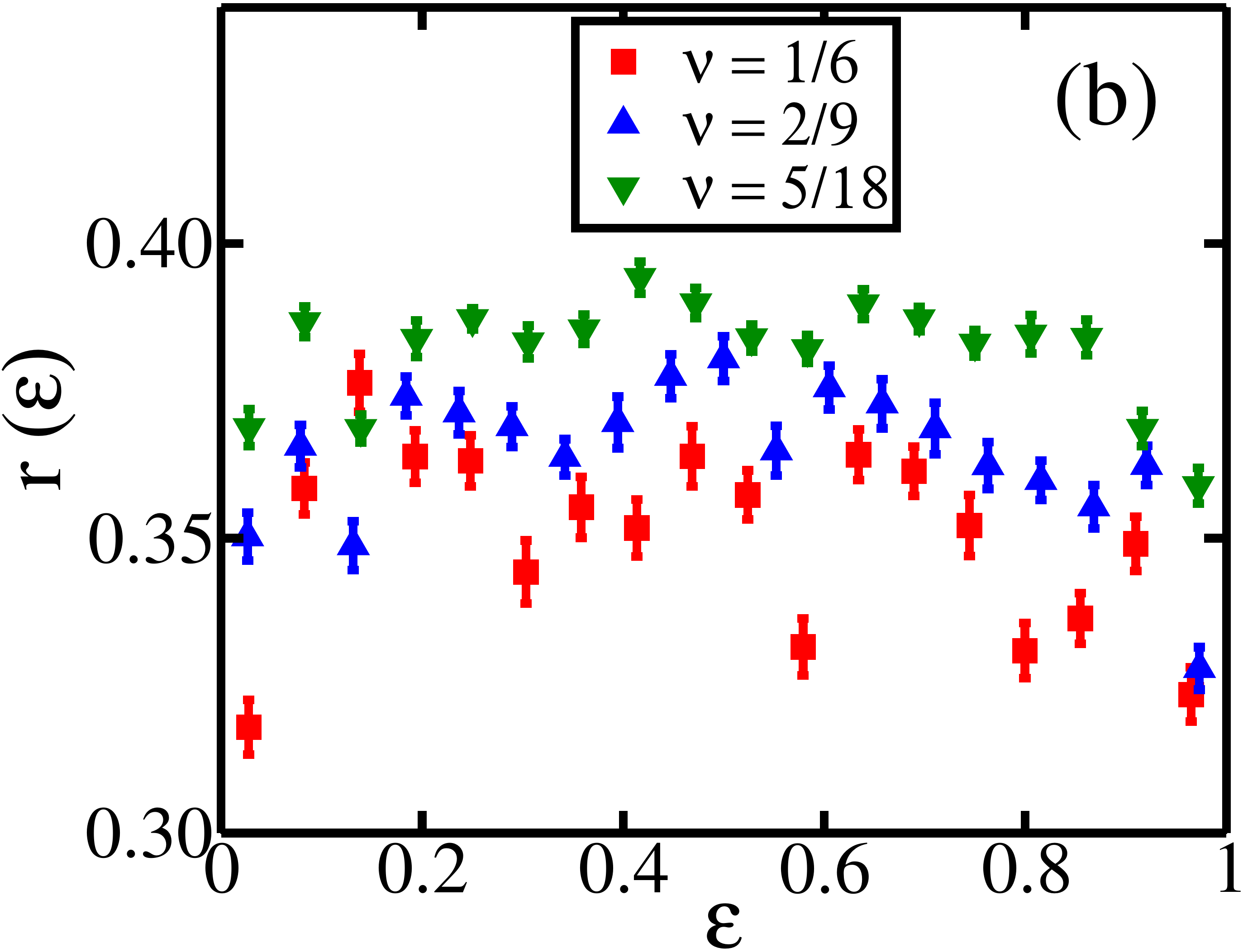}
  \caption{(a) The energy resolved $r(\varepsilon)$ for increasing values of filling fraction $\nu=1/6,2/9,5/18$ for $N=18$ and $V=\Delta=0.01J$. (b) Similar plots but for $V=1.0J$ and $\Delta=0.01J$. For all the plots, number of disorder realizations is $200$.}
  \label{resolved_rvsnu}
\end{figure}

We first discuss the energy resolved level-spacing ratio
$r(\varepsilon)$ in the small disorder $\Delta<<J$ regime. In
Fig.~\ref{resolved_rvsnu}(a) we show $r(\varepsilon)$ for
increasing filling fraction $\nu$ for $V=\Delta=0.01 J$. With
increasing $\nu$, $r(\varepsilon)$ increases and shifts away
from $0.386$ implying more delocalization for $V=\Delta$ as
predicted by other analysis done in the paper. In
Fig.~\ref{resolved_rvsnu}(b) $r(\varepsilon)$ for increasing filling
fraction $\nu$ is shown for $V=1.0 J$ and $\Delta=0.01 J$. Here, with
increasing $\nu$, $r(\varepsilon)$ approaches $0.386$
except at the edges of the spectrum. This is a signature of
localization in the system. However from the $MIPR\propto
1/D^{\gamma}$ scaling one concludes that it is a mixed phase as
$\gamma=0.5$.
\begin{figure}
  \includegraphics[width=0.494\columnwidth]{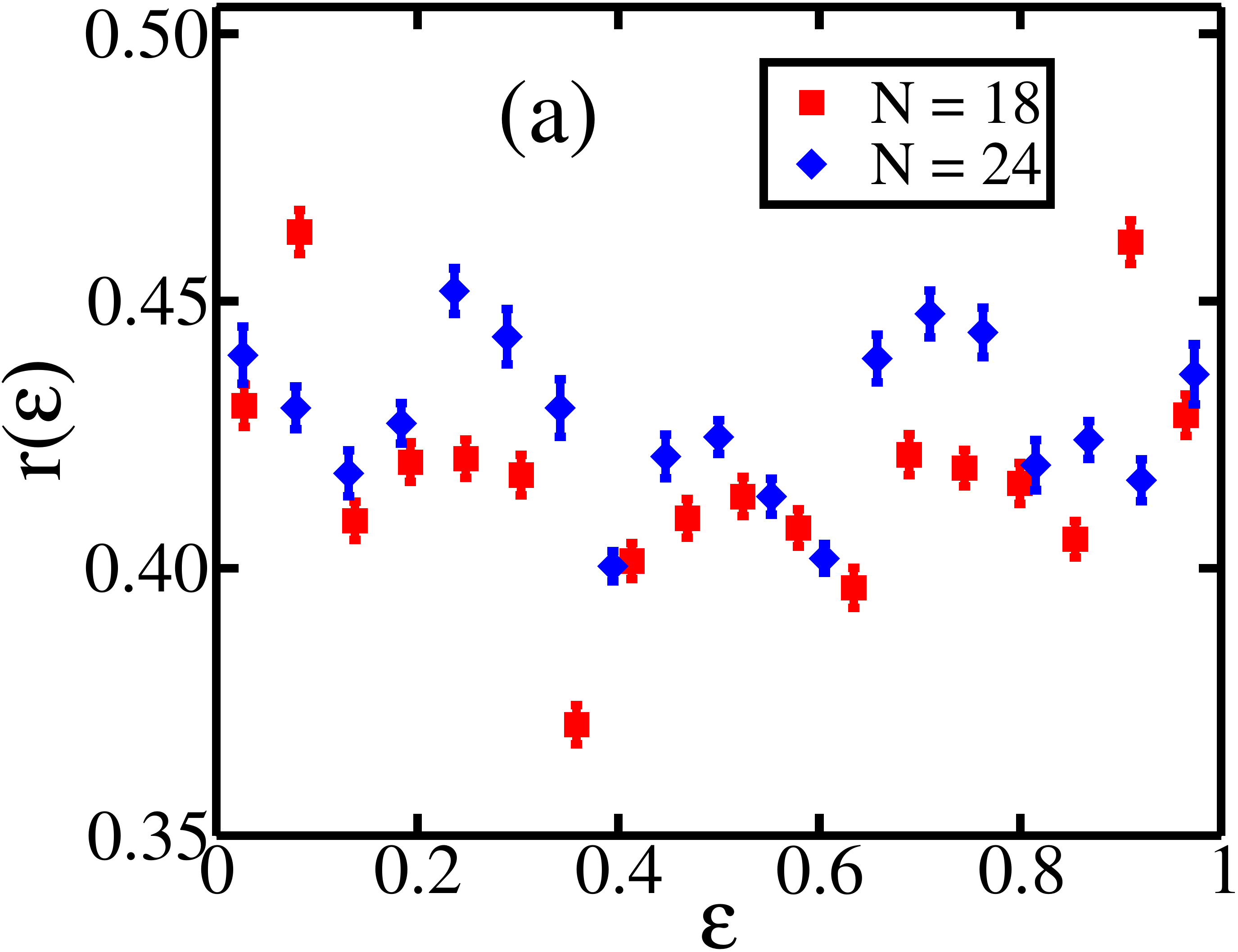}
  \includegraphics[width=0.494\columnwidth]{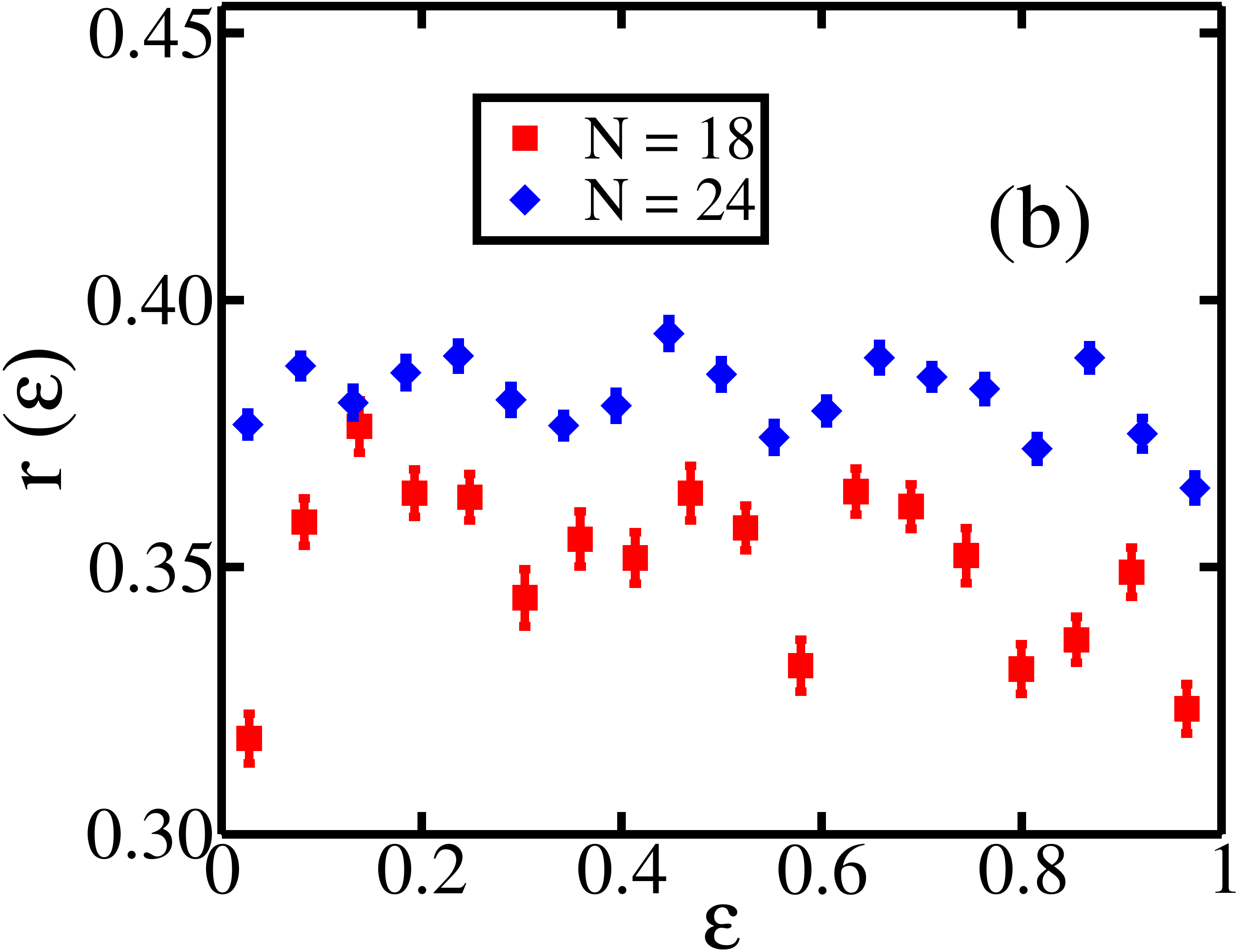}
  \caption{(a) The energy resolved $r(\varepsilon)$ for increasing values of system size $N=18,24$ for $\nu=1/6$ and $V=\Delta=0.01J$. (b) Similar plots but for $V=1.0J$ and $\Delta=0.01J$. For all the plots, number of disorder realizations is $200$.}
  \label{resolved_rvsNs}
\end{figure}
We also show $r(\varepsilon)$ with increasing system size $N$ for
fixed $\Delta=0.01 J$ and $V=0.01 J$ and $1.0 J$ in
Fig.~\ref{resolved_rvsNs}(a) and Fig.~\ref{resolved_rvsNs}(b)
respectively. When $V=\Delta=0.01 J$, $r(\varepsilon)$ increases from
$0.386$ indicating a delocalizing effect in the system. When $V=1.0 J$
and $\Delta=0.01 J$, $r(\varepsilon)$ is consistently spread around
$0.386$ for $N=24$, indicating an effective localization in the
system. Also from $MIPR$ scaling we find $\gamma=0.38$ implying the
dominance of the localized eigenstates. Finite-size effects are rather
severe here and make it difficult to make a conclusive statement on whether
the system will actually become many-body localized in the
thermodynamic limit.

\begin{figure}
  \includegraphics[width=0.8\columnwidth]{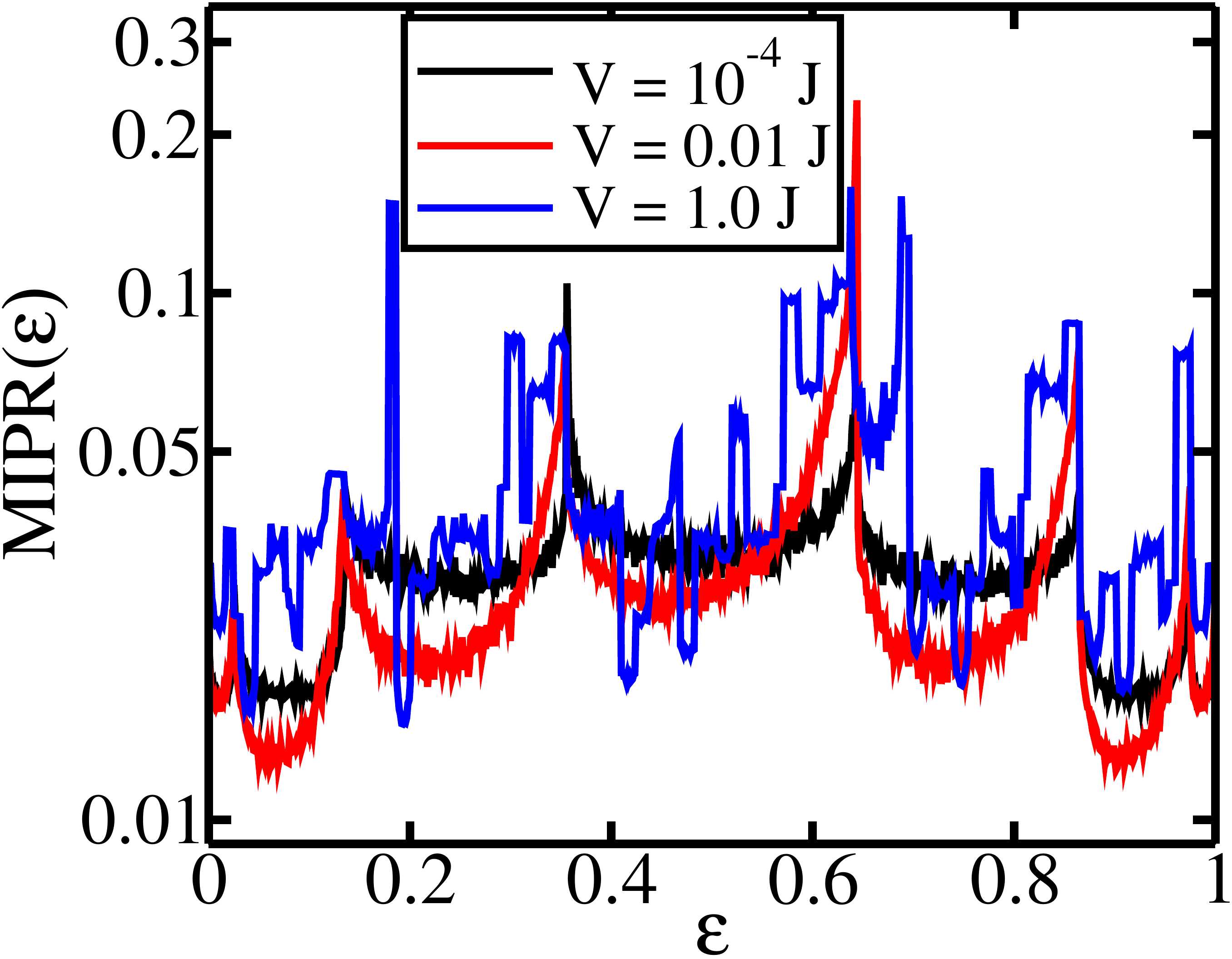}
  \caption{The $MIPR(\varepsilon)$ of all the eigenstates as a
    function of the fractional eigenstate index $\varepsilon$ for
    $\Delta=0.01J$ and different interaction strengths $V$. For all
    the plots, $N=18$, $\nu=1/6$ and the number of disorder
    realizations is $200$.}
  \label{resolved_ipr}
\end{figure}

The $MIPR$ of all the many-body eigenstates is shown in
Fig.~\ref{resolved_ipr} for $\Delta=0.01J$ and different interaction
strengths $V$, while keeping $N=18$ and $\nu=1/6$. The figure captures the
localization properties of all the eigenstates in the small $\Delta$
regime. We can see that for all values of $V$, the $MIPR$ of all the
eigenstates are significantly higher than $1/D$ but not close to $1$
implying that the system is in a mixed phase. Although when
$V=\Delta$, the majority of the eigenstates show delocalization tendency that
amounts to an overall delocalizing effect in the system. As $V>\Delta$
many of the eigenstates show high values of $MIPR$ with occasional
presence of the low-$MIPR$ eigenstates. This keeps the system still in
a mixed phase with a greater proportion of localization.
\begin{figure}
  \includegraphics[width=0.494\columnwidth]{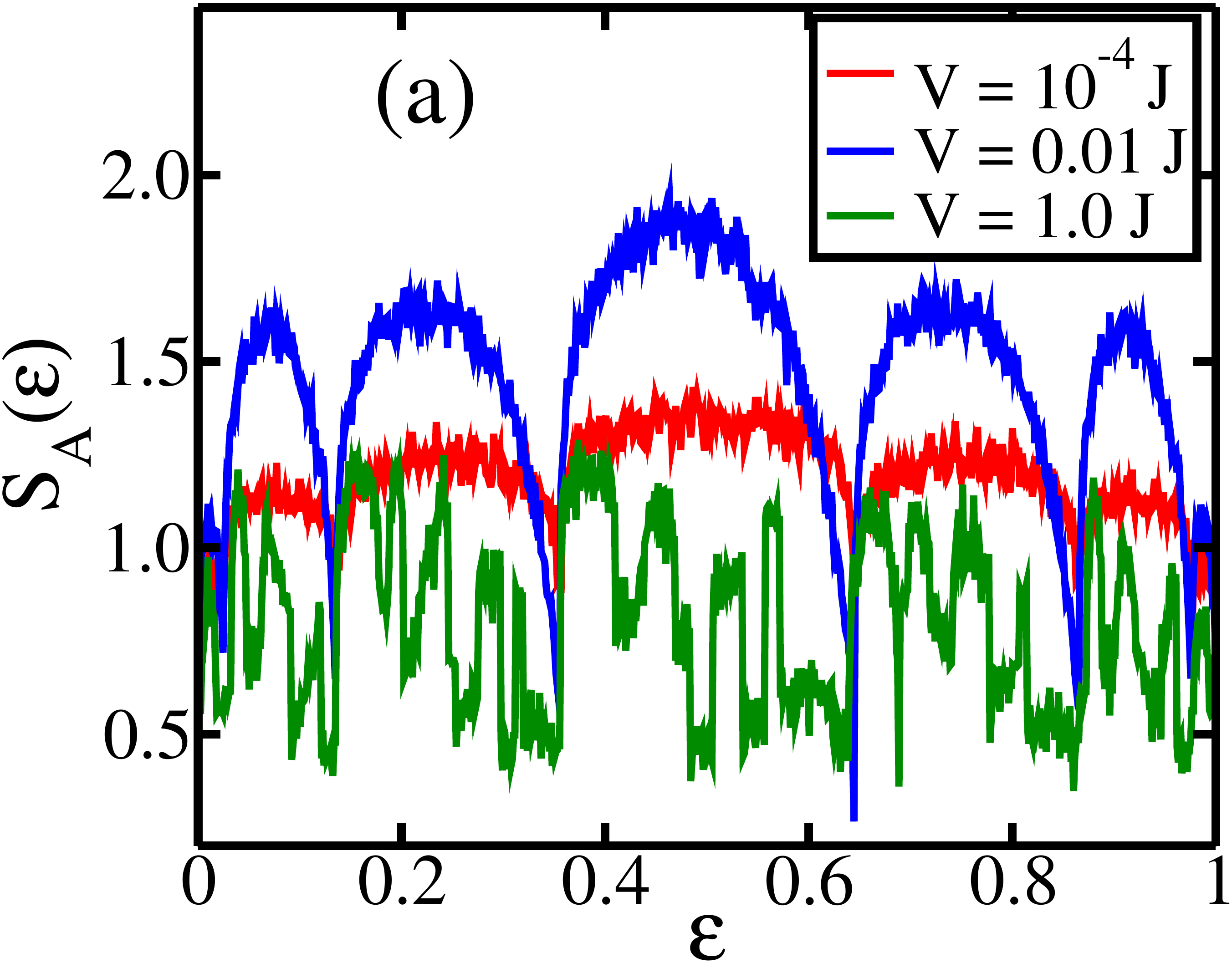}
  \includegraphics[width=0.494\columnwidth]{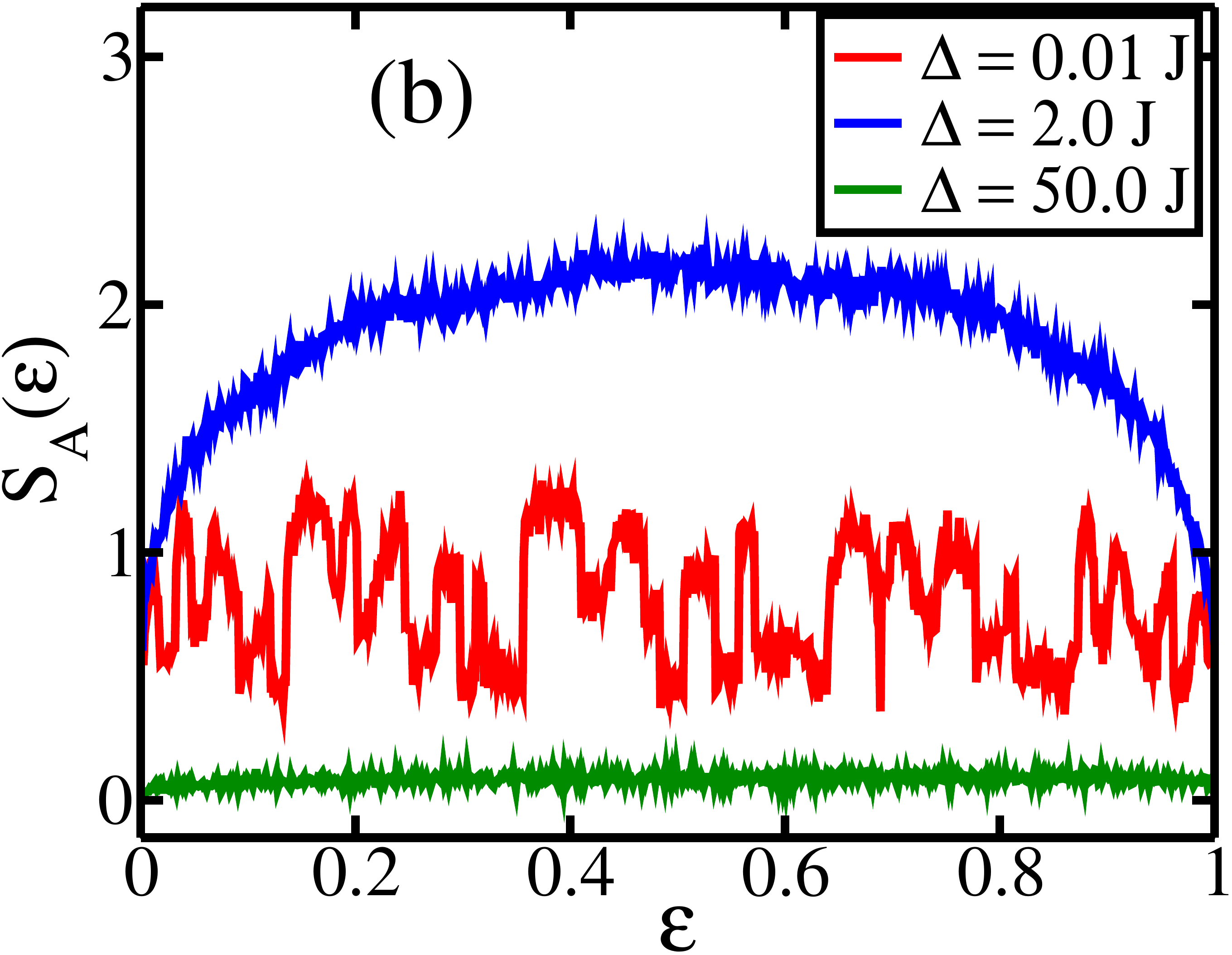}
  \caption{The half-chain entanglement entropy $S_A(\varepsilon)$ of
    all the eigenstates as a function of the fractional eigenstate
    index $\varepsilon$ (a) for increasing values of $V$ and fixed
    $\Delta=0.01J$ and (b) for increasing values of $\Delta$ and fixed
    $V=1.0J$. For all the plots, $N=18$, $\nu=1/6$ and number of
    disorder realizations is $200$.}
  \label{resolved_ent}
\end{figure}  

As an accompanying quantity the half-chain entanglement entropy $S_A$
of all the many-body eigenstates is shown in
Fig.~\ref{resolved_ent}(a) for $\Delta=0.01J$ and different
interaction strengths $V$, while once again keeping $N=18$ and
$\nu=1/6$ fixed. It is noticeable that when $V<<\Delta$ there are
eigenstates with high and occasional low entanglement
entropies. Although when $V=\Delta$, the majority of the eigenstates
has significantly high entanglement entropy whereas the few occasional
eigenstates have significantly low entanglement entropies, indicating
dominance of delocalization in the mixed phase. As $V>\Delta$ majority
of the eigenstates shows low entanglement entropy implying a dominance
of localization. In Fig.~\ref{resolved_ent}(b) half-chain $S_A$ of all
the eigenstates are shown for interaction $V=1.0J$ and $\Delta=0.01J,
2.0J, 50.0J$ such that the system is in the mixed, thermal and $MBL$
phases respectively. As can be seen from the figure, the thermal and
$MBL$ phases give rise to a smooth dependence of $S_A$ on eigenstates
with very high and very low $S_A$ respectively. In the mixed phase
eigenstate entanglement entropies have many humps and dips of
intermediate strength.

\subsection{Nonequilibrium dynamics of interacting spinless fermions}
\begin{figure}
  \includegraphics[width=0.494\columnwidth]{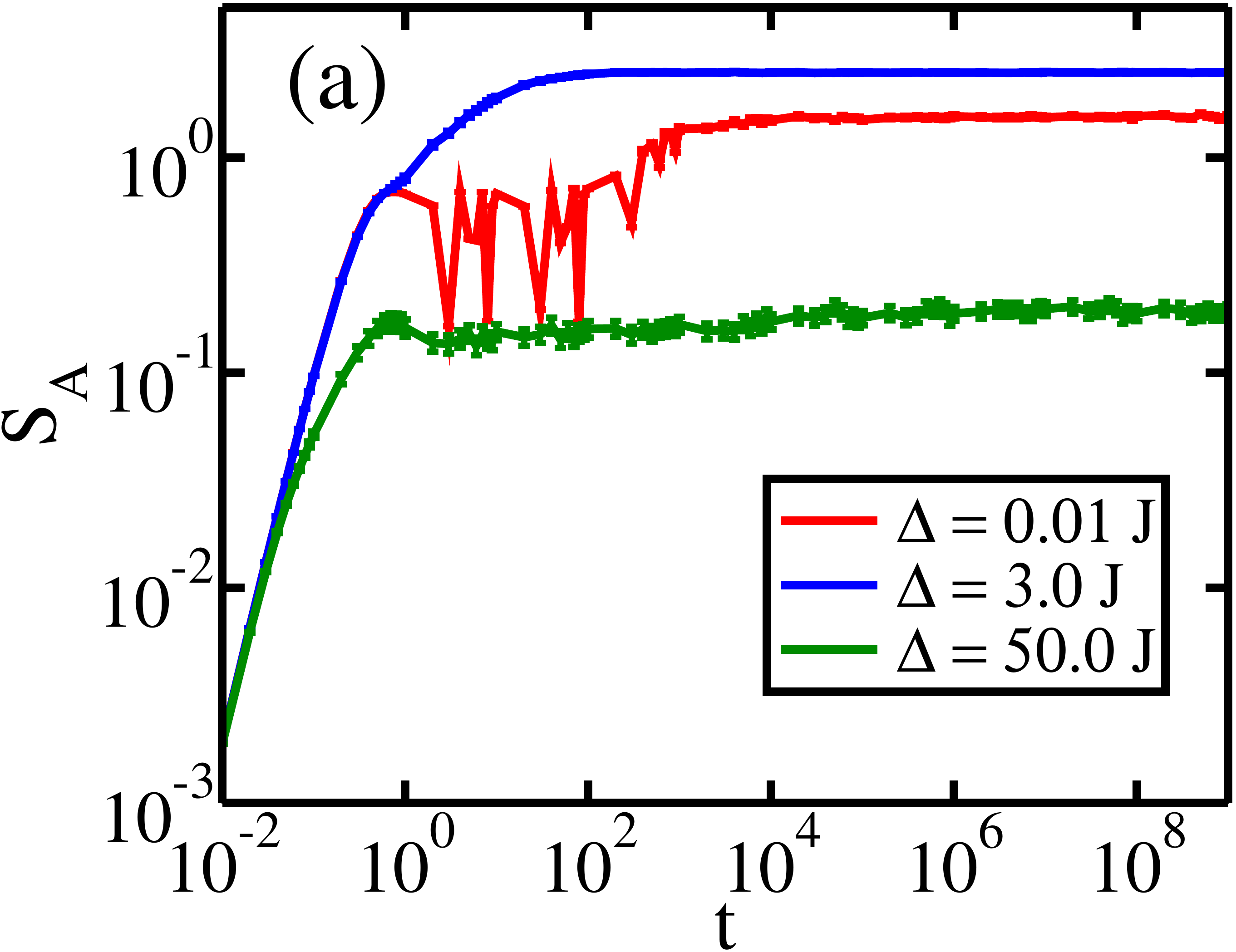}
  \includegraphics[width=0.494\columnwidth]{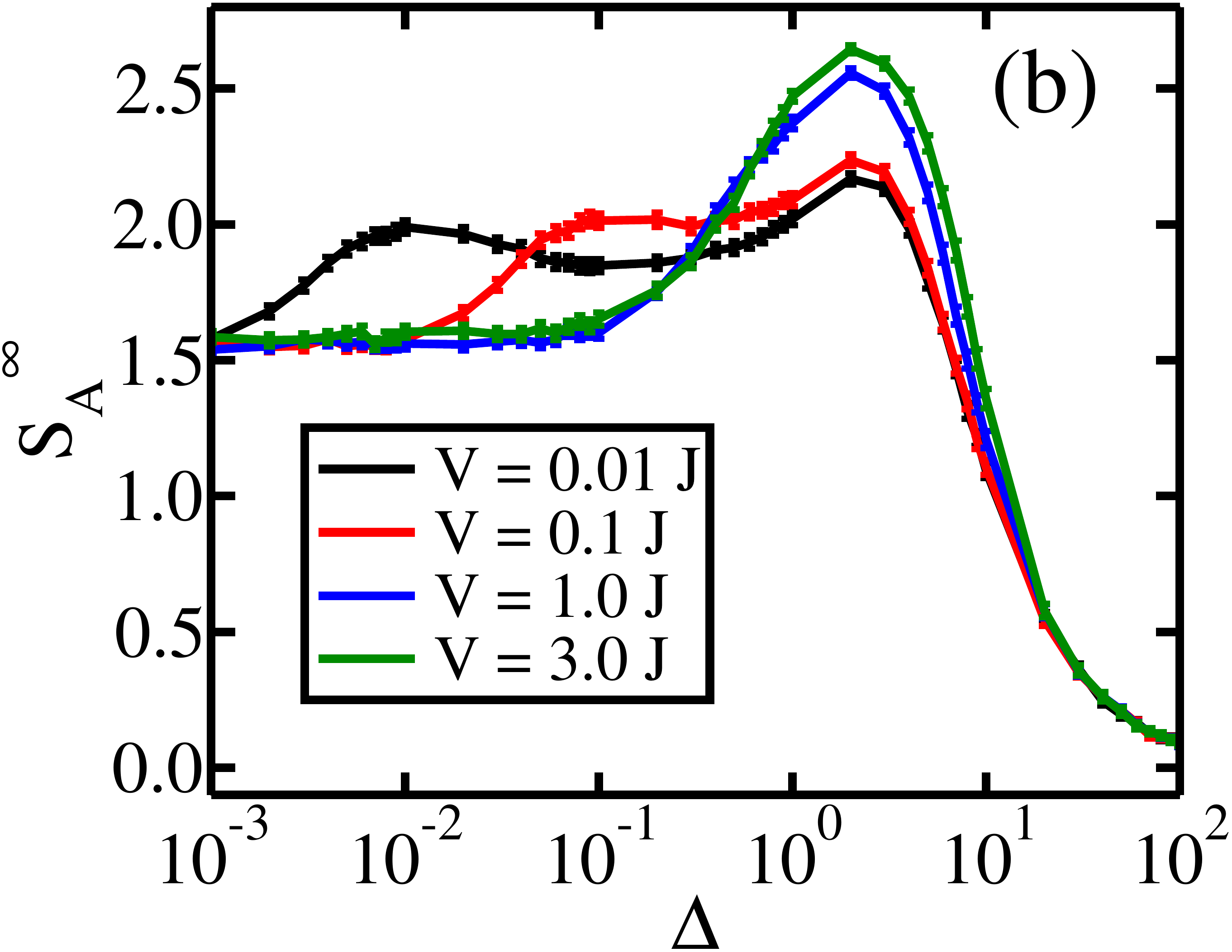} 
  \caption{(a) The half-chain entanglement entropy $S_A$ as a function of time $t$ (in units of $J^{-1}$) for $N=18$, $V=1.0J$ and $\Delta=0.01J,3.0J,50.0J$. (b) The saturation value $S_A^{\infty}$ as a function of $\Delta$ (in units of $J$) for increasing values of $V$ and $N=18$. For all the plots, number of disorder realizations is $200$.}
  \label{int_eedyn}
\end{figure}

In this subsection, we discuss many-body non-equilibrium dynamics,
keeping track of the entanglement entropy, the return probability and
imbalance parameter. We choose the initial state as an experimentally
relevant, density-wave type of state as mentioned previously
$\ket{\Psi_{in}}=\prod\limits_{i=1}^{N/3}\hat{c}_{2i}^\dagger\ket{0}$,
which is a product state with the filling fraction of fermions
$\nu=1/6$ with $N=18$. The dynamics of the entanglement entropy $S_A$
(the subsystem A is the first half of the full system) is calculated
via exact diagonalization for $V=1.0$ and is shown in
Fig.~\ref{int_eedyn}(a) for different disorder strengths. For low
$\Delta=0.01J$ after a super-ballistic transient, there is a damped
oscillatory behavior (see Appendix~\ref{appA} for zero-disorder case)
followed by a substantially sub-diffusive ($t^{0.05}$) regime till
$S_A$ reaches saturation. The little post-oscillatory increment in
$S_A$ is the effect of non-zero interaction $V$ leading to a mixed
phase with a small amount of delocalization in the system. For
intermediate disorder $\Delta=3.0J$, after the transient, there is a
sub-diffusive ($t^{0.3}$) increment in $S_A$ before it quickly
saturates to a large value reflecting delocalization in the many-body
system. The oscillatory part is absent in this case. For large
$\Delta=50.0J$, after the transient, $S_A$ saturates to a lower value
indicating many-body localization. A finer analysis of the dynamical
behavior of the MBL system is given in Appendix~\ref{app_MBL}. The
dependence of the saturation value of the entanglement entropy
$S_A^{\infty}$ on disorder strength $\Delta$ for $N=18$, $\nu=1/6$ and
different $V$ is shown in Fig.~\ref{int_eedyn}(b). Typically the
entanglement entropy in the many-body delocalized phase is
substantially higher as compared to the same in the $MBL$ phase. For
$V\approx J$, we observe the presence of three phases: $MBL$ for large
$\Delta$, delocalization for intermediate $\Delta$ and the mixed phase
for small $\Delta$. For $V<J$ another signature of delocalization is
noted when $V\approx\Delta$. All these results are consistent with the
results of $MIPR$ and $r$ as discussed in the previous
subsection. Plots similar to Fig.~\ref{int_eedyn}(b) are also obtained
for the spectrum averaged entanglement entropy (~Fig.~\ref{app_ee} in
Appendix~\ref{appB}).

\begin{figure}
  \includegraphics[width=0.494\columnwidth]{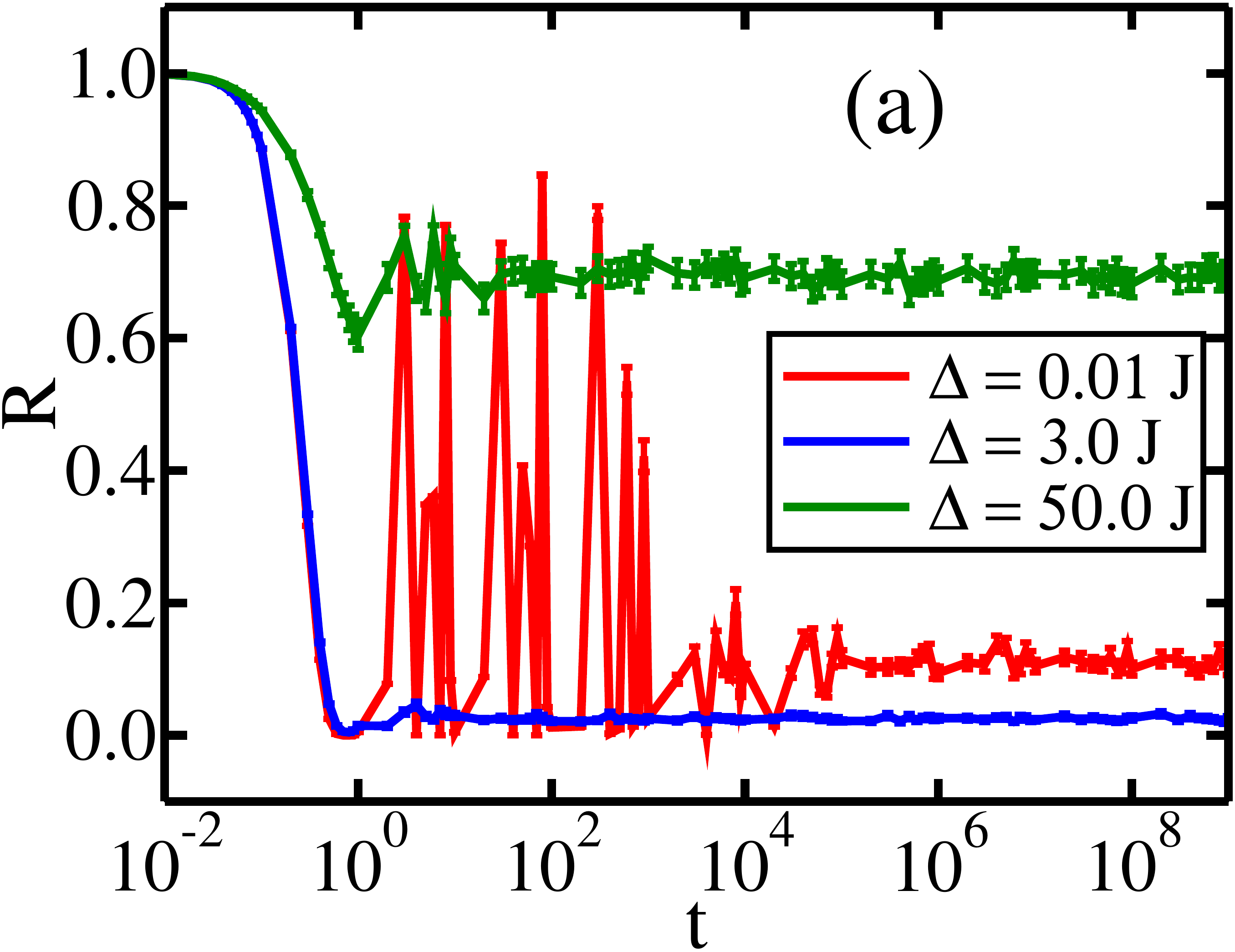}
  \includegraphics[width=0.494\columnwidth]{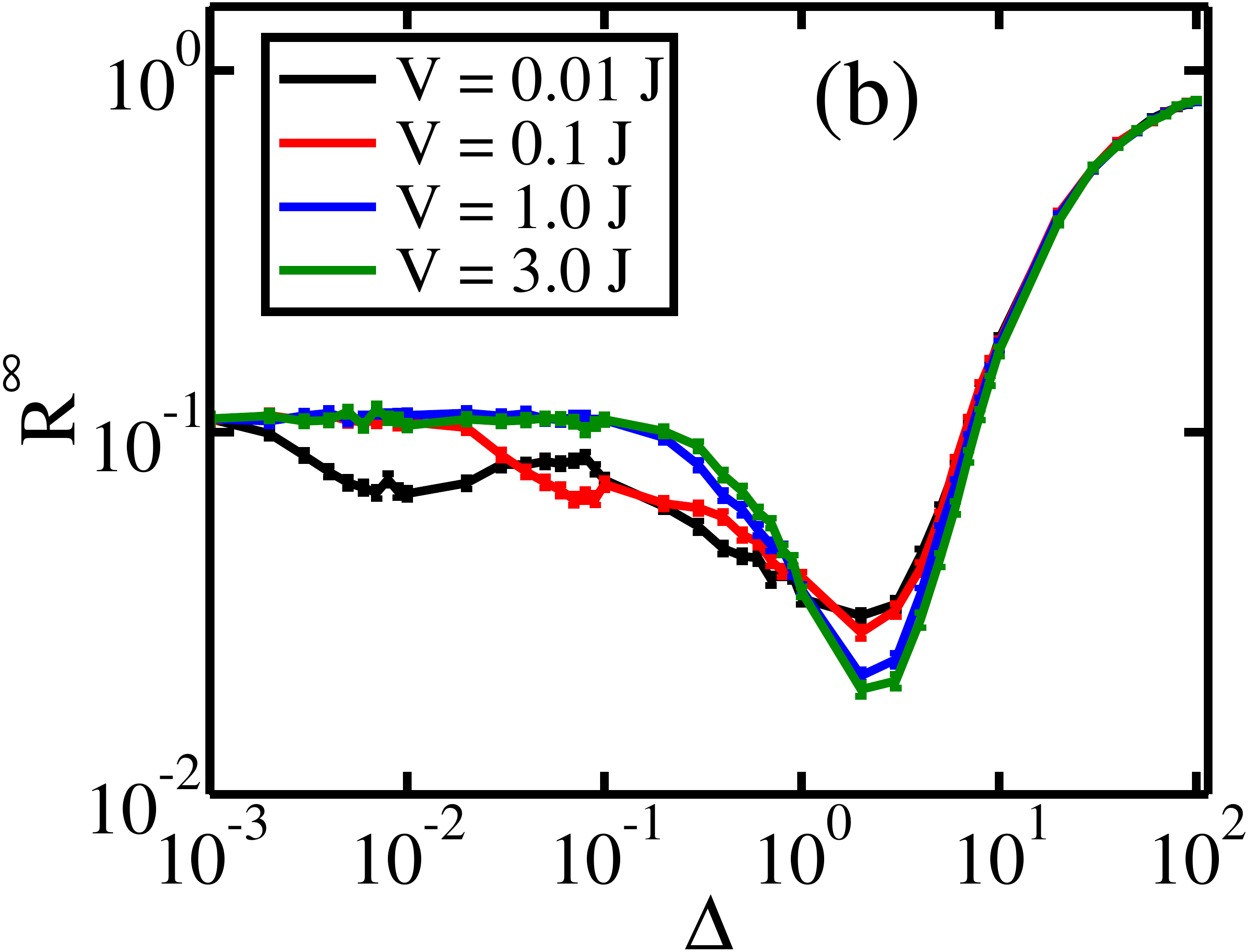} 
  \caption{(a) The return probability $R$ as a function of time $t$ (in units of $J^{-1}$) for $N=18$, $V=1.0J$ and $\Delta=0.01J,3.0J,50.0J$. (b) The saturation value of the return probability $R^{\infty}$ as a function of $\Delta$ (in units of $J$) for different values of $V$ and $N=18$. The number of disorder realizations is $200$ for both the plots.}
  \label{int_revivdyn}
\end{figure}
Although entanglement entropy offers useful insights, it is not easily
measurable in ongoing experiments. In this context, return probability
and imbalance parameter, being experimentally accessible, are
interesting and we study them next.
The return probability is defined
as $R(t) = |\braket{\Psi_{in}|\Psi_t}|^2$. In the perfectly
delocalized phase $R=1/D$ whereas in the perfectly localized phase $R$
is unity. 
In Fig.~\ref{int_revivdyn}(a) the dynamics of $R$ is
shown for $V=1.0J$ for different $\Delta$. For $\Delta=0.01J$, just
like $S_A$, $R$ also shows oscillatory behavior indicating recurrence
before it saturates to a finite value. 
For $\Delta=3.0J$ the
oscillatory behavior is absent and the saturation value is
sufficiently low indicating delocalization. For $\Delta=50.0J$, $R$
saturates to a significantly higher value indicating strong
localization. 
The saturation value $R^{\infty}$ vs $\Delta$ plots for
increasing values of $V$ are shown in Fig.~\ref{int_revivdyn}(b). 
From this figure one can infer the presence of three phases: $MBL$ for
large $\Delta$, delocalization for intermediate $\Delta$ and the mixed
phase for small $\Delta$. For $V<<J$ another signature of
delocalization is noted when $V\approx\Delta$: $R^{\infty}$ shows dips
at those points, but these dips are not as deep as those corresponding
to $V\approx J$.

\begin{figure}
   \includegraphics[width=0.494\columnwidth]{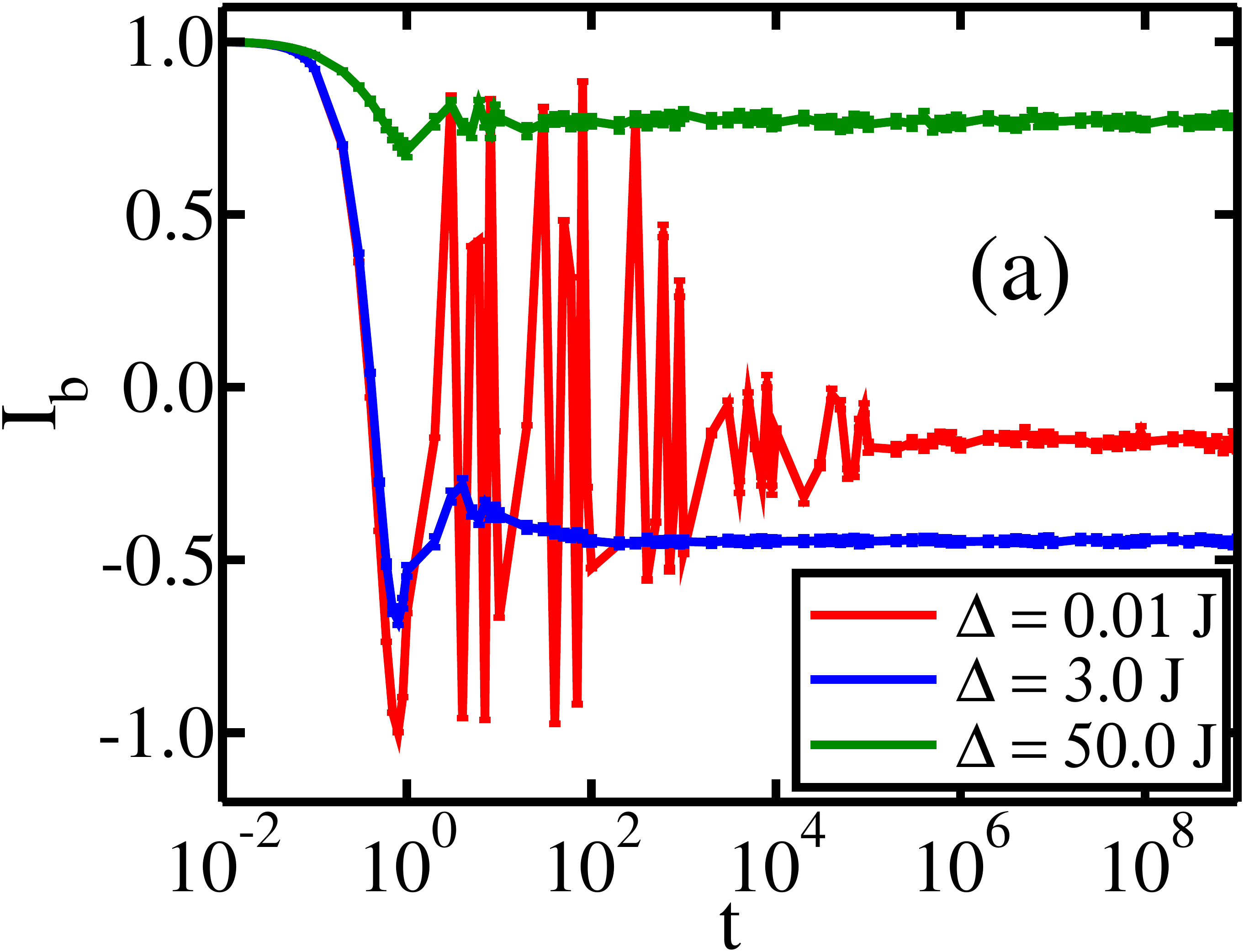}
   \includegraphics[width=0.494\columnwidth]{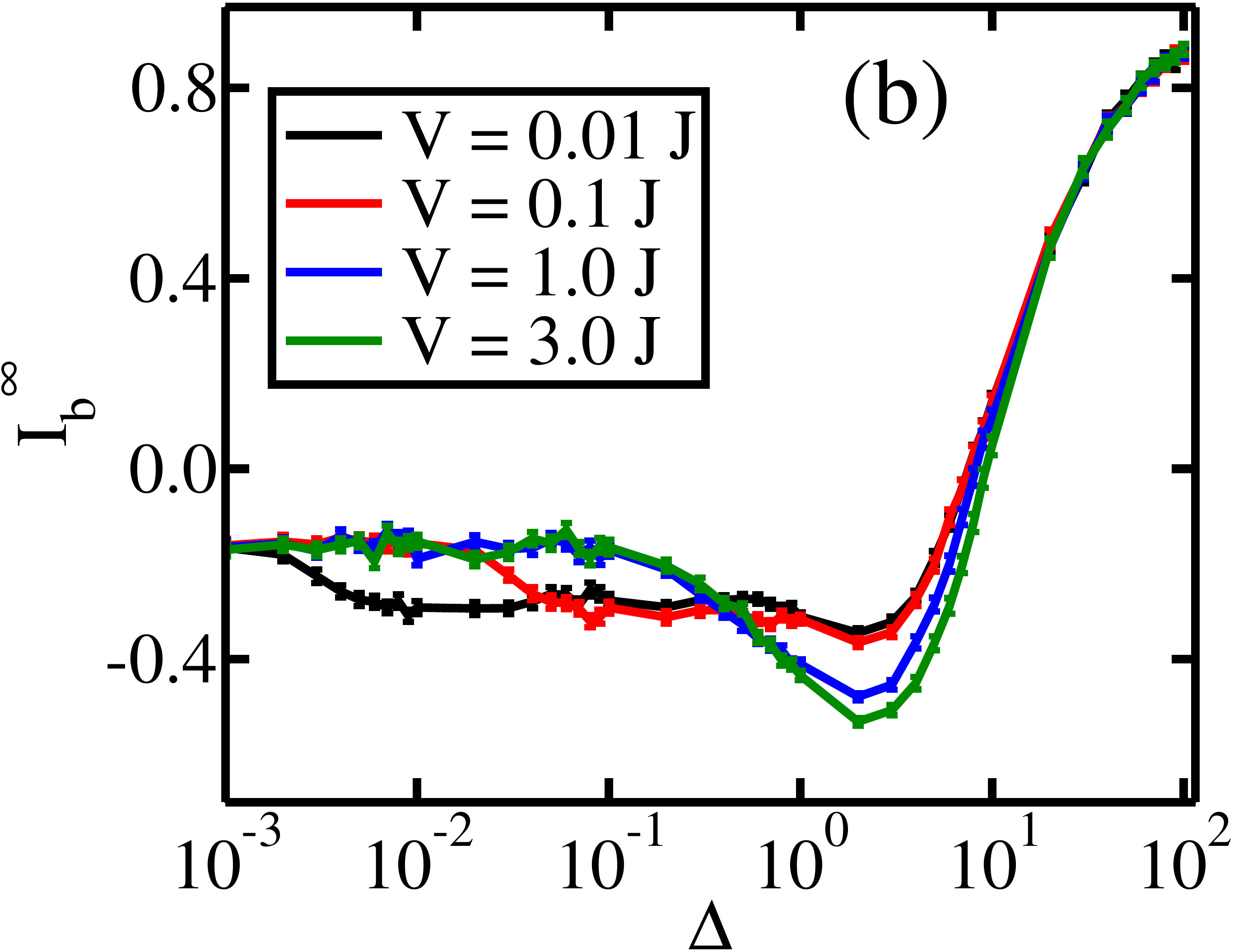} 
   \caption{(a) The imbalance parameter $I_b$ as a function of time $t$ (in units of $J^{-1}$) for $N=18$, $V=1.0J$ and $\Delta=0.01J,3.0J,50.0J$. (b) The saturation value $I_b^{\infty}$ as a function of $\Delta$ (in units of $J$) for different values of $V$ and $N=18$. For all the plots, number of disorder realizations is $200$.}
   \label{int_imbdyn}
 \end{figure}

 The imbalance parameter $I_b$ is already defined in Eq.~\ref{imb} in
 the previous section. In the thermalized and $MBL$ phases respectively,
 we expect the saturation values to be $I_b=-2/3$ and $I_b=1$
 respectively. In Fig.~\ref{int_imbdyn}(a) the dynamics of $I_b$ is
 shown for $V=1.0$ and different $\Delta$. For $\Delta=0.01J$,
 $I_b$ saturates to $-0.19$, which is higher than $-2/3$, after an
 oscillatory behavior due to recurrence. For $\Delta=3.0J$ the
 recurrence is absent and the saturation value is low indicating
 delocalization whereas for $\Delta=50.0J$, $I_b$ saturates to a much
 higher value implying localization in the many-body system. The
 saturation value $I_b^{\infty}$ vs $\Delta$ plots for different $V$
 are shown in Fig.~\ref{int_imbdyn}(b). 
 Similar to $S_A^{\infty}$ and
  $R^{\infty}$, $I_b^{\infty}$ confirms the presence of three phases when $V=\mathcal{O}(J)$:
  $MBL$ for $\Delta>>J$, delocalization for $\Delta\approx J$ and the
  mixed phase for $\Delta<<J$. For $V<<J$ an increment in the
  amount of delocalization is notable in the mixed phase when
  $V\approx\Delta$ as small dips are visible at those points.
To reaffirm our findings we have studied the system size dependence of the time-evolved many body state in the long time limit (Appendix~\ref{app_NPR}), which also indicates the presence of three distinct phases as observed in the study of imbalance parameter.

\section{Conclusions}
We report a systematic investigation of the effect of disorder and
interactions on a flat-band supporting diamond chain.  Disorder
detunes and hybridizes the compact localized states associated with
flat bands and as the strength of disorder is increased, flat-band
based localization and conventional Anderson localization are
observed. Observables obtained from both eigenvalues and eigenfunctions
shine light into the distinct features of the different phases.
We observe a hierarchy of localization: {\it compact localization $>$
  Anderson localization $>$ flat band based localization}. 

Single-particle wavepacket dynamics complements the results obtained via statics.
In the disorder-free limit, a persisting oscillatory recurrence which is attributed to the compact localized states
is observed.  A damped
oscillatory recurrence is observed in the flat band based localization
phase and no oscillatory recurrence is observed in the Anderson
localization phase. The on-site occupation probability of the single
particle in the long time limit confirms that the localization
strength can be classified as {\it compact localization $>$ Anderson
  localization $>$ flat band based localization}. We note that in
other all band flat lattices like one-dimensional cross-stitch
lattice~\cite{tovmasyan2018preformed} and two-dimensional dice
lattice~\cite{vidal2001disorder} the same hierarchy is maintained.

Non-interacting many particle fermion states also show characteristics of flat band based
localization and Anderson localization in the presence of low
and high disorder strengths. A study of entanglement entropy and
imbalance reveals a delocalizing tendency for intermediate
disorder strengths, reflecting the effect of fermionic statistics
in the system. The low-disorder and high-disorder phases are distinguished by the presence and absence of the characteristic damped oscillatory recurrence in the non-equilibrium dynamics of the entanglement entropy and imbalance. 

The interplay of disorder and interaction in a finite system can lead to three distinct
phases: $MBL$ for high disorder strength, thermal phase for
intermediate disorder strength and nonergodic `mixed phase' for low disorder
strengths. We observe that, in the mixed phase, the eigenstates tend
to get delocalized when interaction strength is of the order of
disorder-strength. A detailed analysis of the scaling of the
spectrum-averaged many-body inverse participation ratio with the
dimension of the Hilbert space is carried out by varying the filling
fraction and system size.  We observe that at higher
  filling fractions, the mixed phase in the finite system tends to
  have more dominance of delocalization as long as interaction-strength
  is less than the hopping-strength. However, an increment of
  localization in the mixed phase is observed for high enough interaction-strength. When the interaction strength is comparable to the hopping strength, for the intermediate and higher ranges of
  disorder-strength, one obtains the crossovers from the nonergodic `mixed'-to
  thermal and thermal-to-$MBL$ phases respectively, which become more
  distinct as the filling fraction and system size increase. Nonequilibrium dynamics
  supports the main findings from the above study involving statics. A
  characteristic damped oscillatory behavior is found in the mixed
  phase whereas no such oscillations are found in the thermalized and
  $MBL$ phases. In order to comment on the thermodynamic existence of
  the three phases, especially on the fate of the mixed phase, one
  needs a further study which is not limited by the Hilbert-space
  constraint of numerical exact diagonalization. We think that our work will
  help motivate such studies on the diamond chain in the future.

The joint presence of disorder and interactions in flat-band based
systems give rise to rich and unconventional phases obtained both from
single-particle and many-particle properties of the system. These
phases could potentially be realized in cold-atom based
experiments. This kind of a systematic study of flat-band supporting
systems is lacking in the literature. We believe that our work will
trigger more work on flat-band based systems.

\section*{Acknowledgements}
We are grateful to the High Performance Computing (HPC) facility at IISER Bhopal, where large-scale computations of this
project were run. N.R is grateful to the University Grants Commission (UGC), India for his Ph.D fellowship. A.S is grateful to SERB for the grant (File
Number: CRG/2019/003447), and the DST-INSPIRE Faculty Award
[DST/INSPIRE/04/2014/002461].

\appendix
\section{Oscillatory dynamics of the revival probability: single particle, noninteracting and interacting fermions}\label{appA}
\begin{figure*}
  \includegraphics[width=0.6\columnwidth]{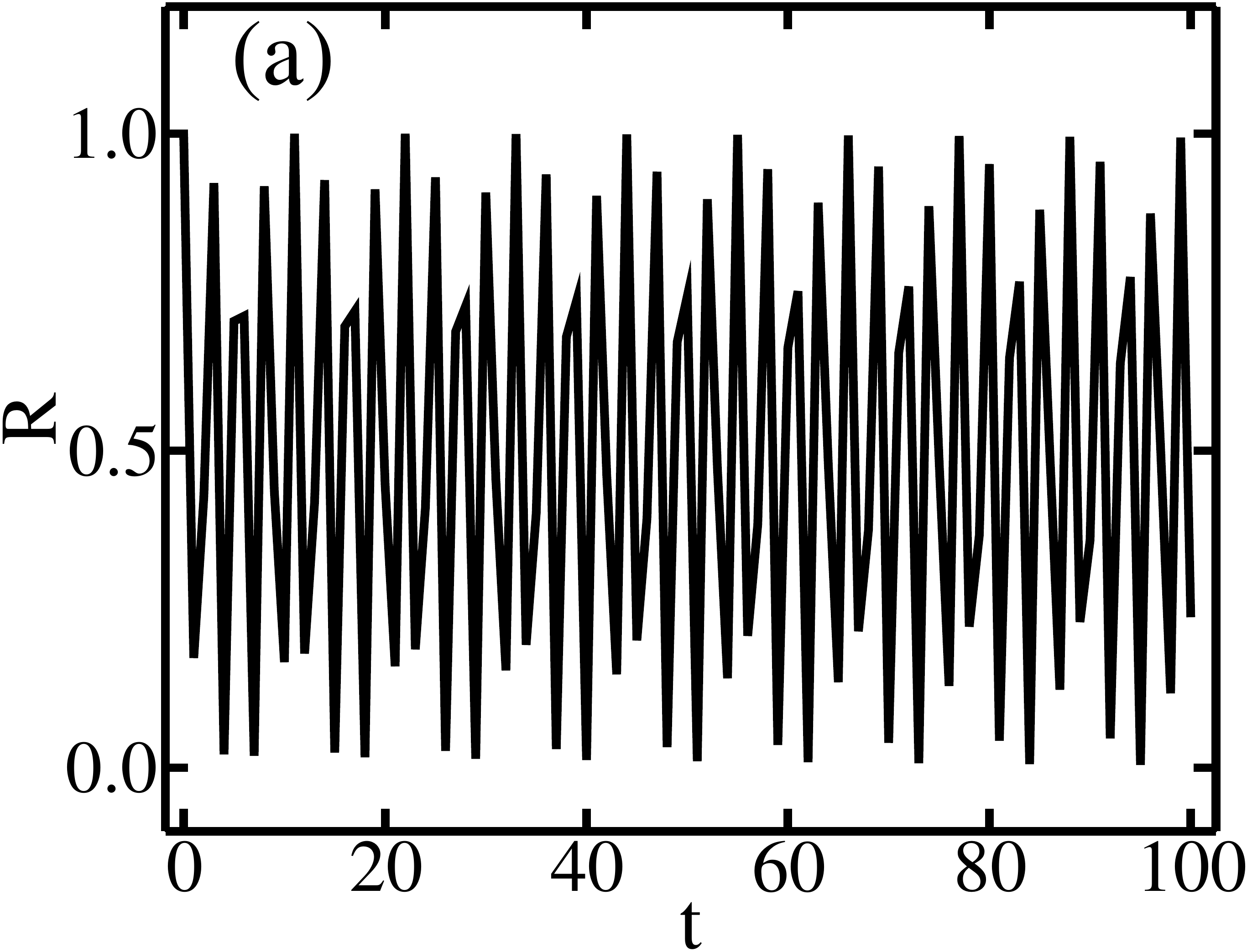}
  \includegraphics[width=0.6\columnwidth]{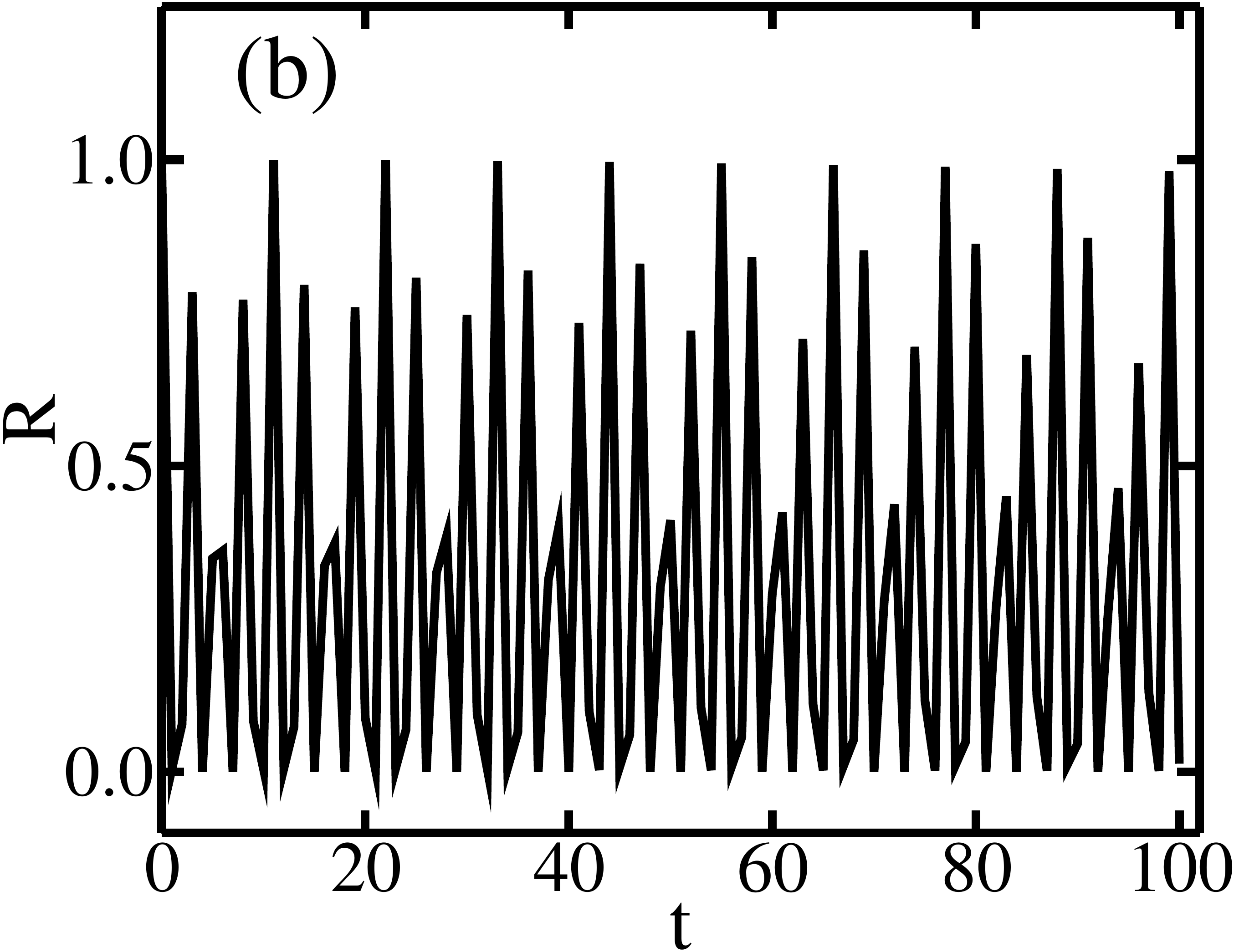}
  \includegraphics[width=0.6\columnwidth]{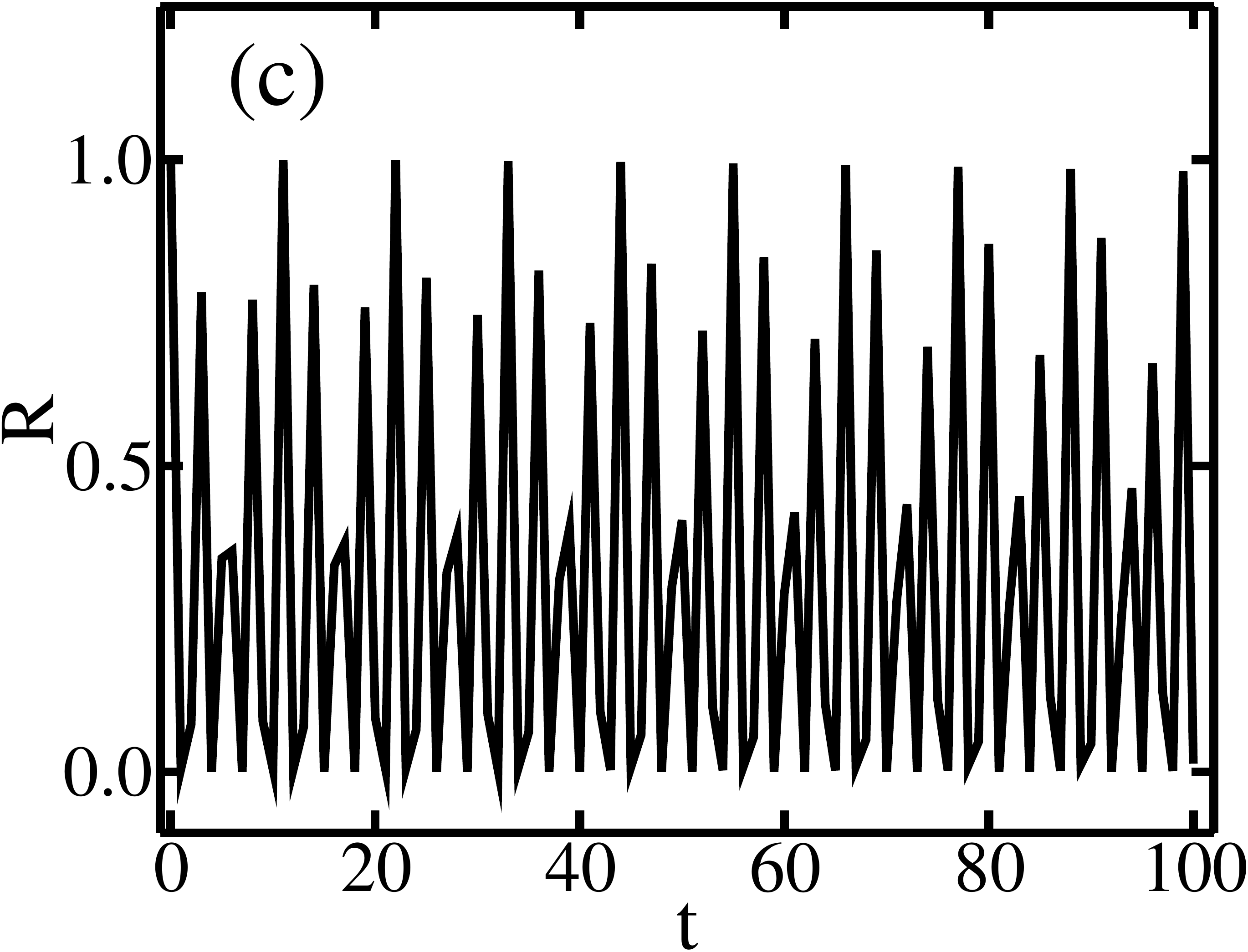}
  \includegraphics[width=0.6\columnwidth]{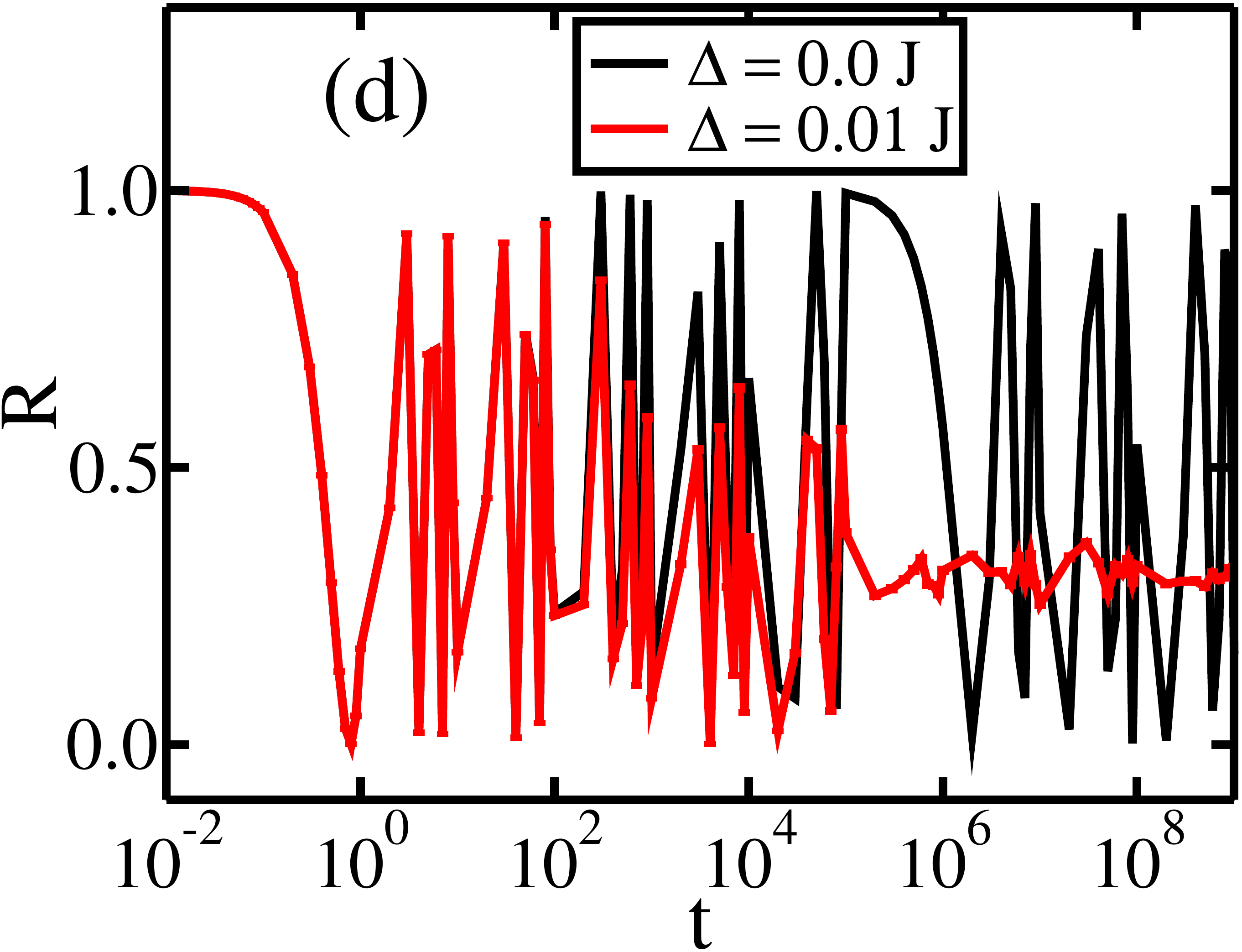}
  \includegraphics[width=0.6\columnwidth]{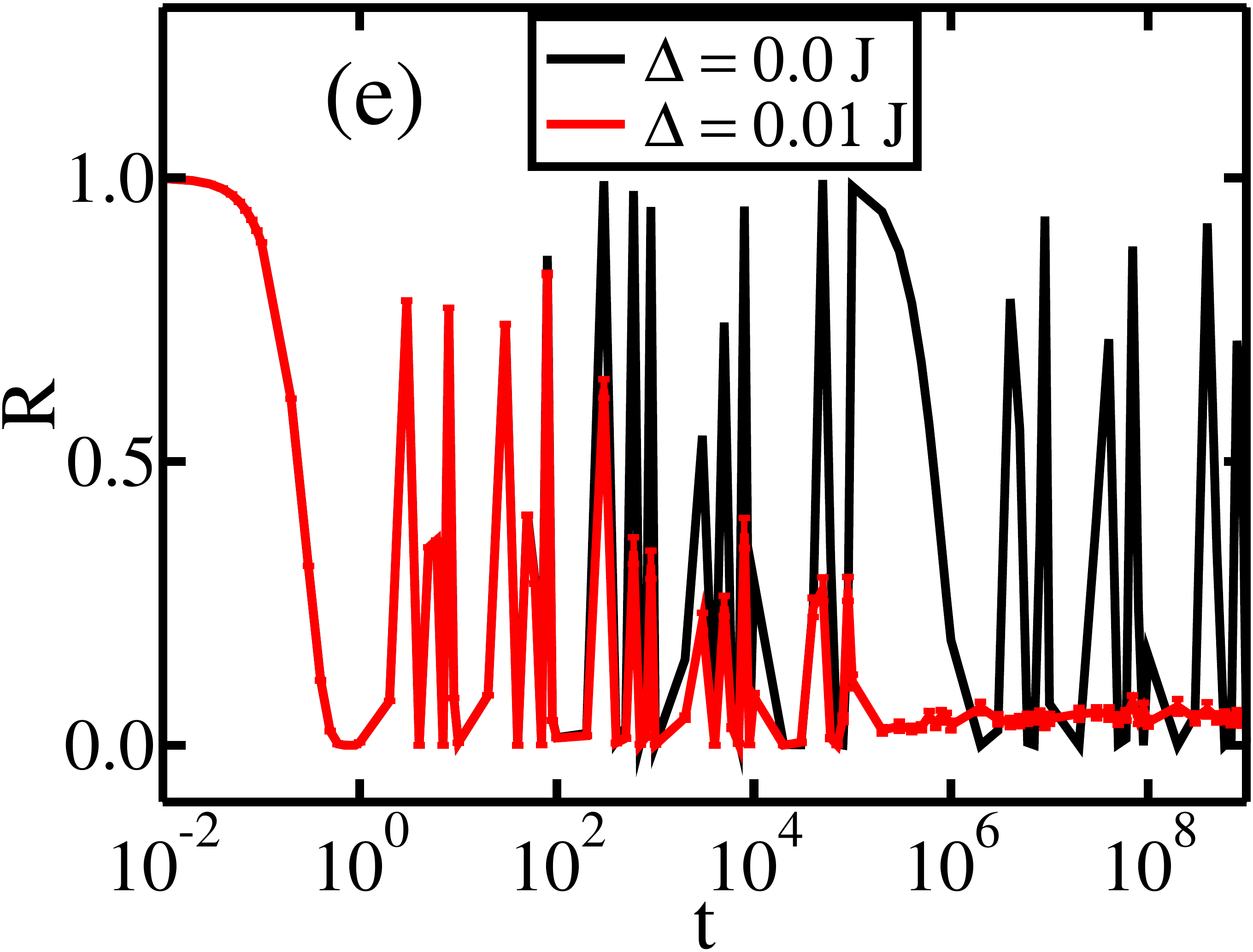}
  \includegraphics[width=0.6\columnwidth]{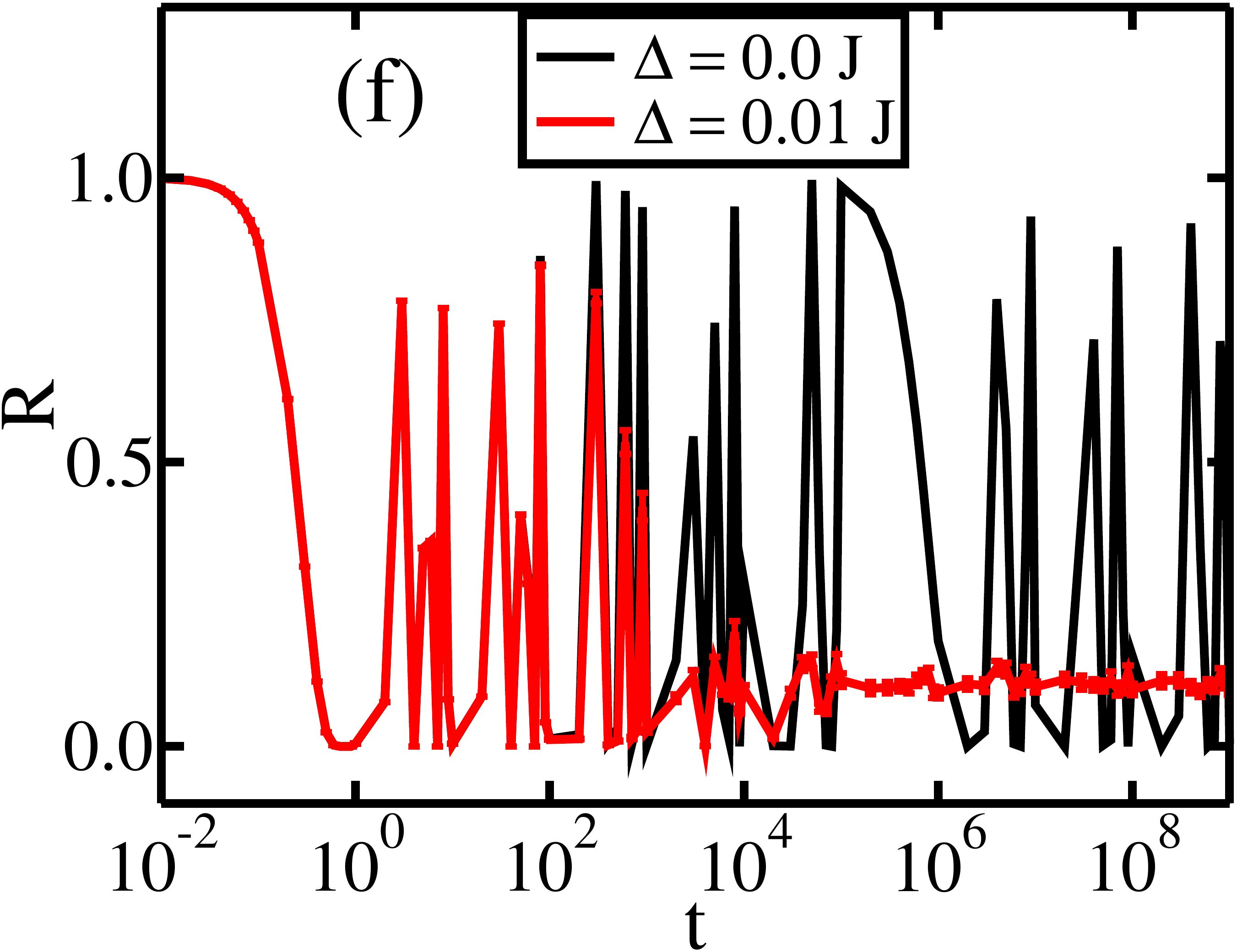}
  \caption{(a-c) The oscillatory behavior of the revival probability
    $R$ as a function of time (in units of $J^{-1}$) for a single particle, noninteracting
    fermions and interacting fermions respectively for
    $\Delta=0.0$. (d-f) The full dynamics of $R$ for $\Delta=0.0$ is
    compared to the that for small $\Delta=0.01J$ for a single
    particle, noninteracting fermions and interacting fermions
    respectively. Here, the time axis is in log-scale. For a single
    particle $N=600$ whereas for both noninteracting and interacting
    fermions $N=18$ with filling of fermions $\nu=1/6$. Interaction
    strength $V=1.0J$ for interacting fermions.}
  \label{app_dyn}
\end{figure*} 
Here we show the results for the dynamics of the revival probability of
a single particle, for non-interacting fermions and for interacting
fermions in the clean system. We also compare results for the clean
system with those for the system in the low-disorder regime. The
zero-disorder $(\Delta=0.0)$ dynamics of the revival probability $R$
is shown in Fig.~\ref{app_dyn}(a), (b) and (c) for a single particle,
noninteracting fermions and interacting fermions respectively. It is
interesting to see that $R$ is oscillatory both in real space and in
the particle-number constrained Hilbert space. This oscillation never
stops and can be observed in the other dynamical quantities (such as
calculated in the main text) as well. Similar recurring behavior has
been reported in a study of interacting bosons in the clean diamond
chain~\cite{di2019nonlinear}. The full dynamics of $R$ for $\Delta=0$
is compared with that for small $\Delta=0.01J$ for a single particle,
noninteracting fermions and interacting fermions in
Fig.~\ref{app_dyn}(d),(e) and (f) respectively. However the
oscillatory behavior is not visible in these figures due to the
variable step size in time and hence lack of data points. For
$\Delta=0.01J$ the amplitude of the oscillation keeps decreasing and
vanishes at some point as $R$ saturates to a finite value for all
three cases.

\section{Spectrum-averaged entanglement entropy for the interacting system}\label{appB}
\begin{figure}
  \includegraphics[width=0.75\columnwidth]{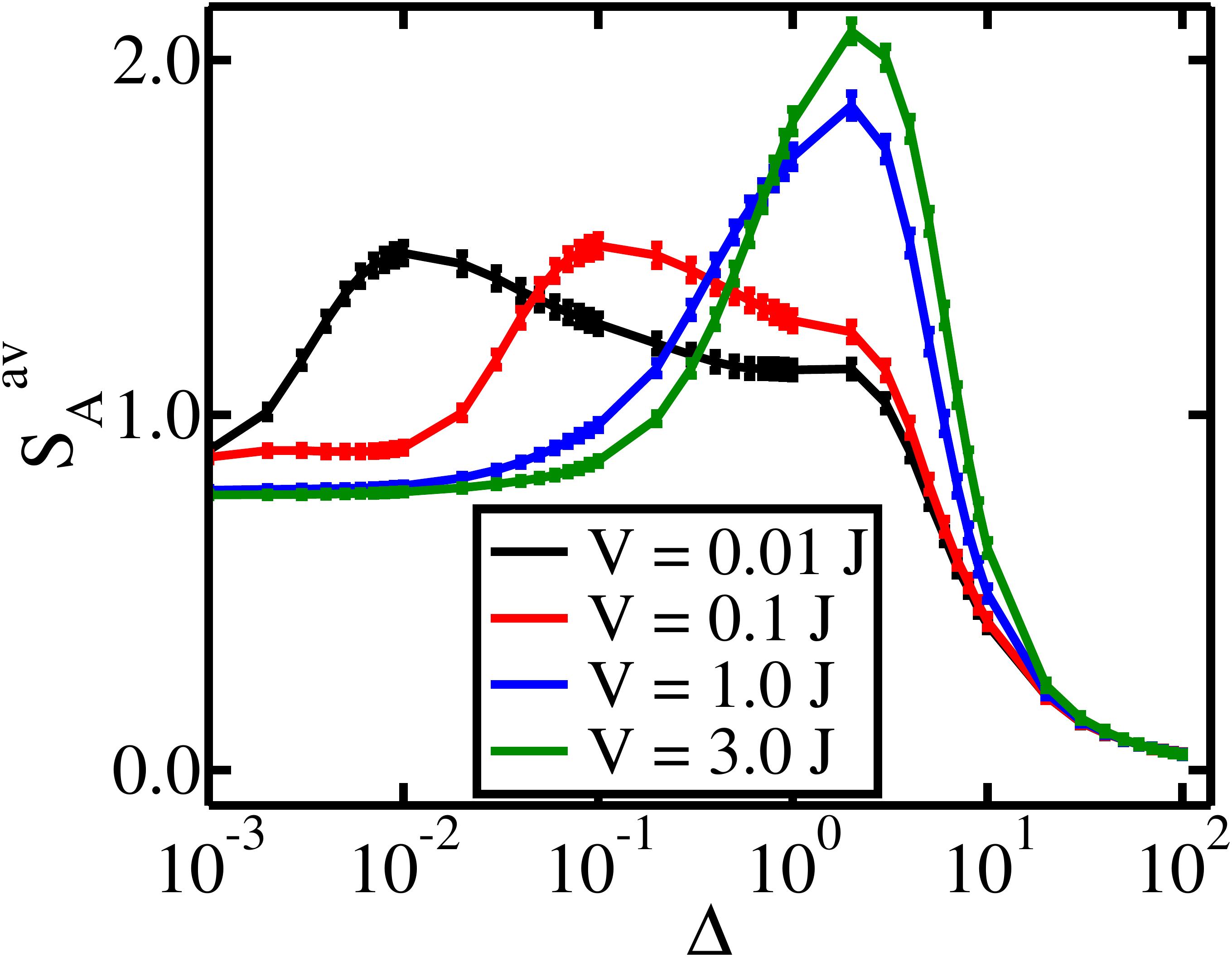}
  \caption{ The spectrum averaged entanglement entropy $S_A^{av}$ as a function of disorder strength $\Delta$ (in units of $J$) for increasing values of $V$ for $N=18$ and $\nu=1/6$. For all the plots, number of disorder realizations is $100$.}
  \label{app_ee}
\end{figure}

Here, we discuss the half-chain entanglement entropy $S_A^{av}$,
averaged over all the eigenstates. It is a useful quantity to help characterize different many-body
phases in the presence of disorder $\Delta$ and interaction
strength $V$. In Fig.~\ref{app_ee} the $S_A^{av}$ vs $\Delta$ plots
are shown for increasing values of $V$ for $N=18$ and $\nu=1/6$.  In the
delocalized phase, $S_A^{av}$ should attain a higher value as compared to
that for the localized phase. From the figure, we see the presence of
three phases: $MBL$ phase for $\Delta>>J$, mixed phase for $\Delta<<J$
and delocalized phase for intermediate $\Delta=\mathcal{O}(J)$ when
$V=\mathcal{O}(J)$ . For $\Delta<<J$ a delocalizing effect is found
when $V\approx\Delta$. All these results complement the results
obtained from $MIPR$ and $r$ as discussed in the main text. Also the
plots of $S_A^{av}$ resemble the plots in Fig.~\ref{int_eedyn}(b).

\section{Further insights into the entanglement growth in the MBL phase}\label{app_MBL}
\begin{figure}
\stackunder{\includegraphics[height=3.3cm,width=4.275cm]{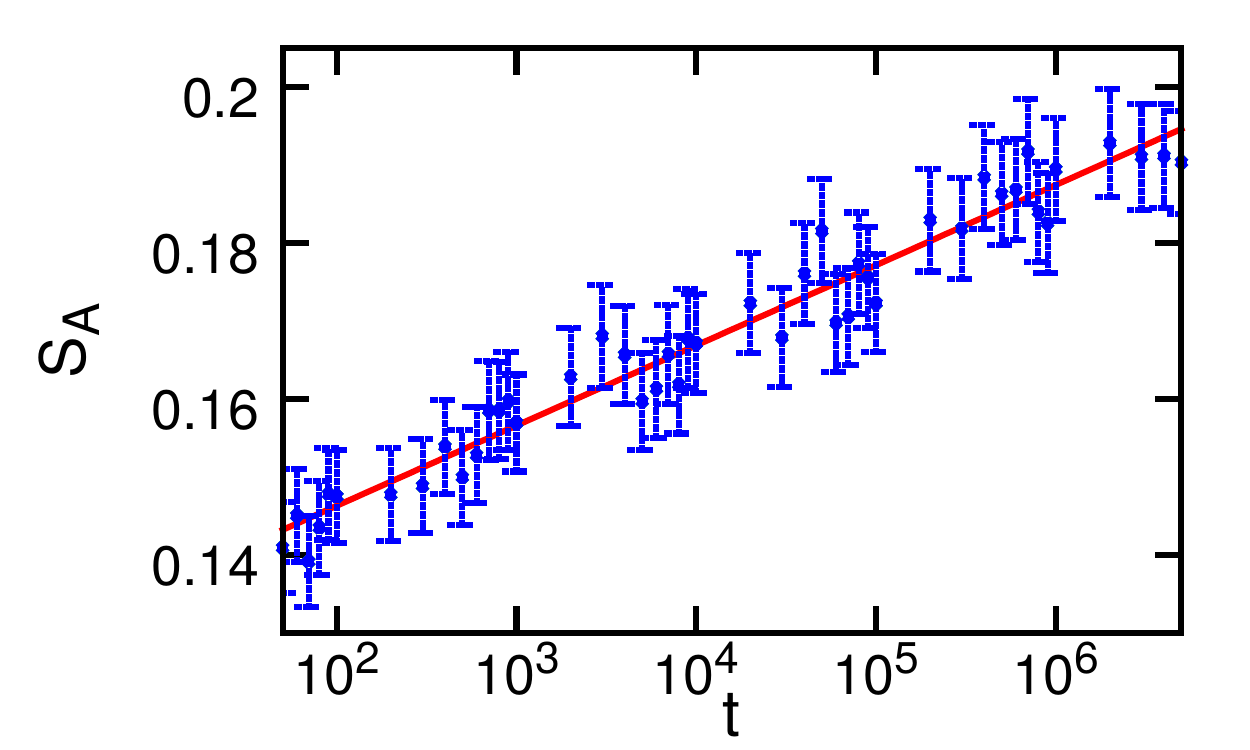}}{(a)}
\stackunder{\includegraphics[height=3.3cm,width=4.275cm]{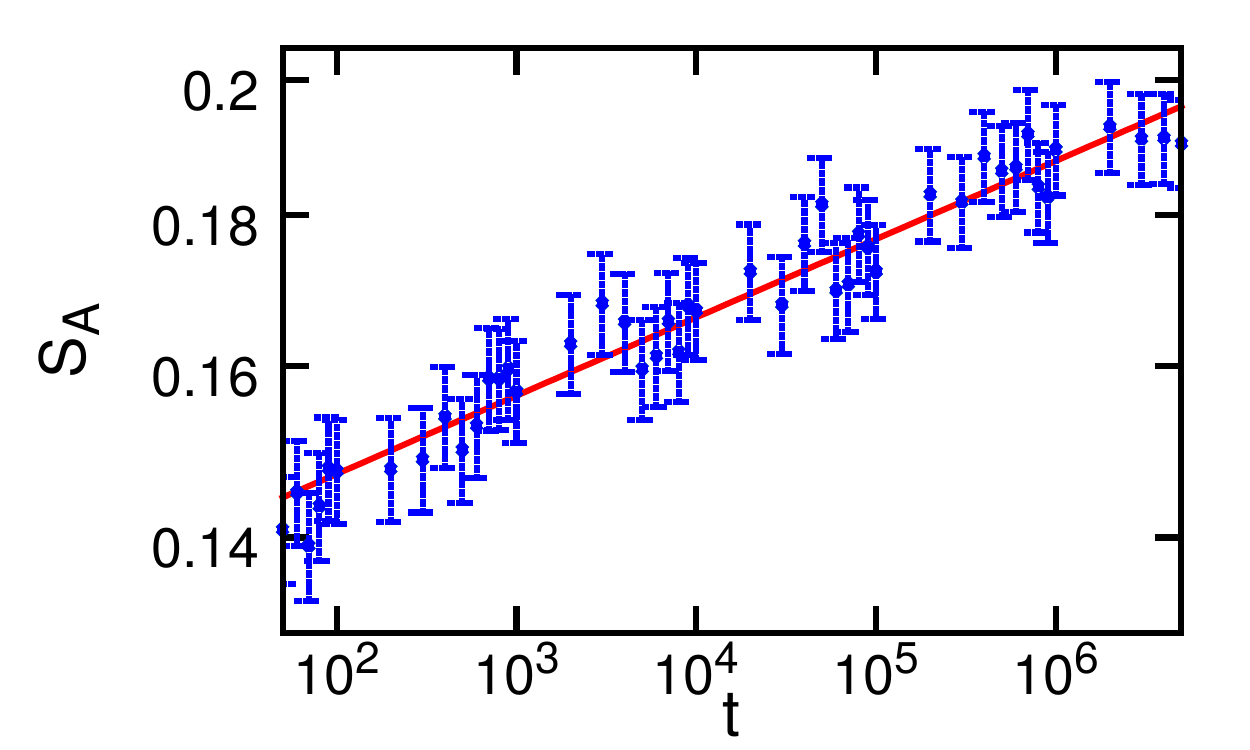}}{(b)}
\caption{Plots of entanglement entropy ($S_A$) growth with time (in units of $J^{-1}$) in the MBL phase, fitted with (a) $y=a\ln(x) + b$ with $a=0.00447; b=0.12569$ and (b) $y=ax^b$ with $a=0.13022; b=0.02650$.  Fitted curves are shown in solid red lines whereas the original data-points are shown in blue color. Figure (a) is a semi-log plot whereas figure (b) is a log-log plot. For all the plots, the disorder strength $\Delta=50.0J$, interaction strength $V=1.0 J$ for $N=18$ and $\nu=1/6$ with number of disorder realizations used being $1000$.}
\label{app_mbl}
\end{figure}

 We fit the curve representing the growth of
  entanglement entropy $S_A$ from an initial product state of the
  system in the MBL phase. The parameters $\Delta=50.0 J$, $V=1.0 J$,
  $N=18$ and $\nu=1/6$ are chosen in the regime where the
  level-spacing ratio $r\approx0.386$ establishing the MBL phase.  For
  a system of the spinless fermions with nearest-neighbor interactions
  in a one-dimensional linear chain, $S_A$ grows logarithmically with
  time in the MBL
  phase~\cite{vznidarivc2008many,abanin2019colloquium}. Here we have
  obtained the time dependence of $S_A$ for the system in the quasi
  one-dimensional diamond chain. From Fig.~\ref{app_mbl}(a), the
  growth of $S_A$ is fitted with a logarithmic function $y=a\ln(x) +
  b$ with $a=0.00447; b=0.12569$. One cannot rule out the possibility
  of obtaining a more-than-logarithmic dependence for systems having
  higher dimensions. We also attempted a power-law fit for the same
  data (Fig~\ref{app_mbl}(b)): $y=ax^b$ with $a=0.13022; b=0.02650$
  with the power-law exponent being tiny. Recently in
  Ref.~\cite{wiater2018impact,doggen2020slow} the possibility of MBL
  in quasi-1D has been indicated. This appendix is an attempt towards
  investigating the dynamical signatures of the quasi-1D MBL. While
  our results are certainly consistent with MBL, much larger system
  sizes would be required to conclusively address the question of
  whether logarithmic or power-law behavior holds. The consequences
  of the simultaneous presence of interaction and higher-dimensionality
  effects deserve a thorough separate investigation.

\section{Normalized participation ratio}\label{app_NPR}
The dependence of time evolved quantum many-body state on system size
in the long-time limit has turned out to be handy to decipher
many-body phases ~\cite{iyer2013many}. In this section, we stick to
the filling fraction $\nu=1/3$ and consider the initial state
$\ket{\Psi_{in}}=\prod\limits_{i=1}^{N/3}\hat{c}_{i}^\dagger\ket{0}$. Any
time evolved state can then be generically written as
$\ket{\Psi(t)}=\sum\limits_{n=1}^{D} C_n(t)\ket{n}$, where $\ket{n}$
stands for the $n\textsuperscript{th}$ configuration and $D$ is the size
of the Hilbert space. The normalized participation ratio $(NPR)$ of
the many-body state in the long-time limit $(t\rightarrow \infty)$ is
defined as~\cite{iyer2013many}
\begin{eqnarray}
\eta=\frac{1}{D \sum_{n} |C_n(t\rightarrow \infty)|^4}.
\label{npr}
\end{eqnarray}
In the ergodic phase phase, $\eta$ must be independent of system size
$N$ whereas in the non-ergodic phase, $\eta$ must depend on $N$. It
has been shown that $\eta\propto e^{-\kappa N}$ with
$\kappa\approx0.5$ for the many-body localized phase in a
quasi-periodic chain~\cite{iyer2013many}.

\begin{figure}
\stackunder{\includegraphics[height=3.3cm,width=4.275cm]{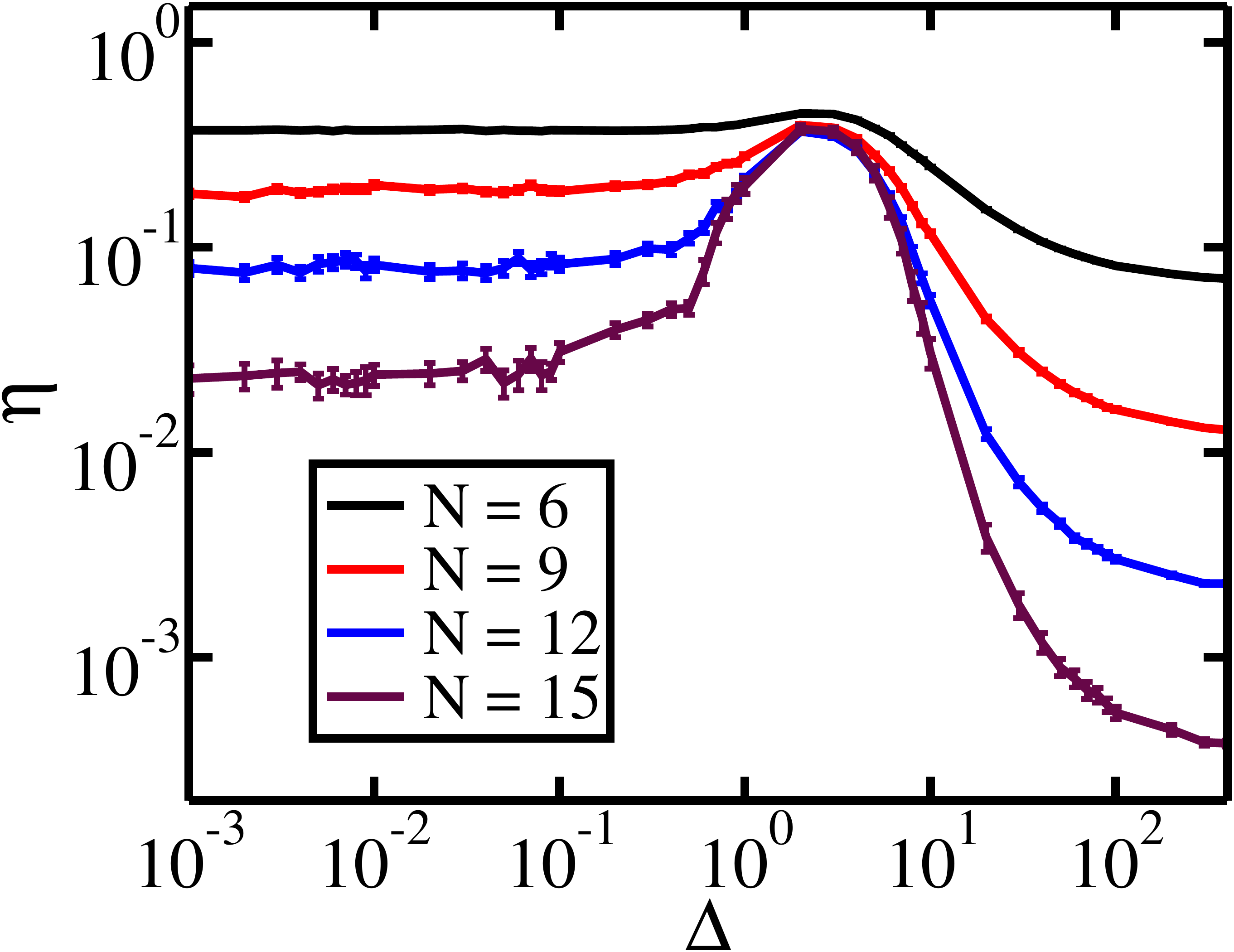}}{(a)}
\stackunder{\includegraphics[height=3.3cm,width=4.275cm]{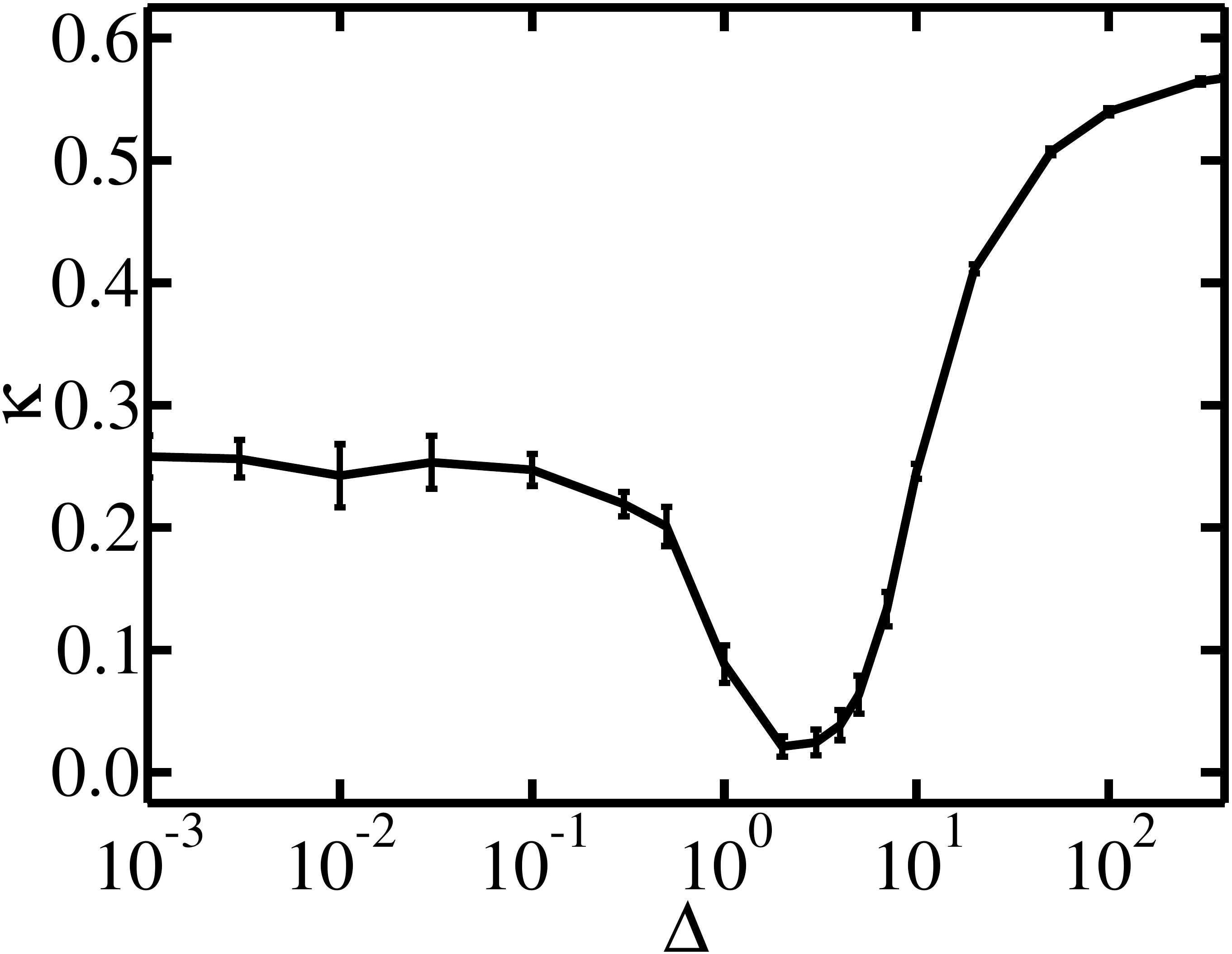}}{(b)}
\caption{ The scaling analysis of satuaration of $NPR$. (a) A log-log plot showing the values of $\eta$ as a function of disorder strength $\Delta$ (in units of $J$) for increasing system size $N$. (b) A semi-log plot of the scaling exponent $\kappa$ as a function of $\Delta$ (in units of $J$). For all the plots the interaction strength $V=1.0 J$ and filling fraction $\nu=1/3$. The number of disorder realizations varies for different system sizes, but all of them have at least $100$ samples of disorder.}
\label{app_npr}
\end{figure}

In Fig.~\ref{app_npr}(a) we show the dependence of
  $\kappa$ on disorder strength $\Delta$ for fixed interaction
  strength $V=1.0 J$ and increasing system size $N$. In the large and
  small disorder regimes, $\eta$ changes with system size, while for
  the intermediate range of $\Delta$, $\eta$ seems to be system size
  independent. For each value of $\Delta$, the exponent $\kappa$ can
  be extracted using the relation $\eta\propto e^{-\kappa N}$. The
  exponent $\kappa$ as a function of $\Delta$ is shown in
  Fig.~\ref{app_npr}(b). In the large $\Delta$ regime,
  $\kappa\approx0.5$ implying the nonergodic many-body localized phase
  whereas for intermediate values of $\Delta$, $\kappa\approx0$ implying
  the ergodic phase. In the small $\Delta$ regime $\kappa\approx0.25$
  which definitely is a sign of non-ergodicity. However, since
  $\kappa$ is not as high as $0.5$ this `mixed' phase
  may not be as nonergodic as the $MBL$. In addition, we have verified that the scaling analysis
  of the spectrum averaged $NPR$ in the static case leads to similar results as shown in Fig.~\ref{app_npr} .
  Also, the analyses reported in the main text point towards a similar 
  conclusion. The low-disordered nonergodic phase could be a remnant of a new 
  kind of MBL phase recently found in the clean limit of such 
  systems~\cite{danieli2020many}.

\bibliography{refs}

\end{document}